\documentclass[twocolumn,aps,pre,showpacs,amsmath,amssymb]{revtex4-1}
\usepackage{epsfig}
\usepackage{graphicx}
\usepackage{dcolumn}
\usepackage{bm}
\usepackage[colorlinks=true,dvipdfm]{hyperref}

\begin{document}
\bibliographystyle{nature}
\title{Phases fluctuations, self-similarity breaking and anomalous scalings in driven nonequilibrium critical phenomena}
\author{Weilun Yuan}
\author{Fan Zhong}  \email{Corresponding author: stszf@mail.sysu.edu.cn}
\affiliation{State Key Laboratory of Optoelectronic Materials and Technologies, School of
Physics, Sun Yat-sen University, Guangzhou 510275, People's
Republic of China}
\date{\today}

\begin{abstract}
We study in detail the dynamic scaling of the three-dimensional (3D) Ising model driven through its critical point on finite-size lattices and show that a series of new critical exponents are needed to account for the anomalous scalings originating from breaking of self-similarity of the so-called phases fluctuations. Considered are both heating and cooling which are found to exhibit qualitatively different behavior in a finite-time scaling (FTS) regime. Investigated are two sets of observable quantities, the order parameters and their fluctuations either using or not using absolute values, which may display different behavior. We find that finite-size scaling (FSS) at fixed driving rates and FTS on fixed lattice sizes are only satisfied for one set of the observables in their respective scaling regimes in conformity with the standard theory, provided that other sub-leading contributions and corrections to scaling can be ignored. They are violated for the other set of the observables even in their respective scaling regimes. However, if a scaled variable which determines the scaling regimes is fixed, the violated scalings are completely restored.

In order to account for these results, we separate critical fluctuations into magnitude fluctuations, which associated with the forming of large clusters of spin-up and spin-down phases, and the phases fluctuations that are the flipping of the large clusters. Moreover, the self-similarity of the phases fluctuations is divided into intrinsic self-similarity originating from the criticality and extrinsic self-similarity imposed by fixing the external scaled variable. Further, the two observables in a set that violates the scaling exhibit different leading behavior and are divided into primary and secondary observables. The breaking of the extrinsic self-similarity introduces four breaking-of-(extrinsic-)self-similarity exponents that are responsible for the violations of either FSS or FTS in either heating or cooling. These exponents lead to different leading behavior of the primary observables in either the ordered phase or the disordered phase where the scalings are violated, in stark contrast to equilibrium critical phenomena in which the leading behavior of the two phases is identical, only their amplitudes differ. The exponents are found to be different for heating and cooling and in FTS and FSS and have different expressions in the 2D and the 3D Ising models, except for FSS in cooling. This implies that they are most likely exponents that are different from the known extant ones but produce the found 2D and 3D values. Crossovers from the self-similarity-breaking-controlled regimes to the usual FSS and FTS regimes are also discussed.

We also study cooling in an externally applied field whose magnitude fixes a scaled variable pertinent to it. The field is applied in three protocols in which either only below or only above the critical point besides during the whole process. The most unexpected result discovered is that the breaking-of-(extrinsic-)self-similarity exponent of FTS in heating describes remarkably well the cases of field cooling in which scaling is poor. In addition, the scaling behavior of the three different protocols shows that the phases fluctuations at and just above the critical point are more important than those below it. A sufficient order can suppress the phases fluctuations in cooling and acts like the heating with an ordered state. Therefore, the essence of the difference between heating and cooling is the symmetry of the system states, namely whether they are ordered or disordered. We also confirm that there exists a revised FTS regime---the regime in which both the lattice sizes and driving rates are indispensable---in between the two end regimes of FTS and FSS even in the presence of an external field. Its effect on the scaling is revealed and its crossover to an FTS regime in a field is estimated. Further, our results show that the revised FTS is never needed for the usual order parameter without an absolute value. Besides, difference between the FTS regime in a field on large lattices in cooling and in heating is revealed.

In addition, from the quality of entire curve collapses for different dynamic critical exponent $z$ in heating and in the absence of an external field, we find that the 3D $z$ values in heating and in cooling appear identical but the 2D ones different.

Our results demonstrate that new exponents are generally required for scaling in the whole driven process once the lattice size or an externally applied field are taken into account. These open a new door in critical phenomena and suggest that much is yet to be explored in driven nonequilibrium critical phenomena.
\end{abstract}

\maketitle
\section{\label{intro}Introduction}
Experimental advances in manipulating real-time evolution of ultracold atoms~\cite{Greiner,Kinoshita,Hofferberth,Zhang} have stimulated renewed interest in driven nonequilibrium critical phenomena~\cite{Feng}. The well-known Kibble-Zurek (KZ) mechanism studies nonequilibrium topological defect formations during a continue phase transition. First proposed in cosmology~\cite{KZ1,Kibble2} and then applied in condensed-matter physics~\cite{KZ2,KZ3}, the KZ mechanism divides the cooling process from a disordered phase to a symmetry-broken ordered phase into adiabatic, impulse, and another adiabatic stages. In the nonequilibrium impulse stage, the evolution of the system is assumed to be frozen. The correlation length of the system then ceases to grow exactly at the boundary between the first adiabatic stage and the impulse stage. This frozen correlation length thus determines the density of the topological defects formed during the transition. Using just equilibrium relations of the correlation length and correlation time in critical phenomena, one can find the dependence of the frozen correlation length and hence the defect density on the cooling rate. This relationship is known as KZ scaling. Although it seems to agree with numerical simulations~\cite{KZp1,KZp2,KZp3,KZp4,KZp5,delcamp,Gomez}, most experimental results require additional assumptions for interpretation of their consistency with the theory~\cite{inexper4}. This leads to a reconsideration of the process from a statistical context.

One way is to consider phase-ordering kinetics~\cite{Bray}. This occurs when a system is quenched instantaneously from a disordered phase at high temperatures into a two-phase coexistence region at low temperatures. It has been found that at long times when the characteristic size of the ordered phases is large enough, this size scales with time with a dynamic ordering exponent that is different from the dynamic critical exponent. Upon connecting the two kinds of dynamics, it was concluded that phase ordering was important below the critical temperature~\cite{Biroli}.

In statistical physics, the divergent correlation time in critical phenomena results in the notorious critical slowing down, which is a stringent situation for accurate estimates of critical properties. Upon noticing the spatial analogue of the divergent correlation length and the well-known finite-size scaling (FSS)~\cite{fss,Barber,Cardy,Privman} to circumvent it, finite-time scaling (FTS) was independently proposed~\cite{Zhong1,Zhong2} on the basis of a renormalization-group theory for driving~\cite{Zhong06}. In this theory, the rate of the linear driving, which drives the system through its critical point, introduces a finite timescale that serves as the temporal analogue of the system size in FSS. Moreover, the system itself can be driven out of equilibrium, because the finite driving timescale becomes inevitably shorter than the diverging correlation time once the system is close enough to its critical point. The renormalization-group theory of such driven nonequilibrium critical phenomena can be generalized to a weak external driving of an arbitrary form and a series of nonequilibrium phenomena, such as negative susceptibility and competition of various equilibrium and nonequilibrium regimes and their crossovers, as well as the violation of fluctuation-dissipation theorem and hysteresis, arise from the competition of the various timescales stemming from the parameters of the driving~\cite{Feng}. FTS has also been successfully applied to many systems both theoretically~\cite{Yin,Yin3,Zhong2,Liu,Huang,Liupre,Liuprl,Pelissetto,Xu,Xue,Feng,Cao,Gerster,Li,Mathey} and experimentally~\cite{Clark,Keesling}. Equilibrium and nonequilibrium initial conditions have also been considered~\cite{Feng,Huang2}. Applying to the KZ mechanism, one finds that the impulse stage is just the FTS regime, in which the driven timescale is shorter than the correlation time, and that the KZ scaling results from a special value of the scaling function which describes the entire process~\cite{Huang}.

FTS works well in all the three stages of the driving, although the adiabatic stages obey the usual (quasi-)equilibrium critical scaling governed by the equilibrium correlation length and time while the diabatic impulse stage satisfies the nonequilibrium scaling controlled by the driven length and time. Accordingly, a first question is whether the phase ordering or the FTS dominates the cooling process through a critical point. If the phase ordering matters, it seems that one has to introduce a new exponent, the dynamic ordering exponent, to the cooling process~\cite{Biroli}. This then leads to another more fundamental question.

So far, in the KZ scaling in particular and the driven nonequilibrium critical phenomena in general, the equilibrium critical exponents including the dynamic critical exponent have been found to characterize the scaling well, apart from the possible effect of the phase ordering. No new exponents surprisingly appear to be needed to describe the apparent driven nonequilibrium process which induce a lot of nonequilibrium phenomena~\cite{Feng} including the KZ mechanism for nonequilibrium topological defect formation. Therefore, a fundamental question is that whether this is true or not.

A seemingly possible case appeared when one extended the KZ scaling to a finite-sized system~\cite{Liu}. In contrast with the previous theoretical results of FTS in a finite-sized system~\cite{Zhong2}, a special scaling of a magnetization squared with a complicated exponent was suggested and verified for the two-dimensional (2D) Ising model exactly at its critical point~\cite{Liu}. However, this exponent was shown to be that of the susceptibility and a revised FTS form was proposed and confirmed within the FTS theory~\cite{Huang}. Moreover, the order parameter and its squared at the critical point were found to exhibit distinct scalings in heating and in cooling, though the susceptibility did not~\cite{Huang}. Therefore, no new exponents are needed again. Still, there exist some peculiar features~\cite{Huang}. The dynamic critical exponent $z$ estimated from data collapses right at the critical point in heating and in cooling was also found to be different. In addition, an externally applied magnetic field which lifts the up-down symmetry is expected to suppress the revised FTS. However, it was found to persist surprisingly for a small field in the 2D Ising model.

In a recent letter~\cite{Yuan}, through studying the whole driving dynamics and hence the whole scaling functions rather than just at the critical point of the 2D Ising model, we discovered that the FTS and the FSS of some observable quantities are violated either in heating or in cooling even in their respective FTS and FSS regimes. Such violations of scaling were found to originate from a novel source, the self-similarity breaking of the so-called phases fluctuations. Note the plural form of phase used both to emphasize that at least two phases are involved owing to the symmetry breaking and to distinguish it from the usual phase of a complex field. New breaking-of-self-similarity, abbreviated as Bressy, exponents are then needed. Moreover, these exponents lead to different leading critical exponents for the disordered and the ordered phases rather than identical leading critical exponents but different amplitudes in the usual critical phenomena.

Symmetry breaking is well known and plays a pivotal role in modern physics. Self-similarity is a kind of symmetry and appears ubiquitously in nature~\cite{Mandelbrot,Meakin}. In contrast to rigorous mathematical objects such as fractals that are self-similar on all scales~\cite{Mandelbrot}, in nature, self-similarity holds inevitably only within a certain range of scales~\cite{Meakin}. This might be regarded as a certain kind of self-similarity breaking with the only consequence of a self-similarity limited to a certain range of scales. Our results thus show that self-similarity can indeed be broken with significant consequences~\cite{Yuan}.

Here, using the 3D Ising model, we elaborate on the idea of the phases fluctuations and their self-similarity and its breaking along with division of critical fluctuations into phases fluctuations and magnitude fluctuations, the observables that violate scaling into primary and secondary observables, and the self-similarity into intrinsic and extrinsic self-similarities, show explicitly the observables that satisfy FSS under driving and FTS on finite-size lattices and the observables that violate those scalings in the same regime, and provide details in estimating the four Bressy exponents for FTS and FSS in heating and in cooling from the different behavior of the observables that violate the scalings. Besides, qualitatively different behavior of the fluctuations in cooling and in heating is revealed. However, once self-similarity of the phases fluctuations is in place, the peculiar cooling behavior disappears. Moreover, both FTS and FSS are good down to quite low temperatures in cooling with self-similarity. These indicate that phase ordering cannot be the origin of the peculiar behavior and can only have an effect at further lower temperatures.

We also study the effects of an externally applied magnetic field on the phases fluctuations in cooling using three different protocols in applying the field. The most unexpected result discovered is that the Bressy exponent of FTS in heating describes remarkably well the cases of field cooling in which scaling is poor, noticing that cooling and heating both in zero field are described by completely different Bressy exponents. In addition, the different scaling behavior of the three different protocols shows that the phases fluctuations at and just above the critical point are more important than those below it. A sufficient order can suppress the phases fluctuations in cooling and acts like the heating with an ordered state. Therefore, the essence of the difference between heating and cooling is the symmetry of the system states, namely whether they are ordered or disordered. We also confirm that there exists a revised FTS regime in between the two end regimes of FTS and FSS even in the presence of an external field. Its effect on the scaling is revealed and its crossover to the usual FTS regime is estimated. Further, our results show that the revised FTS is never needed for the usual order parameter without an absolute value. Besides, difference between the FTS regime in a field on large lattices in cooling and heating is uncovered. In addition, from the quality of the entire curve collapses for different dynamic critical exponent $z$ in heating and in the absence of an external field, we find that the 3D $z$ values in heating and in cooling appear identical but the 2D ones different.

\section{Guide to the content}
As the paper is rather long, we provide a somehow detailed layout as a guide to the content in this section. Some symbols appear here will be defined in the following sections in order not to distract the attention on the content.

In Sec.~\ref{theory}, we first review the comprehensive theory for both FTS and FSS and their combined effects owing to the phases fluctuations. The resultant revised FTS for cooling is generalized to a weak externally applied magnetic field. Spatial and temporal self-similarities are specified and observables that violate scaling are classified into two catalogs. The new Bressy exponents are defined for the primary observables that obey a pure power-law. Various kinds of behavior of the secondary observables that possess regular terms are presented.

In Sec.~\ref{phfl}, we clarify the phases fluctuations, their relation to the usual critical fluctuations, and their self-similarity. Critical fluctuations are separated into magnitude fluctuations and the phases fluctuations. The self-similarity of the phases fluctuations is further divided into two classes. The extrinsic self-similarity of the phases fluctuations can be broken by external conditions such as system sizes and external fields and are the primary focus in this paper.

In Sec.~\ref{model}, we define the model and the two sets of observables we measure and provide simulation details. We focus on the 3D Ising model. Only in three places, Figs.~\ref{conf},~\ref{ftsc3d}(d) and~\ref{heatz2d}, are results from the 2D model displayed, the first for the ease of presentation and the last two for comparison.

In Secs.~\ref{FSS}--\ref{nr of ordering}, we present a vast amount of numerical results in a series of figures to examine the theory, to estimate the Bressy exponents, and to differentiate the possible $z$ values in heating and cooling. These results are organized according to whether an externally applied field is absent (Secs.~\ref{FSS} and~\ref{nr of ordering}) or is present (Sec.~\ref{FTSwh}). We employ all known critical exponents for the examination. The only ones that need to define and estimate are the Bressy exponents.

First, in Sec.~\ref{FSS}, as an appreciation of richness of the driven nonequilibrium critical phenomena, we first display the evolution of one set of the observables in the FTS regime in heating and cooling, which exhibits qualitatively different behaviors. Then, to test the theory in the absence of the field, we study sequentially FSS at fixed rates $R$ (Sec.~\ref{fssafr}), FTS on fixed lattice sizes $L$ (Sec.~\ref{ftsofls}), and the full scaling forms (Sec.~\ref{fsf}) by fixing a scaled variable $L^{-1}R^{-1/r}$ or $RL^r$ both in heating and in cooling. In Secs.~\ref{fssafr} and~\ref{ftsofls}, two fixed rates and two fixed lattice sizes are respectively utilized to compare their effects on the scalings, while in Sec.~\ref{fsf}, two fixed values of $L^{-1}R^{-1/r}$, one in the FTS regime and the other in the FSS regime, in heating are considered, whereas in cooling, only the fixed value in the FTS regime is shown, the other has already appeared in Ref.~\cite{Yuan}. Further demonstrated here is the equivalence of FTS and FSS upon fixing $L^{-1}R^{-1/r}$ or $RL^r$ according to the theory. The cases and their details that violate either FSS or FTS in either heating or cooling are summarized in Sec.~\ref{sumvio} and the exponents originate from the Bressy are extracted in Sec.~\ref{bressy} through first identifying the primary observables and then curve collapsing.

Second, in Sec.~\ref{FTSwh}, to reveal the effect of an externally applied field on the phases fluctuations in cooling, we apply the external field $H$ in such a way that $X=HR^{-\beta\delta/r\nu}$ is fixed and in three protocols. In protocol A, Sec.~\ref{hatTca}, the external field is switched on at the critical temperature $T_c$ till the end of the cooling, while in protocol B, Sec.~\ref{hatTcb}, the field is only applied from the beginning of cooling to $T_c$. Only in protocol C, Sec.~\ref{hatTcc}, is the external field is present in the whole cooling process. In each protocol, the FTS of the observables for two $X$ values, $X=2$ and $X=0.05$, is studied. The cases in which scaling is violated are dealt with in Sec.~\ref{cross}, where the crossover from the revised FTS regime to an FTS regime in a field is studied and the Bressy exponent of FTS in heating is invoked to remedy the poor scaling in field cooling. In sec.~\ref{ftshl}, we investigate the scaling behavior of the FTS regime in a field on large lattices. Brief results for heating in a field are presented in Figs.~\ref{hmlrh} and~\ref{ftshh} for comparison.

Third, in Sec.~\ref{nr of ordering}, we study the scaling collapses in the whole heating process instead of just at the critical point in the absence of an external field to investigate the effect originating from the different dynamic critical exponents in heating and cooling.

Section~\ref{sum} contains a detailed summary. A lot of problems for further study are listed.

\section{\label{theory}Theory}
In this section, we will first review the comprehensive theory for both FTS and FSS and their combination, the revised FTS, together with the crossovers between them. Then, it is generalized to account for additional features such as an external field and the new exponents. First, the revised FTS is generalized to the case of a weak externally applied field and sufficiently large lattice sizes to account for an intermediate revised FTS regime between the FSS and FTS regimes for small and large lattice sizes, respectively. Second, we specify the spatial and temporal self-similarities of the phases fluctuations in the theory by extending the picture of self-similarity~\cite{Li} in space to real time. The Bressy exponents are then introduced to rectify the violated scalings when the self-similarity is broken. In this regard, the observables that violate scalings are classified into two catalogs, the primary and the secondary observables, whose behavior is derived.

Consider a system with a lateral size $L$ that is driven from one phase through a critical point to another phase by changing the temperature $T$ with a rate $R\geq0$ such that
\begin{equation}
T-T_c\equiv\tau=\pm Rt,
\label{rate}
\end{equation}
where $T_c$ is the critical point and the plus (minus) sign corresponds to heating (cooling). We have set the zero time at the critical point for simplicity. Accordingly, the time $t$ can be both negative and positive.
To derive the scaling behavior, it is helpful to begin with the scaling hypothesis for the susceptibility $\chi$
\begin{equation}
\chi(\tau, R, L^{-1}, H)=b^{\gamma/\nu}\chi(\tau b^{1/\nu}, Rb^{r},L^{-1}b, Hb^{\beta\delta/\nu}),
\label{RG}
\end{equation}
where $b$ is a scaling factor, $\gamma$, $\beta$, $\delta$, $\nu$, and $r$ are the critical exponents for the susceptibility, the order parameter $M$, the field $H$ conjugated to $M$, the correlation length, and the rate, respectively. In the absence of either $R$ or $L$, Eq.~(\ref{RG}) can be obtained from the renormalization-group theory~\cite{FSS1,FSS2,Zhong06,Zhong1,Zhong2}. Here, we simply combine them as an ansatz as usual~\cite{Zhong2}. We have also replaced the time $t$ with $R$ because they are related by Eq.~(\ref{rate}). Moreover, from the same equation, we find~\cite{Zhong}
\begin{equation}
r=z+1/\nu
\label{rzv}
\end{equation}
because $t$ transforms as $tb^{-z}$ with $z$ being the dynamic critical exponent~\cite{Hohenberg}.

From Eq.~(\ref{RG}), various scaling forms can be readily derived by choosing proper scale factors~\cite{Zhong06,Zhong1,Zhong2,Huang}.
On the one hand, setting $b = R^{-1/r}$ leads to the FTS form
\begin{equation}
\chi=R^{-\gamma/r\nu}\mathcal{F}_T (\tau R^{-1/r\nu},L^{-1} R^{-1/r},HR^{-\beta\delta/{r\nu}}),
\label{FTSX}
\end{equation}
with $\mathcal{F}_T$ being a universal scaling function. On the other hand, assuming $b =L$ in Eq.~(\ref{RG}) results in
\begin{equation}
\chi=L^{\gamma/\nu}\mathcal{F}_S(\tau L^{1/\nu},RL^r, HL^{\beta\delta/\nu} ),
\label{FSSX}
\end{equation}
which is the FSS form under driving, where $\mathcal{F}_S$ is another scaling function. Note that we have neglected constants in the scaled variables of the scaling functions for simplicity.

With the scaling forms Eqs.~(\ref{FTSX}) and~(\ref{FSSX}), various regimes controlled by their dominated length scales can be defined. The FTS scaling form Eq.~(\ref{FTSX}) dominates if the scaling function $\mathcal{F}_T$ is analytic when its scaling variables are vanishingly small~\cite{Zhong06,Zhong1,Zhong2,Huang}. This implies $R^{-1/r}\ll|\tau|^{-\nu}$, $R^{-1/r}\ll L$, and $R^{-1/r}\ll H^{-\nu/\beta\delta}$. These relations dictate that the driving length scale $\xi_R\sim R^{-1/r}$, be the shortest among the correlation length $\xi\sim |\tau|^{-\nu}$, the system size $L$, and the effective length scale induced by the externally applied field $\xi_H\sim H^{-\nu/\beta\delta}$ in the FTS regime. Because the correlation time $\zeta$ is asymptotically proportional to $\xi^z$, viz., $\zeta\sim\xi^z$, associated with a finite rate $R$ is then a driven timescale $\zeta_R\sim R^{-z/r}$ beyond which the driving changes appreciably and thus comes the name of FTS. Accordingly, the above conditions can be rephrased as the driven timescale is the shortest among the other long timescales. Similarly, the FSS regime requires $L$ is the shortest among the other long length scales. One can readily envision other regimes governed by the equilibrium correlation length and the external field, which are the quasi-equilibrium regime and the field-dominated regime, respectively.

Moreover, properties of the regimes can be deduced. Because the scaling functions are analytic at vanishingly small scaled variables, the leading behavior of each regime is thus determined by the factor in front of the scaling function. Therefore, in the FTS regime, the leading behavior of $\chi$ is proportional to $R^{-\gamma/r\nu}$. This means that as $R$ decreases, $\chi$ must increase. When $R\rightarrow0$, equilibrium is recovered and $\chi$ diverges for $L=\infty$ and $H=0$. Moreover, $\chi R^{\gamma/r\nu}$ versus $L^{-1}R^{-1/r}$ at $\tau=0$ and $H=0$ must be a horizontal line in the FTS regime, because $L^{-1}R^{-1/r}$ can be neglected in the regime~\cite{Huang}. Similarly, in the FSS regime, the leading behavior of $\chi$ is $L^{\gamma/\nu}$ and so $\chi L^{-\gamma/\nu}$ versus $RL^r$ must be a horizontal line, in conformity with the negligibility of $RL^r$ in the regime. In the thermodynamic limit $L\rightarrow\infty$, $\chi$ again diverges for $R=0$ and $H=0$ as expected.

In addition, crossovers between different regimes and their characteristics can also be identified~\cite{Huang}. As all the regimes are governed by the same fixed point, every scaling form can also describe the other regimes besides its own one, though, in this case, the scaling function is singular~\cite{Huang}. As a result, all the scaling functions are related. For example, if $R$ or $L$ are changed such that $\xi_R$ becomes longer than $L$, the system crosses over from the FTS regime to the FSS regime and thus
\begin{equation}
\mathcal{F}_T =\left(L^{-1} R^{-1/r}\right)^{-\gamma/\nu}\mathcal{F}_S.\label{fttofs}
\end{equation}
This gives rise to a slope change from the FTS regime to the FSS regime~\cite{Huang}. Examples will be given below shortly.

All observables must show similar scaling forms to the susceptibility with their own exponents in the critical region. For example, the order parameter $M$ must behave
\begin{equation}
M=R^{\beta/r\nu}F_T (\tau R^{-1/r\nu},L^{-1} R^{-1/r},HR^{-\beta\delta/{r\nu}})
\label{FTSM}
\end{equation}
in the FTS regime, while in the FSS regime, it becomes
\begin{equation}
M=L^{-\beta/\nu}F_S(\tau L^{1/\nu},RL^r, HL^{\beta\delta/\nu} ),
\label{FSSM}
\end{equation}
where $F_T$ and $F_S$ are again scaling functions. They are related by $F_T=(L^{-1} R^{-1/r})^{\beta/\nu}F_S$. Accordingly, exactly at $\tau=0$ and $H=0$, $MR^{-\beta/r\nu}$ versus $L^{-1}R^{-1/r}$ ought to be a horizontal line for such $R$ and $L$ that $L^{-1}R^{-1/r}\ll1$, viz., in the FTS regime; whereas it changes to an inclined line with a slope of $\beta/\nu$ in the FSS regime in which $L^{-1}R^{-1/r}$ is large~\cite{Huang}.

The above scalings are all standard forms of FTS and FSS. However, it was found that, exactly at $\tau=0$ and $H=0$, Eqs.~(\ref{FTSM}) and (\ref{FSSM}) both give rise to different leading characteristics in heating and cooling in the FTS regime although they describe the scalings in both cases well~\cite{Huang}. In particular, the above slope and its change in the frame of $MR^{-\beta/r\nu}$ versus $L^{-1}R^{-1/r}$ is only valid in heating. In cooling, however, the slope in the FTS regime is not zero (horizontal) but $d/2$~\cite{Huang}, where $d$ is the spatial dimensionality. These different leading characteristics indicate different leading exponents for heating and cooling. To account for this behavior, an additional ingredient is needed.

This ingredient is the phases fluctuations. In the FTS regime, the system divides into regions of a size $\xi_R$ on average. During cooling from a disordered phase to a symmetry-broken ordered phase, the ordered direction of each such region fluctuates freely among all possible directions because of the absence of a symmetry breaking direction. Each direction is just one possible phase of the ordered phase and thus the fluctuations between different directions are just the phases fluctuations. The average of the magnetization of these regions ought to be vanished in the thermodynamic limit owing to the central limit theorem~\cite{Huang}. Therefore, in Eq.~(\ref{FTSM}), $L$ cannot be neglected no matter how small the scaled variable $L^{-1}R^{-1/r}$ is. To satisfy the central limit theorem, $F_T$ must behave as
\begin{equation}
F_{T}=(L^{-1}R^{-1/r})^{d/2}F_{TS}\label{ftfts}
\end{equation}
for small $L^{-1}R^{-1/r}$~\cite{Huang}, where $F_{TS}$ is anther scaling function. As a result,
\begin{equation}
M =L^{-d/2}R^{-\gamma/2r\nu}F_{TS}(\tau R^{-1/r\nu},L^{-1}R^{-1/r},HR^{-\beta\delta/{r\nu}}),
\label{rescaling_huang}
\end{equation}
upon application of the scaling laws~\cite{Mask,Fisher,Cardyb}
\begin{equation}
\alpha=2-d\nu,\qquad\alpha+2\beta+\gamma=2.\label{law}
\end{equation}
Comparing Eq.~(\ref{rescaling_huang}) with~(\ref{FTSM}), one sees that the leading FTS behavior of $M$ in cooling is now qualitatively different from heating. Because of Eq.~(\ref{ftfts}), the slope of $MR^{-\beta/r\nu}$ versus $L^{-1}R^{-1/r}$ is now $d/2$ instead of 0 in the FTS regime. In order to have a 0 slope, one has to use $ML^{d/2}R^{\gamma/2r\nu}$ at $\tau=0$ and $H=0$ from the revised FTS form, Eq.~(\ref{rescaling_huang}). In this way, the slope of the FSS regime then becomes $-d/2+\beta/\nu=-\gamma/2\nu$ in cooling~\cite{Huang}. Note that in the FSS regime $L$ is shorter than the correlation length in principle. Accordingly, the system itself is just one phase on average and thus no anomaly equivalent to Eq.~(\ref{rescaling_huang}) is needed for FSS.

In Eq.~(\ref{rescaling_huang}), we have generalized the previous equation valid at the critical point~\cite{Huang} to $\tau\neq0$ and $H\neq0$. This would appear at first sight to contradict with the condition of no symmetry breaking direction to affect the phases fluctuations. However, if the field strength is so small that its associated length scale $\xi_H\gg\xi_R$ and hence $HR^{-\beta\delta/{r\nu}}\ll1$, or $H\ll R^{\beta\delta/{r\nu}}$---a driving induced effective field, the real field $H$ can only have a negligible effect on the phases fluctuations. We will find in Sec.~\ref{cross} below that this is indeed true for $L\gtrsim\xi_H$ or $LH^{\nu/\beta\delta}\gtrsim1$. However, there is an important difference here. Equation~(\ref{rescaling_huang}) is only valid for $HR^{-\beta\delta/{r\nu}}\ll1$, the field-dominated regime. This is in stark contrast to the previous scaling forms that can describe both their own regimes and their opposite regimes when their scaled variables are small and large, respectively, as pointed out above.

However, if $L\gg\xi_H\gg\xi_R$ or $LH^{\nu/\beta\delta}\gg1$ and again $HR^{-\beta\delta/{r\nu}}\ll1$, the lattice size is too large to be important and its role is replaced by $\xi_H$. In this case, the sizes of the clusters of the phases fall well within the field correlation length $\xi_H$. The phase that directs along the field must win the competition among the phases. This lifts the equal probability of the fluctuating phases and thus the revised FTS is invalid. Therefore, there can exist three regimes in the case of a sufficiently weak external field in contrast to just two in its absence. This is an additional FTS regime for $LH^{\nu/\beta\delta}\gg1$ and $HR^{-\beta\delta/{r\nu}}\ll1$ besides the two regimes of the FSS for $LR^r\ll1$ and the revised FTS regime of $L^{-1}R^{-1/r}\ll1$ in the absence of the external field. A quantitative estimate of the crossover will be given in Sec.~\ref{cross} and the properties of the additional FTS regime are to be investigated in Sec.~\ref{ftshl}.

In practical verification of the scaling forms, when there exist more than one scaled variable in the scaling functions, one needs to fix the other variables for exactness. If they are not fixed, for an approximation, good scaling collapses can also be obtained if the leading behavior has been extracted and the other scaled variables are sufficiently small, provided that corrections to scaling~\cite{Wegner} that have not been considered are negligible. However, there are cases that such an approximate approach does not work no matter how small the other scaled variables are. The revised FTS introduced above is just such a case. Moreover, as revealed in Ref.~\cite{Yuan}, the standard FTS, Eq.~(\ref{FTSM}), and FSS, Eq.~(\ref{FSSM}), and even the revised FTS, Eq.~(\ref{rescaling_huang}) itself, all can surprisingly fall into this class in either heating or cooling. In these cases in which the phases fluctuations are strong, the other scaled variables have to be fixed. Note, however, that this is different to the case in which the scaled variables are not small and thus competition among the variables originating from several length scales has to be considered and high-dimensional parameter spaces have to be invoked~\cite{Cao}.

\begin{figure}
  \centerline{\epsfig{file=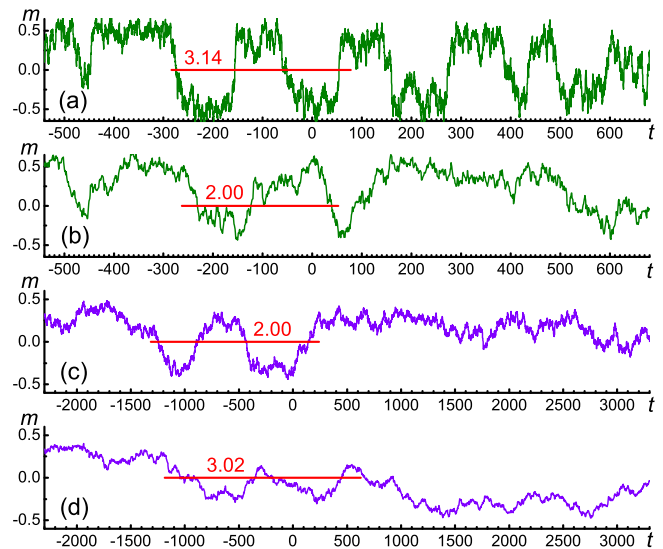,width=1.0\columnwidth}}
    \caption{\label{tss} Time evolution of $m$ for (a) $L=10$ and $R=0.000~03$, (b) $L=10$ and $R\approx0.000~136$, (c) $L=20$ and $R\approx0.000~011$, and (d) $L=20$ and $R=0.000~03$ of the 3D Ising model during heating, the values of (b) and (c) fixing $RL^{r}=0.0005\times7^{r}\approx0.598$, which are used in Fig.~\ref{ftshlr1} below, while those of (a) and (d) are used in Fig.~\ref{fssh403} below. The lengths of the red line segments, which are placed exactly at $m=0$, are the driven timescale $R^{-z/r}$ in (a), $4R^{-z/r}$ in (d) and $1.5R^{-z/r}$ in (b) and (c). The red numbers over the lines are the results of the corresponding timescales divided by $L^z$. $t=0$ at $T_c$. The abscissa scales are identical in (a) and (b) and in (c) and (d). (b) and (c) are temporally self-similarity while (a) and (d) are not.}
\end{figure}
Fixing a scaled variable fixes, in fact, the ratio of the two scales involved in the variable and therefore ensures self-similarity. If one fixes $L^{-1}R^{-1/r}$ for instance, one fixes the ratio between $L$ and $\xi_R$. In the FTS regime, on the one hand, this means that a series of systems of different sizes $L$ always contain the same number of $\xi_R$ regions because of their different $R$ values used, as was schematically illustrated with checkerboards in Fig.~1 of Ref.~\cite{Yuan}. This spatial self-similarity of the phases fluctuations is indispensable for some observables to have good scaling collapses. In the FSS regime, on the other hand, $\xi_R$ might appear to be irrelevant especially for large values of $L^{-1}R^{-1/r}$, because the whole finite-size system is just a phase on average. However, fixing $L^{-1}R^{-1/r}$ also implies fixes $L^{-z}R^{-z/r}$, the ratio of a finite-size relaxation time over the driven time. This dictates that the fluctuating phases on different lattice sizes survive for the same ratio of duration under driving and thus ensures the self-similarity of the phases fluctuations in time. In such a way, good scaling collapses ensue provided that corrections to scaling can be ignored. An example of the temporal self-similarity is given in Fig.~\ref{tss}(b) and~\ref{tss}(c), where the lattices of sizes $L=10$ and $L=20$ share identical $R^{-z/r}/L^z\approx2/1.5$ due to their different heating rates $R$.

If the spatial or temporal self-similarity is broken, it is found that some observables violate the standard FSS or FTS and even the revised FTS either in the ordered phase or the disordered phase~\cite{Yuan}. The violated scaling of an observable can be rectified by a breaking-of-self-similarity, or Bressy in short, exponent $\sigma$ that is defined as~\cite{Yuan}
\begin{eqnarray}
f(L^{-1}R^{-1/r})&\propto&\left(L^{-1}R^{-1/r}\right)^{\sigma}, ~{\rm for~FTS,} \nonumber\\
f(RL^r)&\propto&\left(RL^r\right)^{\sigma/r}, ~{\rm for~FSS,}
\label{bressyexp}
\end{eqnarray}
where $f$ is the scaling function of the observable. The definitions factor out the rate exponent $r$. For a non-integer $\sigma$, the scaling function then depends on the scaled variable singularly in the phase. With such an exponent $\sigma$, the leading behavior of the observable in the phase in which its scaling is violated is then qualitatively different from its usual one in the other phase~\cite{Yuan}. This is in stark contrast to the usual equilibrium critical phenomena.
There, one can also define different critical exponents above and below $T_c$~\cite{Mask,Cardyb,Fisher}. However, they ought to be equal because of the absence of singularities across the critical isotherm~\cite{Fisher}. Only the amplitudes of the leading singularities and thus the scaling functions above and below $T_c$ differ.

In Eq.~(\ref{bressyexp}), we have not written a scaling function on the right-hand side similar to Eq.~(\ref{ftfts}). This is because, in the case of $R=0$ and $L=\infty$, or $RL^r=0$ and $L^{-1}R^{-1/r}=0$, respectively, the usual equilibrium FSS and FTS, respectively, must recover. This means that there is a crossover to the usual behavior under the same condition. Moreover, for the large values of the scaled variables, Eq.~(\ref{bressyexp}) is poor owing to the crossover of the FSS and FTS regimes. As a consequence, Eq.~(\ref{bressyexp}) is only valid in some ranges of the variables. Except for $RL^r=0$ and $L^{-1}R^{-1/r}=0$, the Bressy-exponent dominated regime is believed to be valid for arbitrarily small $RL^r$ and $L^{-1}R^{-1/r}$ for the FSS in both heating and cooling and the FTS in cooling. Accordingly, the crossovers from this regime to the usual FSS and FTS occur abruptly at the two special points. However, for FTS in heating, we will see in Sec.~\ref{bressy} that the crossover happens abruptly at a finite $L^{-1}R^{-1/r}$.

The violations of scaling appears always in one set of observables, either $\chi$ and $\langle m\rangle$ or $\chi'$ and $\langle |m|\rangle$. The remaining set then exhibits good scaling. In every case, there are a primary observable and a secondary one. The primary observable can be rectified directly by Eq.~(\ref{bressyexp}), while the secondary one can also be accounted for by the same $\sigma$ but needs a constant regular term. This can be achieved from the following relation
\begin{equation}
\langle m\rangle^2+L^{-d}\chi=\langle |m|\rangle^2+L^{-d}\chi'\label{mma}
\end{equation}
between the two sets of the observables obtained from the definitions of $\chi$ and $\chi'$ in Eq.~(\ref{modelII}) below. In the following, we present the behavior of the secondary observables for all the four cases.

The violated scalings are always exhibited in the set with absolute values in all cases but FSS in heating and the primary observables are all $\chi$ except for FTS in cooling. Details are summarized in Table~\ref{tab1} in Sec.~\ref{sumvio} below.
In the case of FSS in heating, the primary observable is $\chi$~\cite{Yuan}, whose scaling function then behaves singularly as the second equation in Eq.~(\ref{bressyexp}), viz., $\mathcal{F}_{S\chi-}\propto(RL^r)^{\sigma/r}$, where we have inserted explicitly the observable and the phase ($+$ for the high-temperature phase and $-$ for the low-temperature phase) in which the violation occurs in the subscript for definiteness. Using Eq.~(\ref{mma}), one finds
\begin{equation}
F_{S\langle m\rangle-}^2=\mathcal{F}_{S\chi'}+F_{S \langle |m|\rangle}^2-\mathcal{F}_{S\chi-}\propto{\rm constant}-(RL^r)^{\sigma/r}\label{msigma}
\end{equation}
from Eqs.~(\ref{FSSX}),~(\ref{FSSM}), and~(\ref{law}) as well as~(\ref{bressyexp}). In other words, the secondary observable $\langle m\rangle$ is also singular but with a regular term because the first two terms on the right-hand side of Eq.~(\ref{msigma}) associated with the pair $\chi'$ and $\langle |m|\rangle$ are analytic. Note that an observable obeying the standard FSS or FTS under driving needs only a single scaling function no matter whether under heating or cooling. This is in stark contrast to the usual critical phenomena in which two scaling functions are needed for the two phases. Accordingly, we have not written $\mathcal{F}_{S\chi'-}$ and $F_{S \langle |m|\rangle-}$ for the corresponding parts in $F_{S\langle m\rangle-}$. To verify Eq.~(\ref{msigma}), one has to move the two regular terms to the right. The result is obviously $-\mathcal{F}_{S\chi-}$ from the first equality of the same equation. Therefore, if $\chi$ collapses well, $\langle m\rangle$ will naturally collapse well too. This same reasoning applies to the following cases as well.

For the case FSS in cooling, it is $\chi'$ that is the primary observable. A similar equation to Eq.~(\ref{msigma}) with the two pairs of observables interchanged follows.

In the case of FTS in heating, it is again $\chi'$ that is the primary observable. The secondary observable is then
\begin{eqnarray}
F_{T\langle |m|\rangle+}^2&=&\mathcal{F}_{T\chi}+\left(L^{-1}R^{-1/r}\right)^dF_{T \langle m\rangle}^2-\mathcal{F}_{T\chi'+}\nonumber\\
&\propto&{\rm constant}-\left(L^{-1}R^{-1/r}\right)^{\sigma}\label{msigmat}
\end{eqnarray}
from Eqs.~(\ref{FTSX}),~(\ref{FTSM}),~(\ref{law}), and~(\ref{bressyexp}). The scaling variable $L^{-1}R^{-1/r}$ in front of $F_{T\langle m\rangle}$ does not affect the analyticity of the latter. In cooling, the secondary observable is $\chi'$, whose scaling function becomes
\begin{eqnarray}
\mathcal{F}_{T\chi'-}&=&\mathcal{F}_{T\chi}-\left(L^{-1}R^{-1/r}\right)^{-d}F_{T \langle |m|\rangle-}^2\nonumber\\
&\propto&{\rm constant}-\left(L^{-1}R^{-1/r}\right)^{2\sigma-d}\label{xsigmat},
\end{eqnarray}
since $\langle m\rangle=0$. The exponent $d$ in the second line comes from Eq.~(\ref{ftfts}) for the revised scaling since we define $\sigma$ using $F_{T \langle |m|\rangle}$ instead of $F_{TS \langle |m|\rangle}$. This is because we will see in Secs.~\ref{FSS} and~\ref{FTSwh} that the usual FTS of $\langle |m|\rangle$ becomes valid if the self-similarity of the phases fluctuations is kept and thus the revised scaling is also a result of the self-similarity breaking.

If self-similarity breaking is not taken into account, the above theory yields a single scaling form for $\chi$ either in FTS or in FSS but one more combined form, the revised FTS form, for $M$. However, we will see from Fig.~\ref{hcxm} below that $\chi$ appears to have two different appearances but $M$ seems to be similar in heating and cooling. Nevertheless, we will show in the following that the above theory describes the phenomena well and phase ordering can only play a role at rather low temperatures in cooling provided that the phases fluctuations are properly taken into account.

\section{\label{phfl}Phases fluctuations and their self-similarity}
In this section, we elaborate the idea of the phases fluctuations and their self-similarity, although we have introduced them in the last section and will see more numerical evidences for them in Secs.~\ref{FSS} and~\ref{FTSwh} below. We will argue that critical fluctuations consist of two parts: One is the phases fluctuations and the other is magnitude fluctuations. This may be superficially regarded as a generalization of the critical fluctuations of a continuous phase transition with a broken continuous symmetry in some sense. The phases fluctuations can further possess two kinds of self-similarity: One is intrinsic and the other extrinsic. The former one gives rise to the usual critical scaling while the other is related to external conditions such as the system size and external fields and can thus be broken with significant consequences.

When a system is subjected to a continuous phase transition from a homogeneous disordered state to an ordered phase, its symmetry is broken spontaneously, resulting in a number of dissymmetric phases. For a finite group, this number is given by the index of the group of the new phase in the parent group. Accordingly, the ordered phase can be either a mixture of various such phases forming domains or just a single phase depending on the boundary conditions, defects, or external fields. In particular, for the Ising model, spin down and spin up are the two possible phases at low temperatures because of the broken up-down symmetry. Both can either coexist as domains or appear alone. Of course, at a finite temperature far lower than the critical point, a stable spin-up phase, for example, contains a number of down spins that are fluctuating randomly both spatially and temporally. These fluctuating down spins give rise to the correct equilibrium magnetization of the spin-up phase at the temperature. However, it is apparent that these fluctuating spins forming clusters of various sizes (typified by the correlation length of several lattice sizes) cannot be referred to as the spin-down phase, though they have the same origin. Instead, they are called thermal or spin fluctuations.

Near the critical point, it is well known that the divergent correlation length is responsible for the singular behavior of all measurable quantities and serves as the only relevant length scale. A physical picture of this correlation length scaling hypothesis is droplets of overturned spins in the sea of up spins~\cite{Kogut}. However, a droplet of the large size $\xi$ cannot contain all down spins. Accordingly, a better and frequently-used picture is that there are fluctuations on all length scales up to $\xi$~\cite{Kogut} such that the system is statistically self-similar within $\xi$. In equilibrium, in the thermodynamic limit, and in the absence of an externally applied field, at $T\geq T_c$, both spin directions are equally probable and the above picture of self-similarity may be suitable. However, for $T<T_c$ but close to $T_c$, this picture cannot produce a finite magnetization since each cluster has nearly zero net magnetization $m$. A more realistic picture is then that these large clusters contain predominantly up or down spins such that they have a finite net magnetization. Each such large droplet can be regarded as a dissymmetric phase that contains fluctuating smaller droplets of the other dissymmetric phase. The smaller droplets, in turn, may also contain even smaller fluctuating droplets. However, the averaged magnetization of each large droplet ought to be roughly the equilibrium magnetization of the symmetry-broken phase itself at the temperature with a mean fluctuation of $\sqrt{\chi}/L^{d/2}$ or so. These clusters are not frozen in but are fluctuating unless at low temperatures, at which equilibrium domains of only up or down spins may form after coarsening and coagulation of the clusters. Since the magnetization of the symmetry-broken phase decreases to zero when the temperature is lowered to the critical point, at which each droplet has equal number of up and down spins on average, a situation which persists up to $T>T_c$, such a picture of clusters matches the self-similar one and applies to those temperatures as well. Of course, at too high temperatures, the correlation length is short and the fluctuations become the usual thermal ones. Accordingly, we refer to such large clusters as phases and their fluctuations as ``phases" fluctuations. In nonequilibrium situations, these large ``phases" clusters may not overturn and thus no phases fluctuations. However, the picture of large phases clusters is still valid. The Ising model is a generic example of a system with a discrete symmetry. For a system with a continuous symmetry, the Goldstone modes are also the phases fluctuations, because they just change the directions of the broken symmetry and thus the phases.

\begin{figure*}
\centering
\centerline{\epsfig{file=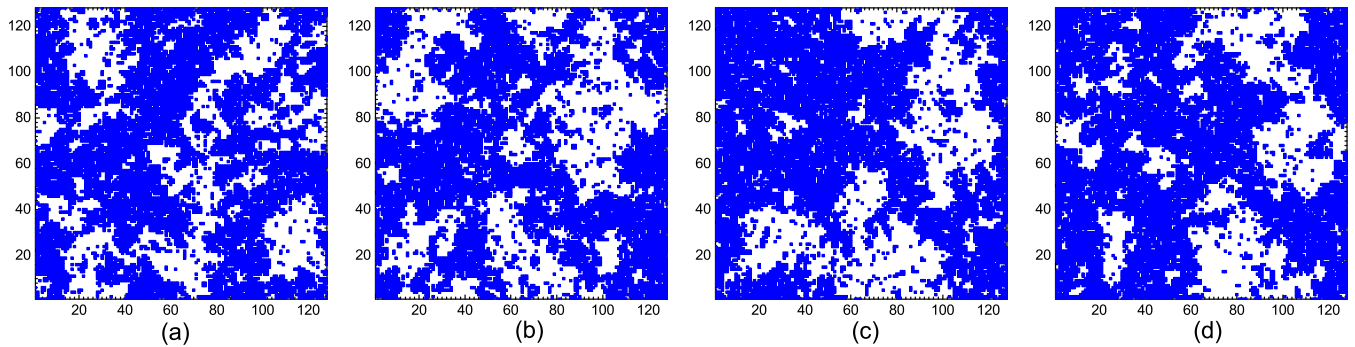,width=1.0\textwidth}}
\caption{(Color online) Configurations of the 2D Ising model in cooling at $R=0.000~075$ on a $128\times128$ lattice at (a) $T=2.3215$, (b) $T=2.3005$, (c) $T=2.2795$, and (d) $T=2.2585$, corresponding to $\tau R^{-1/r\nu}=1.0468$, $0.6266$, $0.2064$, and $-0.2138$, respectively. The time interval between two successive temperature is about $0.4202R^{-z/r}$. Each dot in the configurations denotes an up-pointing spin. $T_c=2/\ln(1+\sqrt{2})\approx2.2692$, $\nu=1$, and $z=2.17$ for the 2D Ising model.}
\label{conf}
\end{figure*}
A most direct evidence of the above picture of the phases fluctuations comes from the time domain rather than the space domain. As pointed out in Sec.~\ref{theory}, the system size is the governing scale in FSS. The system is itself a large cluster and hence a phase on average. Therefore, it must acquire a finite magnetization and turn from up to down and vice versus in a time of $L^z$ on average. This is clearly seen in Fig.~\ref{tss}, in particular, Fig.~\ref{tss}(a). The survival time $L^z$ is, of course, fluctuating widely, as can also be observed from the figure. However, for the same lattice size $L$, it is the same on average, as is manifest upon comparing Fig.~\ref{tss}(a) with~\ref{tss}(b) and Fig.~\ref{tss}(c) with~\ref{tss}(d) for two lattices of identical sizes but different values of $R$. In addition, Fig.~\ref{tss} shows that the magnitude of $m$ decreases as $t$ elapses and hence $T$ increases in heating and finally reaches zero with relatively smaller fluctuations.
In Fig.~\ref{conf} we show four spatial configurations in cooling as an evidence of the phases fluctuations in space domain. Although the picture is not simply that of a checkerboard schematically illustrated in Ref.~\cite{Yuan} because of the large fluctuations in the cluster sizes, their values of the magnetization, their boundaries, their neighbors and so on, and thus not as convincing as the picture in time domain, large clusters of roughly the size $\xi_R\sim R^{-1/r}\approx20$ containing predominantly up or down spins turn over from up to down and vice versus are evident as the temperature is lowered.

Having confirmed the existence of the phases fluctuations, we can thus separate the critical fluctuations into two parts. One is to form large clusters that are the dissymmetric phases with roughly the order parameter at the temperature at which the system sits and the other is the flipping of these large clusters that are accordingly referred to as phases fluctuations. One might argue that such a picture had already implied in the previous picture of self-similarity and that it were not necessary to emphasize the phase nature of the large clusters and to coin a new word. However, as will be seen in Secs.~\ref{FSS} and~\ref{FTSwh} below, this picture is indispensable for accounting the distinctive behavior of two sets of observables, $\langle m\rangle$ and $\chi$ versus $\langle|m|\rangle$ and $\chi'$ to be defined in Sec.~\ref{model} below. In particular, the phases fluctuations manifest themselves as the difference between the two sets of the observables while the magnitude fluctuations are probed by the set of the observables with absolute values that remove the plus and minus signs. This difference can be transparently envisioned from Fig.~\ref{tss}(a): A huge number of samples having different time series similar to Fig.~\ref{tss}(a) but fluctuating at random instances will be averaged to a vanishing $\langle m\rangle$ but a finite $\langle |m|\rangle$.

We note that the difference between $\langle m\rangle$ and $\langle |m|\rangle$ stems from the fluctuations of $m$, the magnetization itself, instead of its spatial distribution, of a single sample at a particular moment or temperature. Accordingly, a fixed spatial checkerboard of clusters containing predominantly up and down spins gives rise to a fixed $m$ independent on time or the temperature, in contrast to a fluctuating $m$ shown in Fig.~\ref{tss}. Different samples may still have their fixed $m$ with different magnitudes and signs. As a result, the difference in the two sets of the observables still follows. However, Fig.~\ref{conf} shows clearly that the clusters are fluctuating rather than fixed in space near the critical point. Only at temperatures sufficiently lower than the critical point can a spatially fixed domain structure exist. Upon taking into account the unambiguously fluctuating phases in time exhibited in Fig.~\ref{tss}, it is therefore justified to emphasize the fluctuating nature of the phases and to single the phases fluctuations out from the critical fluctuations.

The above separation of the critical fluctuations looks like the critical fluctuations of a system with a continuous symmetry breaking. There, the longitudinal fluctuations are
the fluctuations around the symmetry-broken direction, whereas the transverse fluctuations are the Goldstone modes that change the direction of the symmetry-broken phase and are thus the phases fluctuations. However, the longitudinal behavior itself is identical with a scalar theory that describes the critical fluctuations of the Ising model, provided that possible corrections from the transverse directions can be ignored. In other words, the longitudinal fluctuations can themselves be divided into the magnitude fluctuations and the phases fluctuations. Therefore, the similarity is only superficial. Here, the magnitude fluctuations are the critical fluctuations exclusive of the phases fluctuations. They are exhibited by the observables with absolute values that just remove the turnover of the large phases clusters. However, the phases fluctuations also contribute to the magnitude fluctuations via changing the magnitude of $m$. We will see in Secs.~\ref{FSS} and~\ref{FTSwh} below that the magnitude fluctuations are generally far weaker than the phases fluctuations.

Phases fluctuations emphasize directly the phase nature of the fluctuations and thus emphasize the origin of the critical fluctuations, the symmetry-broken dissymmetric phases. They also provide a somehow vivid picture of the critical fluctuations. More importantly, the phases clusters and their fluctuations are crucial in understanding the dynamics of a driven transition from a disordered phase through a critical point to an ordered phase. A simple example is the KZ mechanism of topological defect formation. As the system is cooled towards its critical point, its correlation length increases and thus the size of the fluctuating phases increases. It is the spatial boundaries of the different phases, the domain boundaries, that form the KZ topological defects. Accordingly, this KZ mechanism for the defect formation is transparent from the point of view of phases fluctuations. Upon assuming a frozen correlation length, it thus gives rise directly to the density of the topological defects. However, on the one hand, the phases in neighboring regions have a considerable probability to be identical and thus merge into a larger droplet as evident in Fig.~\ref{conf}, different from the schematic picture of a regular checkerboard shown in Fig.~1 of Ref.~\cite{Yuan}. This substantially reduces the number of the topological defects. On the other hand, each phase cluster also contains fluctuations of droplets of the other phases of various sizes smaller than the domain size and the smaller droplets may contain even smaller droplets of their other phases and thus increasing significantly the number of the topological defects. Consequently, the real density of topological defects can be far different from those reckoned directly from the correlation length. This inevitably leads to disagreement of the KZ scaling with experiments. In addition, phase ordering may occur within the adiabatic region~\cite{Biroli}. Therefore, the density of topological defects is not a good observable to characterize the dynamics.

A less well-known example is the revised FTS form in cooling. Its origin is also the phases fluctuations. Because each fluctuating phase droplet is independent on the other droplets, all these fluctuating droplets must thus obey the central limit theorem as they are identically distributed. This leads to the special volume factor for Eq.~(\ref{rescaling_huang}) in the FTS regime for a finite-sizes system~\cite{Huang}.

Moreover, we will see that the phases fluctuations are crucial not only to cooling but also to heating~\cite{Yuan}. This is related to the self-similarity of the fluctuations.
Self-similarity of a critical system is described by the renormalization-group theory whose consequence is the scale transformation, Eq.~(\ref{RG}). This equation dictates that the so-called self-similarity be the similarity of different scales at different scaled variables such as temperatures and external fields. Self-similarity may then reckon on this similarity~\cite{Suzuki}. This similarity is evidently satisfied by the phases and their fluctuations. Indeed, different distances to the critical point, for instance, mean different values of $\xi$ and thus different sizes of the large droplets with different ordered parameters. All have a statistically similar picture of the phases and their fluctuations. Only the sizes of the fluctuating droplets are apparently different. In Sec.~\ref{FSS} below, we will find that scalings are poor for the set of the observables with absolute values but good for the other set of the observables for the cases of FTS either in heating or cooling and FSS in cooling. This might indicate that the magnitude fluctuations alone did not have self-similarity. However, in the case of FSS in heating, it is the set of the observables without taking absolute values that scales well whereas the other set scales poorly. Moreover, we will see that these violations of scaling appear only either in the ordered phase or in the disordered phase, except for the case of FTS in cooling due to the revised FTS. Therefore, both the magnitude and the phases fluctuations must satisfy the scalings and thus possess self-similarity. In fact, we will argue that the violations of scaling originate from another kind of self-similarity, which we refer to as extrinsic self-similarity.

As pointed out in the last sections, in order to fully describe the scaling behavior, it is essential to maintain an additional self-similarity, the extrinsic self-similarity. This additional self-similarity is to ensure that different-sized lattices are subject to different rates of driving in such a way that all systems contain identical number of either the fluctuating phases clusters~\cite{Li} or their survival time intervals illustrated numerically in Fig.~\ref{tss}. Since both the magnitude fluctuations and the phases fluctuations concern the same clusters, self-similarity of one implies the same to the other, except for the case of FSS in which no phases fluctuations mean no turnover of the clusters and hence no temporal self-similarity at all. Moreover, we will see that when there exist phases fluctuations, violations of scaling do not necessarily occur; however, if there exist no phases fluctuations, there are no violations of scaling, though the latter does not originate from the former but rather from their self-similarity breaking. We therefore associate the additional self-similarity with the phases fluctuations rather than the magnitude fluctuations, though breaking of the self-similarity of the former also implies breaking that of the latter in the case of FTS. Further, violations of scaling in the observables with absolute values may not imply that they stem from the self-similarity breaking of the magnitude fluctuations. In addition, we also note that even if this additional self-similarity is broken, the original self-similarity of the phases fluctuations themselves remains and therefore the scalings of some observables are still exhibited possibly in some ranges.

Therefore, we have two kinds of self-similarity of the phases fluctuations. The first one is the self-similarity of the phases fluctuations themselves. This intrinsic self-similarity is only limited by the criticality. The second kind of self-similarity is additionally limited by external conditions such as the system size and external driving and thus is an extrinsic self-similarity. Both kinds of self-similarity can of course be broken by tuning the system far away from the critical point. However, the second kind can be easily broken by changing the external conditions.

We will see in the following that the differences between heating and cooling, including the anomalous result in cooling exhibited in Fig.~\ref{hcxm} below, stems also from the phases fluctuations. In cooling, the dissymmetric phases can freely fluctuate and thus the phases fluctuations are not restricted. In heating, symmetry is broken from the beginning of the driving due to our initial conditions and thus the phases fluctuations are substantially reduced. Therefore, the phases fluctuations are not only a realistic picture for critical fluctuations, but also play a pivotal role in accounting for the dynamic scaling of critical behavior.

\section{\label{model}Model and method}
We consider the 3D Ising model defined by the Hamiltonian
\begin{equation}
\mathcal{H}=-J\sum_{\langle i,j\rangle} \sigma_{i}\sigma_{j}+H\sum_i \sigma_{i},
\label{modelI}
\end{equation}
where $J>0$ is a nearest-neighbor coupling constant and will be set to $1$ hereafter as an energy unit, $\sigma_i=\pm 1$ is a spin on site $i$ of a simple cubic lattice, and the summation of the first term is over all nearest neighbor pairs. Periodic boundary conditions are applied throughout. The critical temperatures~\cite{exponent2} and the critical exponents~\cite{Tc,exponent1,exponent2} of the 3D simple cubic lattice are
$T_c =1/0.221~659~5(26)=4.511~42(6)$, $\nu = 0.630~1(4)$, $\beta = 0.326~5(3)$, $\gamma =1.237~2(5)$, and $\delta=4.789(2)$.
Most of estimation of the dynamical critical exponent $z$ is near $2.0$. We choose $z=2.055$ here if not mentioned explicitly, which is estimated in the cooling process~\cite{Huang} and close to the previous results~\cite{z3d1,z3d2}.

The observables we measured are the order parameter $M$ and the susceptibility $\chi$ defined as
\begin{equation}
M=\left\{
\begin{array}{l}
\left\langle m\right\rangle =\left\langle\frac{1}{N}\sum_{i=1}^N \sigma_{i}\right\rangle, \\
\\
\left\langle |m|\right\rangle =\left\langle\frac{1}{N}\left|\sum_{i=1}^N \sigma_{i}\right|\right\rangle,
\end{array}
\right.\label{M}
\end{equation}
\begin{equation}
\chi=\left\{
\begin{array}{l}
\chi=L^d\left(\left\langle m^2\right\rangle-{\left\langle m\right\rangle}^2\right),\\
\\
\chi'=L^d\left(\left\langle m^2\right\rangle-{\left\langle |m|\right\rangle}^2\right),
\label{modelII}
\end{array}
\right.
\end{equation}
where the angle brackets represent ensemble averages and $N$ is the total number of spins. The first set of definitions, containing the first lines in Eqs.~(\ref{M}) and~(\ref{modelII}), is the usual definitions of the order parameter and its susceptibility, while the second set including the remaining two equations is usually employed when $\langle m\rangle=0$ in the absence of symmetry breaking and thus absolute values are needed. The absolute values disregard the sign change of $m$ and hence remove the phases fluctuations and ought to probe the magnitude fluctuations. However, as mentioned in the last section, we will see that the phases fluctuations also contribute to the magnitude fluctuations in FTS cooling. Therefore, probing both sets of observables is invaluable to uncover some secrets of phases transitions and their effects both in cooling and in heating.

We note that the real susceptibility defined as the change of magnetization due to a unit change of an externally applied field on the left-hand side of Eq.~(\ref{modelII}) is not equal to the fluctuations on the right-hand side in a nonequilibrium situation. However, their scaling behaviors are identical~\cite{Feng}. For simplicity, we thus simply use the susceptibility defined in Eq.~(\ref{modelII}) to represent the fluctuations. We will generally refer to the order parameter $M$ and the susceptibility $\chi$ for both definitions, which accounts for the somewhat odd expression in the first line of Eq.~(\ref{modelII}), and stipulate to a particular one when so indicated, except for Sec.~\ref{nr of ordering}, where only the first set of the observables is employed.

The single-spin Metropolis algorithm~\cite{MC} is employed and interpreted as dynamics~\cite{Glauber,landaubinder}. The time unit is the standard Monte Carlo step per site, which contains $N$ randomly attempts to update the spins. For both the heating and cooling processes, we prepare the system far away from $T_c$ in an ordered or a disordered initial configuration at a negative initial time and then heat and cool it, respectively, through $T_c$ at $t=0$ according to a given $R$. We check that the initial states create no observable effects once they are sufficiently far away from $T_c$, because the system can then equilibrate quickly due to the short relaxation time there. Various sizes $L$ and rates $R$ are used. All the results are averaged over $10$--$30$ thousand samples.

\section{\label{FSS}FSS and FTS in heating and cooling in the absence of external field}
In this section, we will study the FSS and FTS in heating and cooling in the absence of an external field. However, as an appreciation of how surprising driven nonequilibrium critical phenomena can be, we first of all compare behaviors in heating with cooling in the FTS regime.

\begin{figure}
\centering
\centerline{\epsfig{file=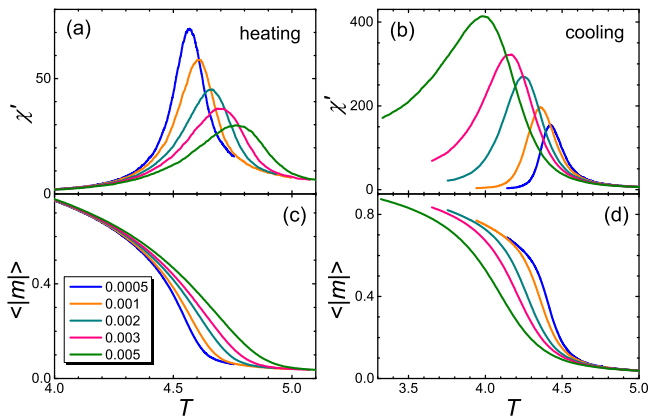,width=1.0\columnwidth}}
\caption{(Color online) (a) and (b) susceptibility $\chi'$ and (c) and (d) order parameter $\langle |m|\rangle$ upon heating, (a) and (c), and cooling, (b) and (d), at the various rates given in the legend on $20\times20\times20$ simple cubic lattices. The two panels in each column share the same axis of temperature $T$ and its scales. All panels share the same legend.}
\label{hcxm}
\end{figure}
In Fig.~\ref{hcxm}, we display the rate dependence of the order parameter and the susceptibility on a fixed lattice size. The most prominent feature is that the susceptibility exhibits a sharp qualitative difference in heating and cooling: The peaks increase in heating whereas decrease in cooling with decreasing $R$. Note that, as pointed out above, exactly at the critical point, the scaling of the susceptibility in heating and cooling was found to be similar, only the order parameter and its squared are different~\cite{Huang}. In contrast, in Fig.~\ref{hcxm}, $\chi$ appears to have two different forms but $M$ seems to be similar in heating and cooling. In fact, one can sees from Figs.~\ref{hcxm}(a) and~\ref{hcxm}(b) that the susceptibility exactly at $T_c\approx4.511$ indeed increases with decreasing $R$ both in heating and in cooling, in consistence with Eq.~(\ref{FTSX}). Only the process after $T_c$ exhibits difference. Also, the order parameter exactly at $T_c$ exhibits different trends with $R$ in heating and cooling in accordance with Eqs.~(\ref{FTSM}) and (\ref{rescaling_huang}), respectively, although the two sets of curves appear not so different. One might think that the difference come as no surprise. However, the usual differences of critical behavior above and below the critical point are not the critical exponents but only the amplitudes~\cite{Mask,Cardyb}. Yet, the difference as exhibited in Fig.~\ref{hcxm} is the different dependence of the peaks on the rate, a difference which clearly cannot be accounted for by the amplitude itself. In addition, the susceptibility in cooling is much larger than that in heating. It does not vanish and still has large fluctuations at low temperatures for large rates. Accordingly, it seems that new ingredients like phase ordering might be needed to account for the cooling beyond the critical point.

Besides the sharp differences in the susceptibility, Fig.~\ref{hcxm} shows that the positions of the peak maxima behave similarly in heating and cooling. As $R$ is lowered, $\xi_D$ becomes longer and hysteresis weaker. As a consequence, the peak positions get closer to the equilibrium transition temperature $T_c$. The same reason leads to the higher peaks for lower $R$, as the result of heating demonstrates. Accordingly, the result of cooling is strange. We will come back to it towards the end of Sec.~\ref{fsf} below.

\subsection{\label{fssafr}FSS at fixed rates}
\begin{figure}
  \centerline{\epsfig{file=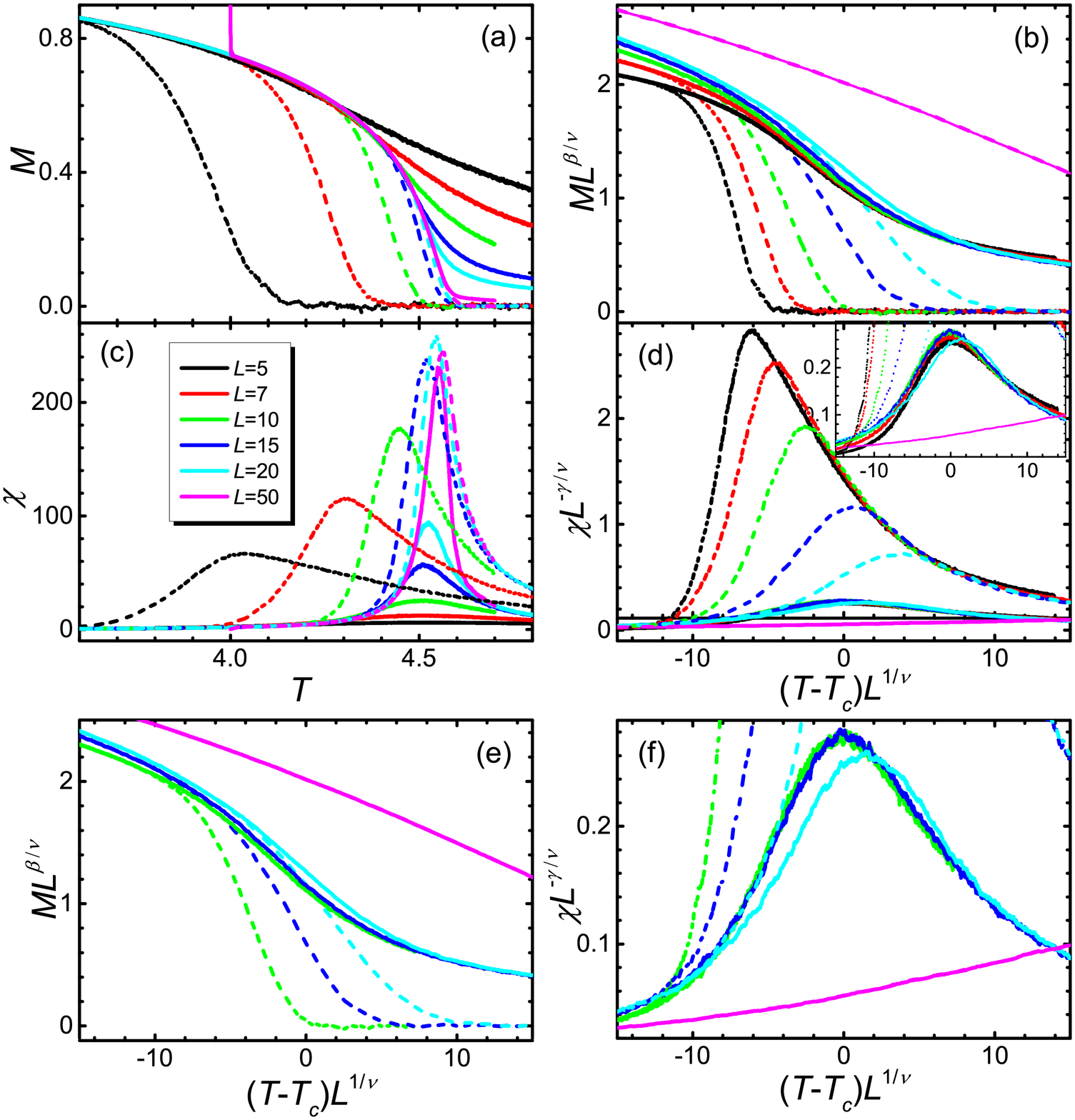,width=1.0\columnwidth}}
  \caption{\label{fssh} (Color online) (a) $M$ and (b) its FSS and (c) $\chi$ and (d) its FSS in heating at a fixed rate $R=0.000~1$ on different lattice sizes given in the legend. (e) and (f) are (b) and (d) in the absence of the two smallest lattice sizes $L=5$ and $L=7$. The dashed lines represent results of $\langle m\rangle$ and $\chi$ and the solid lines those of $\langle |m|\rangle$ and $\chi'$. The inset in (d) zooms in on the scale for $\chi'$. All panels share the same legend.}
\end{figure}
We now study the FSS of both heating and cooling at a fixed rate $R$. This helps to reveal the origin of the differences. We will see that phases fluctuations affect FSS too. In fact, clear evidences are exhibited for the phases fluctuations.

Figures~\ref{fssh}(a) and~\ref{fssh}(c) show $M$ and $\chi$ in heating for a fixed rate $R=0.000~1$. This rate corresponds to a driving length scale of $\xi_R\sim R^{-1/r}\approx12.5$ or timescale of $\zeta_R\sim R^{-z/r}\approx180.7$. Accordingly, for $L\lesssim\xi_R$ or $L^z\lesssim\zeta_R$, FSS shows and thus observables must depend appreciably on $L$ according to the leading behavior of Eqs.~(\ref{FSSM}) and~(\ref{FSSX}); whereas for $L$ sufficiently large, FTS is exhibited and the results from different lattice sizes vary only slightly due to the negligibility of $L^{-1}R^{-1/r}$. One sees from Fig.~\ref{fssh}(a) that $\langle |m|\rangle$ (solid curves) of different lattice sizes separates at different temperatures near $T_c$ from that of the largest lattice size, which is, of course, closest to equilibrium among the curves. The smaller $L$ is, the lower the separating temperature, indicating the earlier transition to the disordered phase because of its shorter correlation time $L^z$, which is about $27.3$, $54.5$, $113.5$, and $261.1$ for $L=5$, $7$, $10$, and $15$, respectively, in the FSS regime. For $L>15$, the separated curves are closer to each other than those of the smaller sizes, reflecting the crossover to FTS regime. The values of $\langle |m|\rangle$ are rather large above $T_c$, in agreement with similar results without driving~\cite{Landau76,landaubinder}. However, they decrease as $L$ increases and hence ought to be a finite-size effect. On the other hand, $\langle m\rangle$ (dashed curves) can deviate from the main curve at a rather low temperature. These different separating temperatures of $\langle |m|\rangle$ and $\langle m\rangle$ give rise to different transition temperatures, which are characterized by the peak temperatures of $\chi'$ and $\chi$, respectively. Moreover, both transition temperatures are lower than $T_c$ for those $L\lesssim15$, though their distances to $T_c$ are much different in magnitude. Furthermore, the peak heights exhibit a different size dependence. While the peak heights of $\chi'$ (solid curves) increase with $L$ rapidly and monotonically, those of $\chi$ (dashed curves) rise somehow mildly for large $L$ and even slightly decrease at $L=50$. These trends indicate that $L=20$ lies possibly in the crossover between the FSS and FTS regimes and $L=50$ in the latter regime. In the FSS regime on the one hand, $\chi$ is proportional to $L^{\gamma/\nu}$ according to Eq.~(\ref{FSSX}). In the FTS regime on the other hand, the dependence on $L$ is slight as mentioned. In addition, in the FSS regime, the phases cluster size is about $L$ itself, whereas in the FTS regime, it is only $\xi_R$. This may contribute to the peak height difference in $L=20$ and $L=50$. Besides, in the FTS regime, fluctuations are suppressed to $\xi_R$ which protects the existing phase and thus the transition can only be delayed rather than advanced.

Now come direct exhibition and evidence of the phases fluctuations. A prominent feature observed from Figs.~\ref{fssh}(a) and~\ref{fssh}(c) is that the two sets of the observables, $\langle m\rangle$ and $\chi$ versus $\langle |m|\rangle$ and $\chi'$, are markedly different. As seen in Fig.~\ref{fssh}(a), $\langle m\rangle$ collapses onto $\langle |m|\rangle$ at low temperatures. As $T$ increases, except for the largest lattice which belongs to the FTS regime, $\langle m\rangle$ first deviates from $\langle |m|\rangle$ at intermediate temperatures and then falls to zero or so at higher temperatures. At the low temperatures, the correlation length is not long enough. One might then imagine that a finite-sized system would comprise a few clusters of predominantly up or down spins with different quantities or with different magnitudes of magnetization. The quantity or the magnetization of the clusters with predominantly down spins would then increase with $T$ while that with up spins would decrease and finally at sufficiently high temperatures both would become equal. However, as pointed out in Sec.~\ref{phfl}, such a picture would at least result in a $\langle |m|\rangle$ different from $\langle m\rangle$ even at low temperatures. Therefore, a more appropriate picture is that the system is just one cluster with spin fluctuations to produce its correct magnetization even if the correlation length is smaller than $\xi$ at the low temperatures and no turnover exists at all. As $T$ increases towards $T_c$, $\xi$ is raised to the size of $L$. $m$ of the cluster can then overturn and fluctuate between up and down as a whole, as exemplified in Fig.~\ref{tss}. This implies that both directions ought to possess similar magnitudes of $m$. In fact, that $\langle |m|\rangle$ of different lattice sizes converges almost to a single envelop for $T\lesssim4.3$ in Fig.~\ref{fssh}(a) means all $m$---no matter whether they are positive or negative---ought to be the same on average and independent on $L$. Because the fluctuations between the two directions with finite values of the magnetization $m$ happen at random instances, averaging different samples thus gives rise to different $\langle m\rangle$ and $\langle |m|\rangle$. Moreover, the frequency of the fluctuations may depend on the temperature. As a result, in the nonequilibrium heating, there can be no time enough for the cluster to overturn freely so that $\langle m\rangle$ can also be finite at some temperatures. This picture thus qualifies us to call the fluctuating clusters as phases, and the deviation of $\langle m\rangle$ from $\langle |m|\rangle$ is thus a direct evidence of the phases fluctuations.
Note, however, that although smaller lattice sizes exhibit more pronounced difference between $\langle m\rangle$ and $\langle |m|\rangle$ in that the averaged magnetization is zero at lower $T$ below $T_c$, this is due to their shorter correlation times or survival times and does not mean that their phases fluctuations are stronger. Rather, lattices of larger $L$ values have larger phases cluster sizes and hence stronger fluctuations, as is evident from the larger $\chi$ peaks in Fig.~\ref{fssh}(c). However, the phases fluctuations are suppressed when the absolute values are employed to average. Only the magnitude fluctuations remain. Consequently, $\chi'$ is much smaller than $\chi$ in the FSS regime and its peak appears near $T_c$ as can be seen from Figs.~\ref{fssh}(a) and~\ref{fssh}(c). We note that for the small lattice sizes, the usual random thermal fluctuations---those that produce the temperature dependence of the magnetization---may also contribute to the turnover of the phases at temperatures quite lower than $T_c$. For a small lattice, with several existing opposite spins, other spins can likely turnover without appreciable energy penalties.

\begin{figure}
  \centerline{\epsfig{file=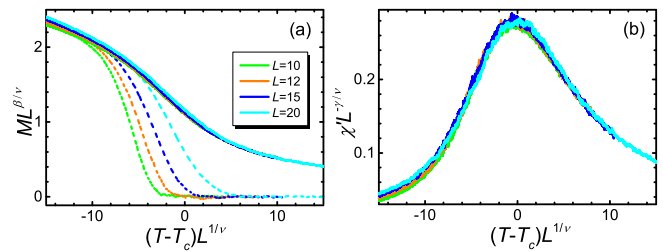,width=1.0\columnwidth}}
  \caption{\label{fssh403} (Color online) FSS in heating at a fixed rate $R=0.000~03$ of (a) $M$ and (b) $\chi'$ on different lattice sizes given in the legend. Solid lines are results for $\langle |m|\rangle$ and dashed lines for $\langle m\rangle$. Both panels share the same legend.}
\end{figure}
Figures~\ref{fssh}(b) and \ref{fssh}(d) show that the FSS of $\langle m\rangle$ and $\chi$ for the small lattice sizes is severely violated, though that of $\chi$ in the disordered phase beyond the peak appears good. However, the FSS of $\langle |m|\rangle$ and $\chi'$ on small lattice sizes is reasonably good noting that corrections to scaling~\cite{Wegner} are likely not small for the small lattice sizes. Indeed, in the absence of the two smallest lattice sizes, the scaling appears quite good even for $L=15$ bigger than $\xi_R\approx12.5$, or $RL^r\approx1.92$, as shown in Figs.~\ref{fssh}(e) and \ref{fssh}(f). This means that $RL^r$ can be ignored even up to a value of about $2$ in the scaling functions and so $L^z$ of $L=15$ can be longer than $\zeta_R$. This is not unreasonable because we have omitted possible multiplying constants for the scaled variables. In Fig.~\ref{fssh403}, we confirm the above conclusions using results at a smaller fixed rate of $R=0.000~03$ whose $\xi_R\approx17.4$. Again, $\langle |m|\rangle$ and $\chi'$ on $L=20$ corresponding to $RL^r\approx1.64$ follows well FSS, since the previous value of $1.92$ leads back to an $L\approx21$ for $R=0.000~03$, while the bad FSS of $\langle m\rangle$ remains. Of course, for sufficiently larger lattice sizes, the system must show FTS instead of FSS.

Therefore, upon heating at a fixed rate, FSS is exhibited for the observables with absolute values such as $\langle |m|\rangle$ and $\chi'$. In fact, they were utilized in the early verification of FSS~\cite{Landau76,landaubinder}. The FSS of $\chi$ in the disordered phase is also good. This is reasonable because $\chi$ is just $\langle m^2\rangle$ when $\langle m\rangle=0$ but $\langle m^2\rangle$ and $\langle |m|\rangle$ which compose $\chi'$ scale well. However, the FSS of $\langle m\rangle$ and $\chi$ in the ordered phase is violated. A possible reason might be the seemingly large deviation from $T_c$ of the $\langle m\rangle$ and $\chi$ curves. One might also imagine that, possibly for very large lattice sizes on which $\langle m\rangle$ and $\chi$ differ negligibly from $\langle |m|\rangle$ and $\chi'$, respectively, their FSS might show. However, comparing Figs.~\ref{fssh}(e) to~\ref{fssh403}(a), we observe that the curve of $L=15$ moves to low temperatures and the associated transition is further advanced as $R$ is lowered. Noting that $L$ must not be too larger than $\xi_R$ for FSS to be fulfilled, it is thus unlikely that the FSS of $\langle m\rangle$ at a fixed $R$ could appear good. In fact, as will be shown in Secs.~\ref{fsf} and~\ref{bressy} below, the violation does not result from the difference between the two sets of observables but rather from the self-similarity breaking.

\begin{figure}
  \centerline{\epsfig{file=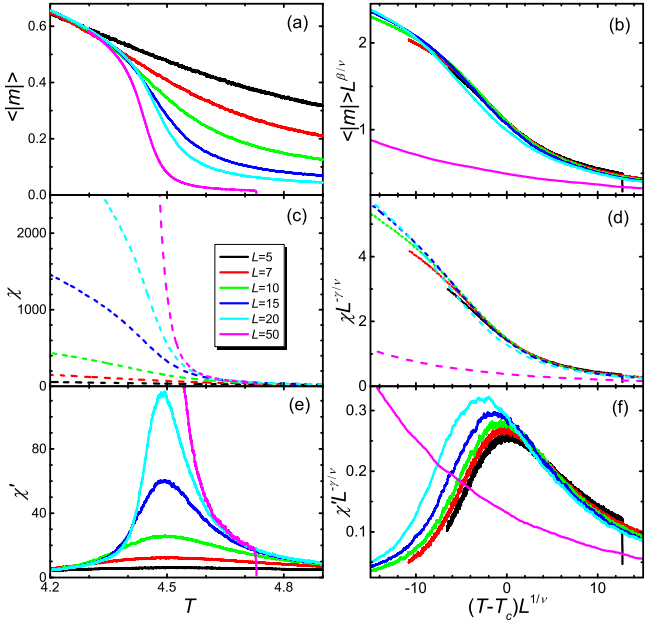,width=1.0\columnwidth}}
  \caption{\label{fssc} (Color online) (a) $M$ and (b) its FSS, (c) $\chi$ and (d) its FSS, and (e) $\chi'$ and (f) its FSS in cooling at a fixed rate $R=0.000~1$ on different lattice sizes given in the legend, which is shared by all panels.}
\end{figure}
Under cooling and in the absence of an externally applied field, $\langle m\rangle$ is vanishingly small. $\langle m\rangle$ and $\langle |m|\rangle$ are thus completely different; each large cluster can assume any direction and fluctuate freely, resulting in the strong phases fluctuations. In fact, it is the large phases fluctuations between the two phases that result in the vanishing $\langle m\rangle$ for not too low temperatures at which are exhibit different $\langle m\rangle$ and $\langle |m|\rangle$ in heating in Fig.~\ref{fssh}. For lower temperatures, the surviving phases can rarely overturn. However, these phases are selected randomly out of the phases fluctuations and thus the averages of many samples again lead to the vanishing $\langle m\rangle$. We thus show $\langle |m|\rangle$ only. Moreover, again due to the phases fluctuations, $\chi$ is far much larger than $\chi'$ and exhibits no peak at all, as seen in Fig.~\ref{fssc}. This is different from the case in heating in which $\chi$ shows a peak similar to $\chi'$. This in turn reflects an important difference between heating and cooling. In heating, we start the driving from an ordered initial state and the symmetry is broken from the beginning. The phases fluctuations can only start from the symmetry-broken phase rather than freely fluctuate between the two dissymmetric phases and hence are substantially reduced. Furthermore, in contrast with Fig.~\ref{hcxm}(b) in the FTS regime, in the FSS regime, $\chi'$ is vanishingly small at low temperatures, a point to which we will come back later on in Sec.~\ref{ftsofls} below. In addition, similar to Fig.~\ref{fssh} in heating, the curves of $L=50$ which does not fall in the FSS regime exhibit qualitatively distinct behavior to the other curves.

\begin{figure}
  \centerline{\epsfig{file=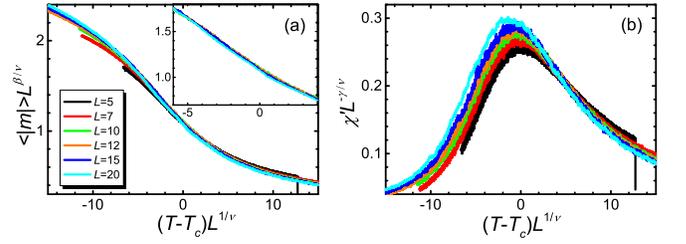,width=1.0\columnwidth}}
  \caption{\label{fssc403} (Color online) FSS in cooling at a fixed rate $R=0.000~03$ of (a) $M$ and (b) $\chi'$ on different lattice sizes given in the legend. The inset in (a) enlarges the curves in the absence of the two smallest lattice sizes.}
\end{figure}
As demonstrated in Figs.~\ref{fssc}(b) and~\ref{fssc}(d), the FSS of $\langle |m|\rangle$ and $\chi$ appear quite good similar to the corresponding figures in Fig.~\ref{fssh} in heating. However, in stark contrast with heating, the FSS of $\chi'$ appears reasonable only for $T>T_c$; the low-temperature side including $T_c$ is violated. This cannot stem from the sample size because the separation of different lattice sizes depends on $L$ systematically. In Fig.~\ref{fssc403}, we show the results for $R=0.000~03$ for which $L=20$ exhibits good FSS in heating. The violation of FSS for $\chi'$ at the low-temperature side remains though the rescaled curves appears closer compared with those for $R=0.000~1$ in Fig.~\ref{fssc}. Corrections to scaling~\cite{Wegner} cannot be attributed to either, although they are responsible for the poor collapses of the curves of the two smallest lattice sizes as can be seen from Fig.~\ref{fssc}(d), \ref{fssc}(f), and \ref{fssc403}. Indeed, the collapse of $\langle |m|\rangle$ becomes better if the two are removed as shown in the inset in Fig.~\ref{fssc403}(a). However, the violation of FSS for $\chi'$ remains even without the two curves.

Summarizing, at a fixed $R$, FSS is valid except $\langle m\rangle$ and $\chi$ in heating and $\chi'$ in cooling, all in the ordered phase. On the one hand, in heating, as the phases fluctuations are reduced when the absolute values such as $\langle |m|\rangle$ and $\chi'$ are employed, FSS is remedied. On the other hand, in cooling, the phases fluctuations are so strong that $\langle m\rangle$ is not a valid order parameter and $\chi$ has no peak. Again, when the absolute value is used, the FSS of $\langle |m|\rangle$ appears good. However, that of $\chi'$ is only good in the disordered phase. This goodness in the disordered phase rules out a global revised scaling form similar to Eq.~(\ref{rescaling_huang}), because it can then be destroyed. Although this might indicate some effects of phase ordering, the good scaling of $\chi$ and $\langle |m|\rangle$ seems to exclude them. As pointed out above, for $R=0.000~03$, $RL^r\approx1.64$ falling within the FSS regime even for the largest $L=20$ and thus corrections to scaling can be safely neglected, as the scaling of $\langle |m|\rangle$ and also $\chi$ show, though the two smallest $L$ values may need. Therefore, the violation of FSS in cooling appears surprising.

\subsection{\label{ftsofls}FTS on fixed lattice sizes}
In this section, we study the FTS of both heating and cooling on fixed lattice sizes.

\begin{figure}
  \centerline{\epsfig{file=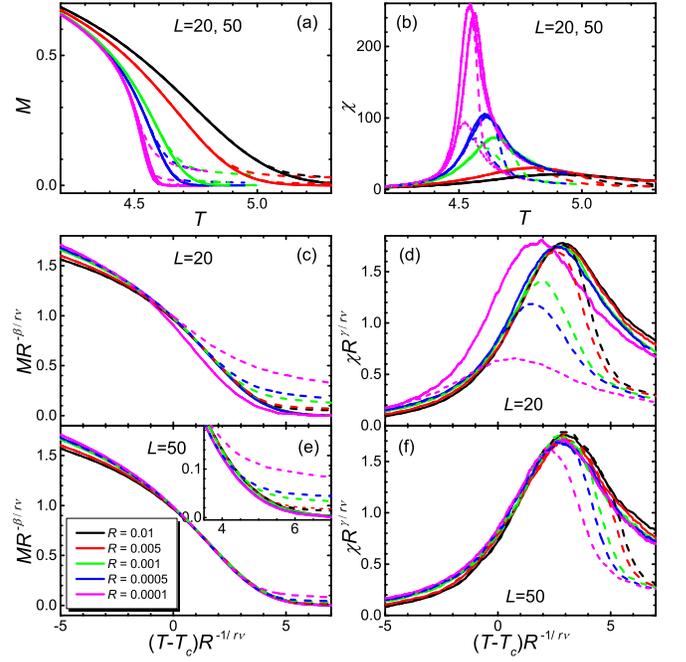,width=1.0\columnwidth}}
  \caption{\label{ftsh} (Color online) (a) $M$ and (c) and (e) its FTS and (b) $\chi$ and (d) and (f) its FTS in heating on fixed lattice size of $L=20$, (c) and (d), and $L=50$, (e) and (f). In contrast with Figs.~\ref{fssh}--\ref{fssc403} and all following ones in this subsection, here dashed lines stand for $\langle |m|\rangle$ and $\chi'$ while solid lines for $\langle m\rangle$ and $\chi$ for clarity of scaling collapses. The inset in (e) magnifies the curves in the disordered phase. In (a) and (b), the curves of both lattice sizes are displayed together. $\langle m\rangle$ and $\chi$ of the different lattice sizes overlap except for the smallest rates. The dashed curves of $\langle |m|\rangle$ and $\chi'$,  deviating from $\langle m\rangle$ and $\chi$, respectively, at higher temperatures have a larger $L$. In (b), for the two smallest rates, the right $\chi$ peaks have a larger $L$. All panels share the same legend. Three curves in both (a) and (b) have already displayed in Fig.~\ref{hcxm}.}
\end{figure}
Figure~\ref{ftsh} shows $M$ and $\chi$ and their FTS in heating on two fixed lattice sizes. In contrast with Fig.~\ref{fssh} in the FSS regime, the transition is now delayed as mentioned there; the larger $R$ is, the stronger the hysteresis. In the FTS regime, the evolution is controlled by $\xi_R$ rather than $L$. This is clearly reflected in Figs.~\ref{ftsh}(a) and~\ref{ftsh}(b) in that the large $R$ $\langle m\rangle$ and $\chi$ curves for the two different lattice sizes overlap, except for the smallest $R=0.000~1$ and almost the second smallest one in Figs.~\ref{ftsh}(b). This means that the smallest rate for $L=20$ is not in the FTS regime, though $\xi_R\approx12.5<L=20$ for $R=0.000~1$, or $L^{-1}R^{-1/r}\approx0.627<1$. In the last section, we also noted that this same parameters violates FSS. It must thus lie in the crossover regime between FTS and FSS, as has already been pointed out there.

From Figs.~\ref{ftsh}(a) and~\ref{ftsh}(b), it can be seen that $\langle m\rangle$ and $\chi$ are again different from $\langle |m|\rangle$ and $\chi'$, respectively. Although the differences for large $R$ values appear at $T>T_c$, opposite to FSS in Fig.~\ref{fssh}, those for small $R$ values can occur before the $\chi$ peaks where the transition is happening. Accordingly, the differences reveal again the phases fluctuations rather than spin fluctuations. Figures~\ref{ftsh}(a) and~\ref{ftsh}(b) show that the deviations of $\langle m\rangle$ and $\chi$ to $\langle |m|\rangle$ and $\chi'$, respectively, for a fixed $L$ are reduced as $R$ increases or $\xi_R$ decreases. They also diminish for a fixed $R$ in the FTS regime on the larger $L$. Both trends are related to more regions of the size $\xi_R$. However, in FSS, the deviations occur at lower temperatures for the smaller $L$ and $R$, see Figs.~\ref{fssh}(a),~\ref{fssh}(c), and ~\ref{fssh403}(a). Here, they are completely opposite to the case of FSS in heating but in accordance with the opposite trend of hysteresis versus advance. This is may be understood as the driving suppresses fluctuations to the scale set by $\xi_R$ or $\zeta_R$, similar to the origin of the hysteresis mentioned above, and thus the phases fluctuations can only occur at high temperatures. Indeed, the fluctuations, as indicated by the $\chi$ peaks, are reduced as $R$ increases for a fixed $L$ due to the smaller $\xi_R$. However, they are hardly affected by $L$ in the FTS regime as the $\langle m\rangle$ and $\chi$ curves for the two different lattice sizes overlap for the large $R$ values. It seems that the size dependence of the deviations of the two sets of the observables is related to the size dependence of $\langle |m|\rangle$. As seen in Figs.~\ref{fssh}(a) and~\ref{fssc}(a), $\langle |m|\rangle$ depends on $L$ significantly for $T>T_c$. It takes on almost the same value for the same $L$ no matter whether in heating or in cooling. This is also true for FTS, as can be seen in Figs.~\ref{hcxm}(c) and~\ref{hcxm}(d) as well as~\ref{ftsh}(a), because far away from $T_c$, a system is controlled by its short correlation length. Accordingly, all curves of a same lattice size collapse independently on their rates. This is not, of course, due to the choice of the initial configurations, peculiarities of the algorithm used, etc.. In fact, such a long tail of finite $\langle |m|\rangle$ at high temperatures was found in the early Monte Carlo simulations~\cite{Landau76} and agrees with its FSS $\langle |m|\rangle\sim L^{-\beta/\nu}$, Eq.~(\ref{FSSM}), and thus, as pointed out in the last section, a finite-size effect.

\begin{figure}
\centering
\centerline{\epsfig{file=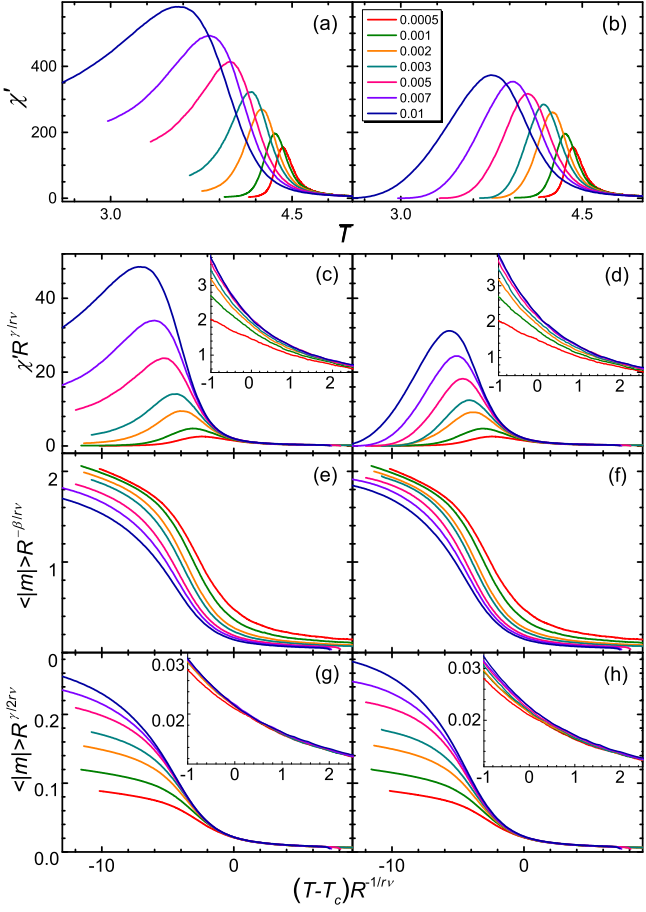,width=1.0\columnwidth}}
\caption{(Color online)\label{FTS_30} (a) and (b) $\chi'$, (c) and (d) its FTS, (e) and (f) FTS of $\langle|m|\rangle$, and (g) and (h) revised FTS, Eq.~(\ref{rescaling_huang}), of $\langle|m|\rangle$ averaged over all samples (left column) and samples with an unsaturated magnetization excluded (right column) upon cooling at the rates given in the legend on $20\times20\times20$ simple cubic lattices. The inset in each panel zooms in on the curves in the hight-temperature regimes. Panels in each row share identical ordinates. All panels share the same legend.}
\end{figure}
One sees from Figs.~\ref{ftsh}(c)--\ref{ftsh}(f) that while $\langle m\rangle$ and $\chi$ show FTS well and $\langle |m|\rangle$ and $\chi'$ also display better FTS as the lattice sizes get larger, the FTS of $\chi'$ is still poor on $L=50$ lattices at high temperatures. In fact, as seen in the inset in Fig.~\ref{ftsh}(e), the FTS of $\langle |m|\rangle$ is also poor at those temperatures. This is different from the FSS in heating in which it is $\chi$ and $\langle m\rangle$ that display bad scaling but similar to the FSS in cooling in which $\chi'$ shows poor scaling though $\langle |m|\rangle$ appears good. Note that the poor scaling here does not arise from the sample size. We checked that more samples only affect slightly the top of the peak.

To study FTS in cooling, we first note that, as seen in Fig.~\ref{hcxm}, in cooling $\chi'$ does not vanish and still exhibits large fluctuations at low temperatures different from the FSS case, as mentioned in the last section. This is because the magnetization of each sample can assume any value between the plus and the minus saturated magnetization in the FTS regime due to the phases fluctuations, whereas it takes on its equilibrium value (with thermal fluctuations around it) at sufficiently low temperatures in the FSS regime. We can thus remove those samples whose magnetization is smaller than a threshold below the peak temperature. As a consequence, the shape of $\chi'$ becomes normal, though the inverse rate dependence as compared to heating remains, as is illustrated in Figs.~\ref{FTS_30}(a) and~\ref{FTS_30}(b), which also show the reduction of the fluctuations (the peak size) with the removal.  However, the method is rather rough in that both the threshold and the temperature chosen are rather ad hoc, though it confirms the origin of the fluctuations.

\begin{figure}
\centering
\centerline{\epsfig{file=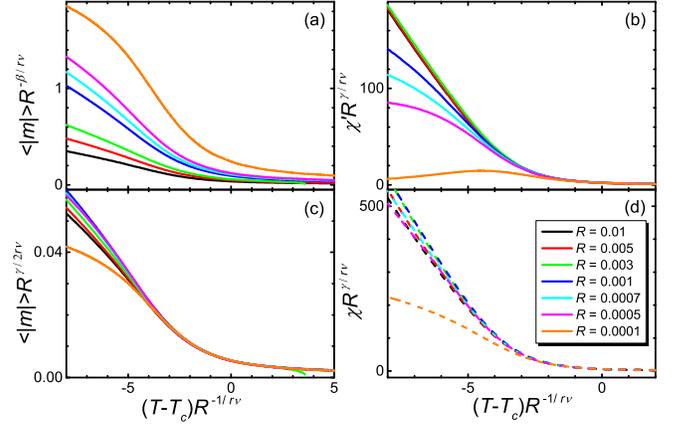,width=1.0\columnwidth}}
\caption{(Color online)\label{FTS_50} FTS of (a) and (c) $\langle|m|\rangle$ and (b) and (d) $\chi$ upon cooling at the rates given in the legend on $50\times50\times50$ simple cubic lattices. All panels share the same legend.}
\end{figure}
In Fig.~\ref{FTS_30}, we also display the FTS of $\chi'$ and $\langle |m|\rangle$ averaged over all samples and samples with a saturated magnetization only. We further depict the scaling of $\langle |m|\rangle$ using Eq.~(\ref{rescaling_huang}) with the $L^{-d/2}$ factor omitted because of the fixed $L$. One sees generally and in particular from the insets that removal of the unsaturated samples does not improve but may even worsen the scaling in Fig.~\ref{FTS_30}(h). As pointed out above, the removal is rough, we therefore do not consider it further in the following. One sees also that the scaling of $\chi'$ is acceptable only for rates bigger than $0.003$, corresponding to $\xi_R\approx4.9$, or $L^{-1}R^{-1/r}\approx0.246$, consistent with the crossover value of $0.627$ found above. Moreover, the scalings only extend to a small range below $T_c$. However, the scaling of $\chi'$ is quite remarkable upon noticing its inverse rate dependence in comparison with heating. In addition, the revised scaling for $\langle |m|\rangle$ is quite good, as seen in Fig.~\ref{FTS_30}(g). To the contrast, without the revised, the original scaling Eq.~(\ref{FTSM}) cannot describe the scaling behavior of $M$ both above and below $T_c$ at all.

These results are confirmed in Fig.~\ref{FTS_50}, in which a larger lattice size is used. One sees again that without taking the phases fluctuations into account, the na\"{\i}ve FTS, Eq.~(\ref{FTSM}), does not work at all even for the large lattice size. As $L$ increases, the range that obeys the scaling extends to further lower temperatures, including that of $\chi'$ despite its peculiar feature. In addition, the scaling of $\chi$ is somehow better than that of $\chi'$.

Therefore, the standard FTS of the order parameter, Eq.~(\ref{FTSM}), cannot at all describe the scaling behavior of $\langle |m|\rangle$ both above and below $T_c$ in cooling on fixed lattice sizes. The FTS of $\chi'$ along with $\langle |m|\rangle$ in heating is also violated in the disordered phase. Nevertheless, the FTS of all other observables studied in both heating and cooling, including the peculiar $\chi'$ and the revised scaling for $\langle|m|\rangle$ in cooling, is quite good and extends to a larger range for bigger lattice sizes. This seems to indicate that phase ordering cannot enter at least to this range provided that a proper scaling form is employed.

\subsection{Summary of the violations of FSS and FTS\label{sumvio}}
\begin{table}
\caption{\label{tab1} Summary of the observables and the phases that violate FSS and FTS in heating and in cooling. The breaking-of-self-similarity exponent $\sigma$ that rectifies the violated scaling of the primary (pri) observable (obs) in each entry is also given, including the 2D results~\cite{Yuan}. The secondary (sec) observables in the parentheses exhibit apparently good scaling because of their regular terms. The figures in the parentheses mark the ranges within which the scaling collapses are similar.}
\begin{ruledtabular}
\begin{tabular}{lcccc}
&\multicolumn{2}{c}{FSS}& \multicolumn{2}{c}{FTS} \\\cline{2-3}\cline{4-5}
\rule{0pt}{10pt}&heating & cooling&heating& cooling \\
\hline
\rule{0pt}{9pt}pri obs&$\chi$ & $\chi'$     &$\chi'$ & $\langle |m|\rangle$ \\
sec obs&$\langle m\rangle$&$(\langle |m|\rangle)$&$\langle |m|\rangle$ &$(\chi')$ \\
phase                    &ordered                 & ordered     & disordered                & ordered\\
$\sigma$~(2D)            &$-2.75(15)$             &$\beta/2\nu$ &$-0.625(25)$               &$\frac{d}{2}+\frac{\beta}{2\nu}\pm\frac{\beta}{4\nu}$\\
\rule{0pt}{10pt}$\sigma$~(3D)            &$-d\pm0.15$             &$\beta/2\nu$ &$-0.584(50)/\nu$           &$\frac{d}{2}+\frac{\beta}{4\nu}\pm\frac{\beta}{8\nu}$\\
\end{tabular}
\end{ruledtabular}
\end{table}
In Table~\ref{tab1}, we summarize the measured quantities that are violated in either FSS or FTS and either in heating or in cooling. The phase that the violation occurs is also given. One sees the violations occur in a set except for the cases of FSS and FTS in cooling whose secondary observables are parentheses. We will see in Sec.~\ref{bressy} below that this is because the primary observables $\chi'$ and $\langle |m|\rangle (L^{-1}R^{-1/r})^{-d/2}$, the latter having been reduced by the revised FTS, of these two cases exhibit relatively weak violations in the sense that their corresponding exponents are relatively small. Consequently, the regular terms similar to Eq.~(\ref{msigma}) and in Eq.~(\ref{xsigmat}) for the two cases, respectively, appear dominant. We point out also that although the phase that the violation occurs for FTS in cooling is designated as ordered in Table~\ref{tab1}, the revised scaling is needed in fact in both the ordered and the disordered phases, as will be discussed in Sec.~\ref{bressy}. Only a further rectification occurs in the ordered phase. Including this special case, one sees that in all but one cases the violations occur in the ordered phase. This is reasonable because the effects of the phases fluctuations are prominent in the ordered phase and, in particular, near the $\chi$ peaks where $\langle m\rangle$ and $\langle |m|\rangle$, together with $\chi$ and $\chi'$, are different. In fact, this rule is also followed by the exceptional case of FTS in heating. In heating, the initial condition is always an ordered state that breaks the up-down symmetry. For FTS on fixed lattice sizes, as pointed out above, the deviation of $\langle m\rangle$ from $\langle |m|\rangle$ and hence the phases fluctuations usually occur near the peak of $\chi$ and onwards and are thus designated as the disordered phase, as Fig.~\ref{ftsh} shows. By contrast, for FSS in heating at fixed rates, the small lattice sizes make the phases fluctuations exhibit even in the ordered phase as seen in Fig.~\ref{fssh}. In fact, except for the special case of FTS in cooling, the primary observables are all $\chi$ or $\chi'$. We will see in Sec.~\ref{bressy} that the ordered or the disordered phase of the violations corresponds to the low- or the high-temperature slope of the $\chi$ or $\chi'$ peak. We also note that the violated scalings are always exhibited in the set with absolute values in all cases but FSS in heating. All in all, the general rule here is that all rules have exceptions and why this is so is yet to studied.

From Table~\ref{tab}, one observes that the violations only display in one set of the observables. Consequently, they can only occur when both sets of the observables behave differently. This happens when the phases fluctuations appear. Consequently, no phases fluctuations, no violation of scaling. However, we will see in the following section that the phases fluctuations themselves are not sufficient for the violations. Once the extrinsic self-similarity is kept, no violation occurs at all even though the phases fluctuations are still there, provided that other sub-leading terms and corrections to scaling that have not been considered in the theory are negligible. Therefore, phases fluctuations are the necessary condition for the violations of scaling.

\subsection{Full scaling forms\label{fsf}}
\begin{figure}
  \centerline{\epsfig{file=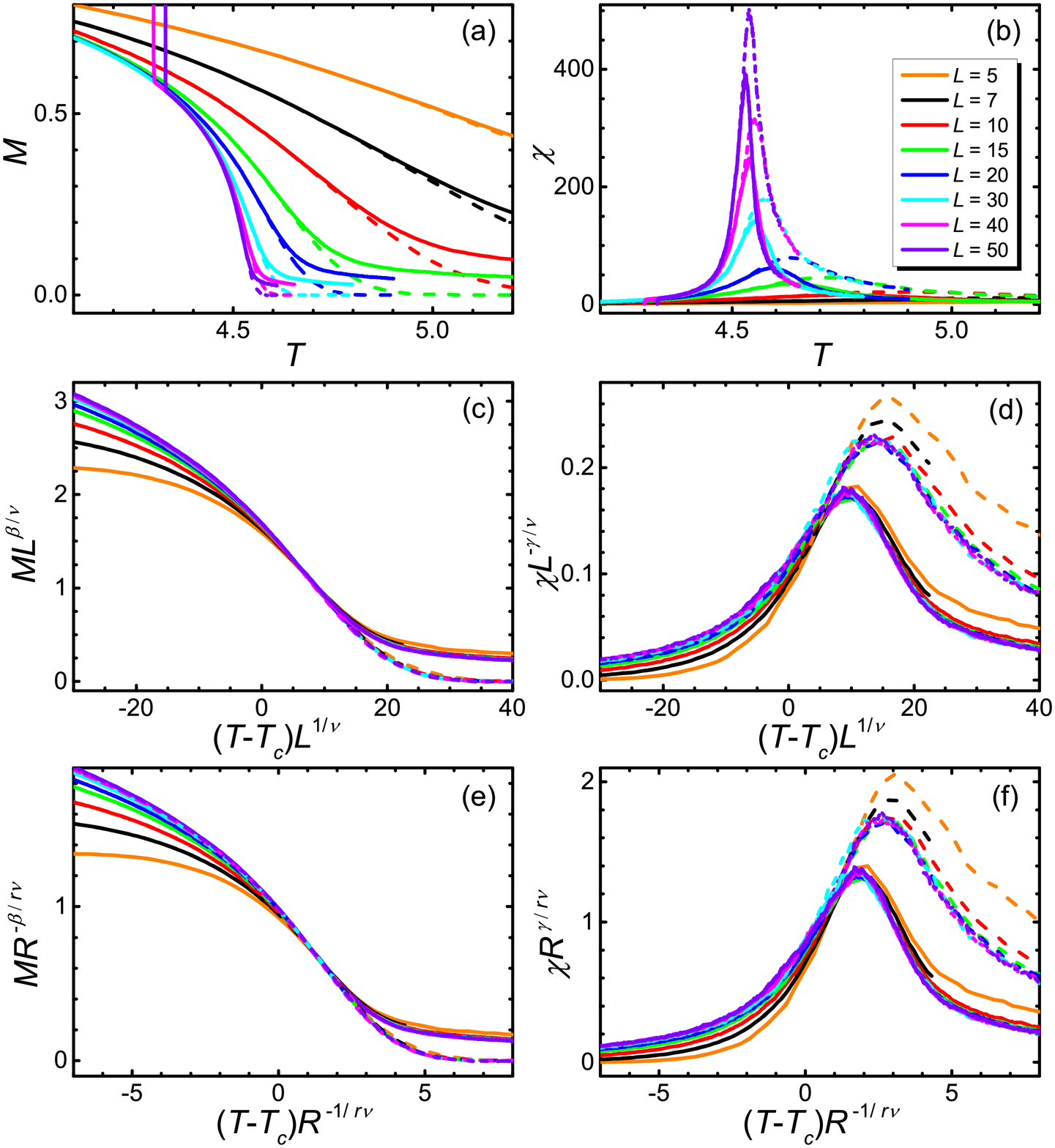,width=1\columnwidth}}
  \caption{\label{ftshlr} (Color online) (a) $M$ and (b) $\chi$ and (c) and (d) their respective FSS and (e) and (f) their respective FTS in heating with fixed $L^{-1}R^{-1/r}\approx0.3541$, which is obtained from $L=10$ and $R=0.01$. Solid lines denote results of $\langle|m|\rangle$ and $\chi'$ while dashed lines of $\langle m\rangle$ and $\chi$. All panels share the same legend.}
\end{figure}
We now study the full scaling forms Eqs.~(\ref{FTSX}) to~(\ref{rescaling_huang}). To this end, we fix $L^{-1}R^{-1/r}$ to a constant and thus a given different $L$ means a different $R$. This is different from the above studies in which either $L$ or $R$ is fixed.

Figure~\ref{ftshlr} shows both the FSS and FTS of $M$ and $\chi$ in heating according to Eqs.~(\ref{FSSM}),~(\ref{FSSX}),~(\ref{FTSM}), and~(\ref{FTSX}). One sees that as $L$ gets larger and correspondingly $R$ smaller, both FSS and FTS become better and extend to a larger range. This is more transparent if one disregards the three smallest lattice sizes. Moreover, owing to the phases fluctuations, $\langle m\rangle$ and $\langle|m|\rangle$ as well as $\chi$ and $\chi'$ are again different in the disordered phase. However, they all satisfy the scalings and the violated scaling collapses for $\langle |m|\rangle$ and $\chi'$ in Fig.~\ref{ftsh} disappear. Therefore, the phases fluctuations themselves are not sufficient for the violations of the scaling. In addition, the scalings in the high-temperature part appear better than those in the low-temperature parts. The deviation of the small lattice sizes must arise from other sub-leading or correction terms~\cite{Wegner} that are not included in the scaling forms. Note that the fixed value of $L^{-1}R^{-1/r}\approx0.3541$ lies on the verge of FTS regime and corresponds to $RL^r$ of about $43.86$, a number much bigger than the previous ones of $1.92$ or even $5.48$ for $R=0.0001$ and $L=20$ and thus the system cannot fall in the FSS regime. For $R=0.0001$, for example, the fixed value leads to $L\approx35.4$, which cannot show FSS as clearly seen in Figs.~\ref{fssh}(e) and~\ref{fssh}(f). However, once $L^{-1}R^{-1/r}$ is fixed, Figs.~\ref{ftshlr}(c) and~\ref{ftshlr}(d) show clearly that FSS can still describe the FTS regime, because one can freely insert a certain power of the constant factor to change the scaling form from one type to the other, as Eq.~(\ref{fttofs}) demonstrated.

\begin{figure}
  \centerline{\epsfig{file=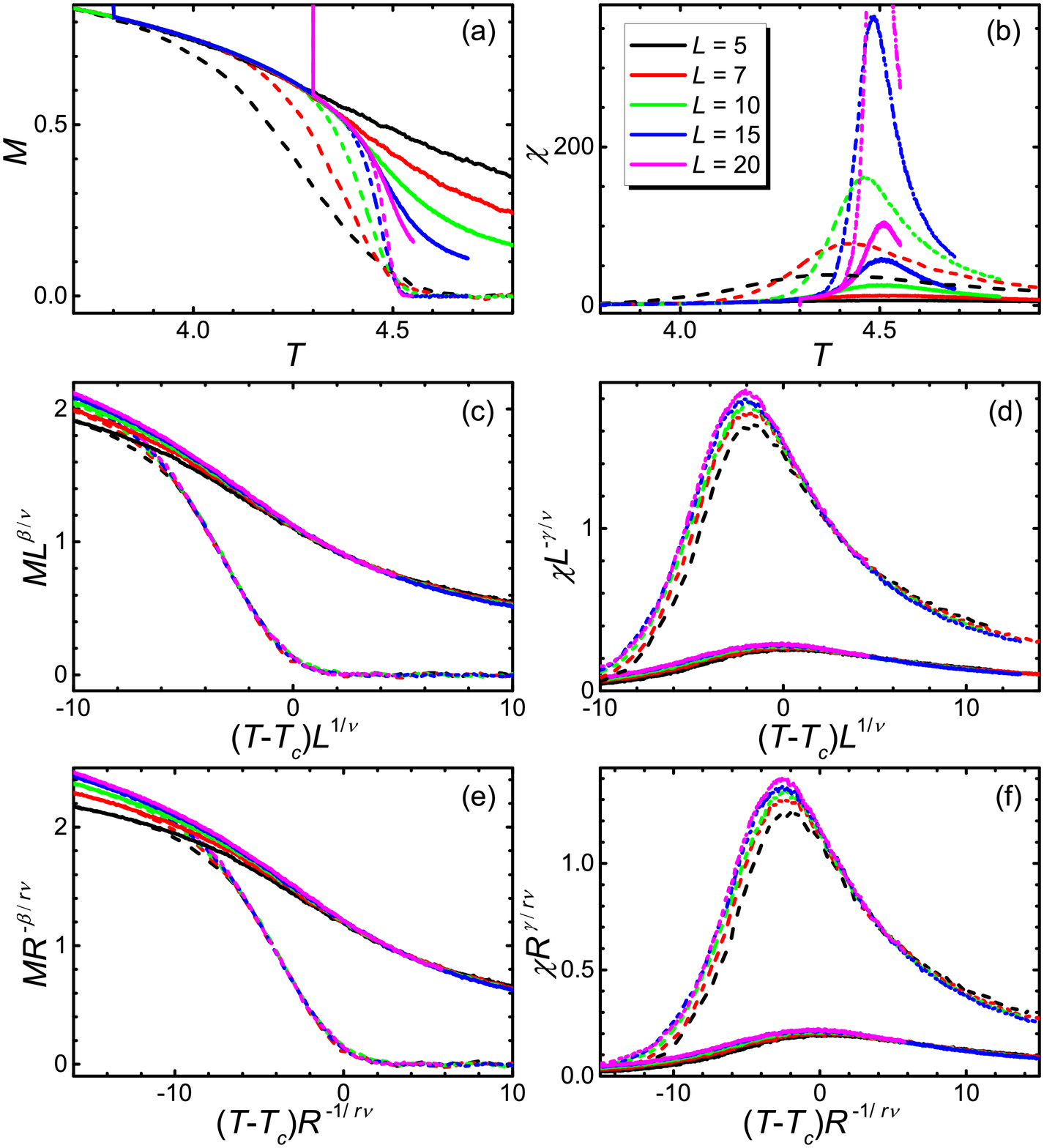,width=1\columnwidth}}
  \caption{\label{ftshlr1} (Color online) (a) $M$ and (b) $\chi$ and (c) and (d) their respective FSS and (e) and (f) their respective FTS in heating with fixed $L^{-1}R^{-1/r}\approx1.152$, which is determined by $L=7$ and $R=0.0005$. Solid lines denote results of $\langle|m|\rangle$ and $\chi'$ while dashed lines of $\langle m\rangle$ and $\chi$. In (b), the top of the $L=20$ peak has been not cut for visibility of other peaks. All panels share the same legend.}
\end{figure}
The good FSS of both $\langle |m|\rangle$ and $\chi'$ might seem usual. For the fixed $L^{-1}R^{-1/r}\approx0.3541$, the rates on $L=5$ and $7$ are about $0.1248$ and $0.0367$, respectively. As mentioned above, these relatively large rates substantially reduce the deviations of $\langle m\rangle$ from $\langle|m|\rangle$, which are clearly exhibited in Fig.~\ref{ftshlr}(a). The driven transitions on these small lattices are thus not advanced compared to Fig.~\ref{fssh}. In Fig.~\ref{ftshlr1}, we depict the FSS and FTS of heating at $L^{-1}R^{-1/r}\approx1.152$ or $RL^r\approx0.5982$, which puts the system firmly in the FSS regime. Figure~\ref{ftshlr1}(a) shows clearly that the transitions are advanced before $T_c=4.511~42(6)$ since all $\langle m\rangle$ curves become almost zero at $4.5<T_c$. Nevertheless, both FSS and FTS of both sets of the observables are all remarkably well once the small lattices and thus large rates are removed, even though the difference between the two sets of the observables is even more evident.

\begin{figure}
  \centerline{\epsfig{file=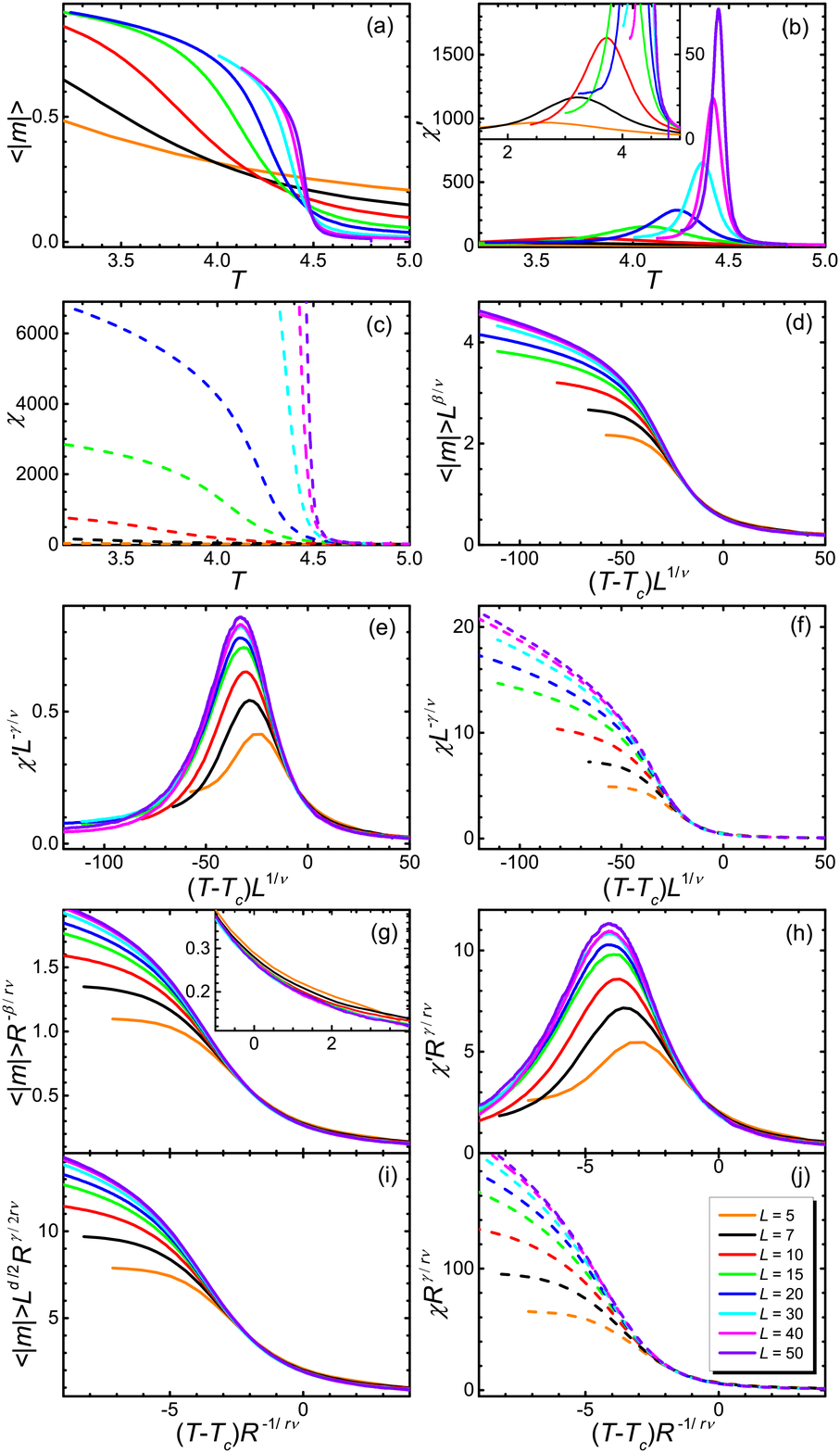,width=1\columnwidth}}
  \caption{\label{ftsclr} (Color online) (a) $\langle|m|\rangle$, (b) $\chi'$, and (c) $\chi$ and (d), (e), and (f) their respective FSS and (g), (h), and (i) their respective FTS in cooling with fixed $L^{-1}R^{-1/r}\approx0.2687$, which is derived from $L=30$ and $R=0.0005$. (j) the revised FTS, Eq.~(\ref{rescaling_huang}), of (a). The insets in (b) and (g) zoom in on the low- and high-temperature parts, respectively. The inset in (g) exhibits the broadening of the rescaled curve by the small $L$ curves. All panels share the same legend.}
\end{figure}
We now turn to the qualitative difference exhibited in Fig.~\ref{hcxm} and the systematic deviations in Figs.~\ref{fssc}(f) and~\ref{fssc403} in cooling. Figure~\ref{ftsclr} displays the FSS and FTS of $\langle|m|\rangle$ and $\chi$ and $\chi'$ in cooling for fixed $L^{-1}R^{-1/r}$ again according to Eqs.~(\ref{FSSM}),~(\ref{FSSX}),~(\ref{FTSM}), and~(\ref{FTSX}), as well as Eq.~(\ref{rescaling_huang}). One first notices upon comparing Fig.~\ref{ftsclr}(a) to~\ref{ftshlr}(a) that different $\langle|m|\rangle$ curves of different $L$ cross each other below $T_c$ in cooling. The large fluctuations of $\chi'$ at low temperatures for small rates can also be observed in Fig.~\ref{ftsclr}(b) and its inset. These again remind us of the uniqueness of cooling. However, with $L^{-1}R^{-1/r}$ being fixed, the rate dependence of the peak heights now resembles that in heating, in sharp contrast with Fig.~\ref{hcxm}, although the $\langle|m|\rangle$ curves now behave distinctively as mentioned. These therefore show the dramatic effects of $L^{-1}R^{-1/r}$ in cooling. Moreover, the FSS and FTS now describe the data well in cooling for sufficiently large $L$ and correspondingly sufficiently small $R$ (so that corrections to scaling can be ignored) for a large region in both sides of $T_c$ similar to heating, as is clearly seen in Fig.~\ref{ftsclr}. Also, both Eqs.~(\ref{FTSM}) and Eq.~(\ref{rescaling_huang}) are equally well because the latter is a special form of the former. So are FSS and FTS regardless of the fact that $RL^r$ is now almost $120$. These results, together with those in heating in this section, should already confirm the validity of the full scaling forms in describing the processes no matter whether in the FTS or the FSS regime. Indeed, a similar plot has been shown in Fig.~3(b) of Ref.~\cite{Yuan} for a fixed $RL^r\approx0.548$ in the FSS regime of the 3D Ising model. It is no doubt that the poor FSS in Figs.~\ref{fssc}(f) and~\ref{fssc403} and the FTS of $\langle|m|\rangle$ in Figs.~\ref{FTS_30}(e) and~\ref{FTS_50}(a) all disappear in such a plot. Although the tops of the $\chi'$ peaks scale not so well even for the large $L$, this must result from the sample size, since fluctuations are violent there. Therefore, FTS itself well describes the cooling transition across the critical point for large $L$ and small $R$ and no phase ordering is needed down to at least $(T-T_c)R^{-1/r\nu}<-5$ as seen in Fig.~\ref{ftsclr}. Note that the frozen temperature $\hat{T}$ at which the KZ scaling is reckoned is defined just at $(\hat{T}-T_c)R^{-1/r\nu}=\pm1$~\cite{Huang}.

We have confirmed the validity of the full forms Eqs.~(\ref{FTSX}) and~(\ref{FSSX}) in describing the FTS and FSS of both sets of the observables in both heating and cooling, including the seemingly large deviation away from $T_c$ of FSS in heating, provided that other sub-leading and corrections to scaling are negligible. The scaled variables $L^{-1}R^{-1/r}$ in FTS or $RL^r$ in FSS cannot be neglected in the cases in which scalings are violated even though they are small. Looking back on the inverse rate dependence in Fig.~\ref{hcxm} and the systematic dependence on $L$ in the ordered phase in Figs.~\ref{fssc}(f) and~\ref{fssc403}(b), one thus convinces oneself that they cannot stem from phase ordering. Rather, it is the scaled variable $RL^r$, albeit small, that gives rise to the qualitative difference and the poor scalings at low temperatures in cooling at both fixed $R$ and fixed $L$.

By definition, $\chi'$ removes the effect of the phases fluctuations and ought to probe the magnitude fluctuations. Fixing $L^{-1}R^{-1/r}$ renders it in cooling normal indicates that the normal trend of the dependence of $\chi'$ on $R$, viz., the increasing $\chi'$ peak with decreasing $R$, indeed reflects the larger cluster sizes and hence the larger magnitude fluctuations. This is in conformity with the same trend in the FSS regime as exhibited in Figs.~\ref{fssh}(c) for heating and~\ref{fssc}(e) for cooling. As phase ordering is ruled out, the qualitatively different dependence on $R$ of $\chi'$ in heating and cooling must thus stem from the self-similarity breaking of the phases fluctuations, since the most important difference between heating and cooling is their phases fluctuations. Accordingly, the phases fluctuations ought to contribute to the special behavior of $\chi'$ in cooling, as mentioned in Sec.~\ref{phfl}. Indeed, one flip of a single large cluster changes $m$ of a sample at an instance or a temperature and hence $\langle|m|\rangle$ and $\chi'$. If self-similarity is present by fixing $L^{-1}R^{-1/r}$, the change for different lattices is the same on average and thus does not alter the dependence of $\chi'$ on $R$. Otherwise, the bigger the number of the large phases clusters, the smaller the contribution of one turnover of a single cluster. This suppresses the magnitude of $\langle|m|\rangle$ and $\chi'$ at a temperature but expands the distribution of $m$ and hence increases $\chi'$, a feature which Fig.~\ref{hcxm}(b) exhibits.

The results of this section solve a superficial conflict in our previous descriptions. We pointed out in Sec.~\ref{ftsofls} that the usual FTS form, Eq.~(\ref{FTSM}), cannot describe the scaling behavior of $\langle|m|\rangle$ at all; only the revised form, Eq.~(\ref{rescaling_huang}) can do. However, in Sec.~\ref{intro}, we also pointed out that both these two scaling forms can describe well the scaling behavior exactly at $T_c$, though their leading singularities differ, as shown in Ref.~\cite{Huang}. We emphasize that these two statements do not conflict. In Sec.~\ref{ftsofls}, we did not fix $L^{-1}R^{-1/r}$. Therefore, Eq.~(\ref{FTSM}) failed, as seen in Figs.~\ref{FTS_30}(e) and~\ref{FTS_50}(a). When we fix it, it works well, as demonstrated in Fig.~\ref{ftsclr}(g). However, in Ref.~\cite{Huang}, we worked exactly at $T_c$ only and plotted $\langle|m|\rangle R^{-\beta/r\nu}$ with respect to $L^{-1}R^{-1/r}$. In this way, the different values of the ordinate at $T_c$ in Figs.~\ref{FTS_30}(e) and~\ref{FTS_50}(a) appear then in different places of the abscissa in the $\langle|m|\rangle R^{-\beta/r\nu}$ versus $L^{-1}R^{-1/r}$ plot because they just have different $L^{-1}R^{-1/r}$ values and hence the scaling is satisfied. In other words, the full scaling function has been considered in this case and thus the results are no doubt good.

\subsection{Bressy exponents\label{bressy}}
In the last section, we have verified unambiguously that the full scaling forms of both FSS and FTS can well describe both sets of observables in both heating and cooling provided that other sub-leading terms and corrections to scaling are negligible. All the violated scalings summarized in Sec.~\ref{sumvio} arise thus solely from the scaled variables $L^{-1}R^{-1/r}$ or $RL^r$ even though they are small. In this section, we study the properties of the violations. We find that new breaking-of-(extrinsic-)self-similarity, or Bressy, exponents $\sigma$ can remedy the violations. These exponents lead to different leading singularities of the ordered and disordered phases for the primary observables, in stark contrast to the other observables and in usual equilibrium and dynamic critical phenomena.

To recapitulate, fixing $L^{-1}R^{-1/r}$ in FTS (or $RL^r$ in FSS) fixes the ratio of $L$ and $R^{-1/r}$ in FTS (or $L^z$ and $R^{-z/r}$ in FSS) and therefore ensures the spatial (or temporal) self-similarity of the phases fluctuations. This kind of self-similarity is related to the system size or the external driving and is thus the extrinsic self-similarity. Accordingly, all the mentioned violated of scalings result from the breaking of the extrinsic self-similarity. Although the violations appear just in one phase and a global revised form like Eq.~(\ref{rescaling_huang}) destroys the good scalings in the other phase, we can find exponents to remedy the violated scalings in just one phase according to the theory in Sec.~\ref{theory}. This, however, leads to different leading behavior in the two phases for the observables.

\begin{figure}
  \centerline{\epsfig{file=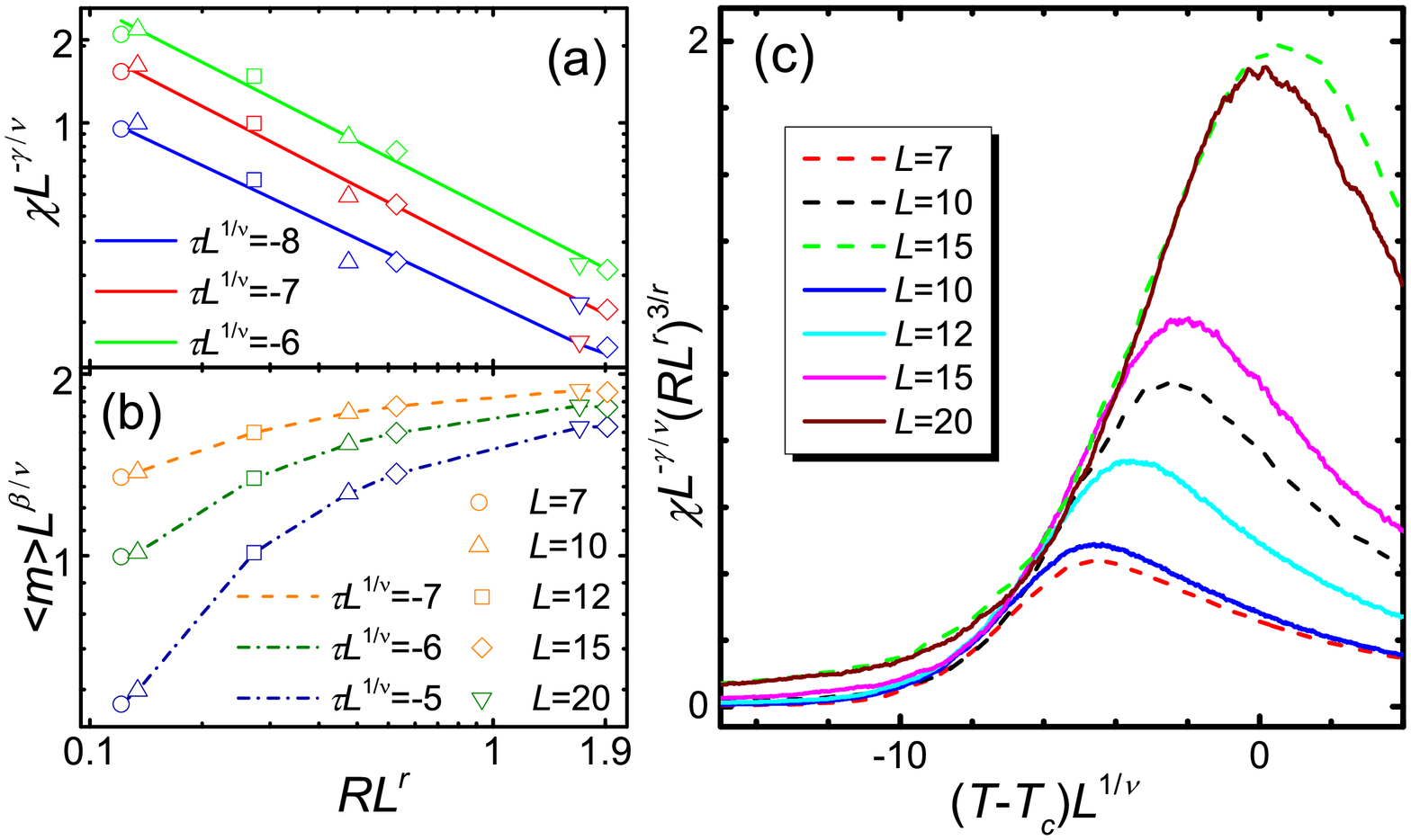,width=1\columnwidth}}
  \caption{\label{fssh3d} (Color online) (a) $\chi L^{-\gamma/\nu}$ and (b) $\langle m \rangle L^{\beta/\nu} $ at fixed values of $\tau L^{1/\nu}=(T-T_c)L^{1/\nu}$ given in their respective legends versus $RL^r$ in double logarithmic scales for FSS in heating. (c) $\chi L^{-\gamma/\nu}(RL^r)^{-\sigma/r}$ versus $(T-T_c)L^{1/\nu}$ for $\sigma=-3$. Solid lines in (a) are linear fits to the data points, chain lines in (b) connecting data points are a guide to the eyes, while, in (c), dashed lines have $R=0.000~1$ and solid lines $R=0.000~03$. The seven symbols in (a) and (b) represent the data that are calculated from the seven curves with the parameters shown in the legend in (c). The error bars are estimated to be about the sizes of the symbols. The slopes of the three fitted lines in (a) are $-0.69(11)$, $-0.75(8)$, and $-0.75(8)$ from up to down.}
\end{figure}
We first consider FSS in heating. In Fig.~\ref{fssh3d}(a), we show the dependence on $RL^r$ of $\chi L^{-\gamma/\nu}$ cut vertically at fixed values of $\tau L^{1/\nu}=(T-T_c)L^{1/\nu}$ from plots like Fig.~\ref{fssh}(d). The apparent linearity of the three lines indicates the power-law relation in consistence with Eq.~(\ref{bressyexp}). By contrast, similar cuts on $\langle m \rangle L^{\beta/\nu} $ shown in Fig.~\ref{fssh3d}(b) exhibit a relation compatible with Eq.~(\ref{msigma}). These indicate that $\chi$ is the primary observable which obeys Eq.~(\ref{bressyexp}) whereas $\langle m \rangle$ is the secondary one. Indeed, it is seen from Fig.~\ref{fssh3d}(c) that a Bressy exponent of $\sigma=-3\pm0.15$ collapses the low-temperature slope of the $\chi$ curves rather well, in comparison with the completely separation in the ordered phase shown in Fig.~\ref{fssh}(d) for just one $R$. Moreover, $\sigma/r\approx0.82$ agrees with the slopes in Fig.~\ref{fssh3d}(a). The range of $RL^r$ displayed in Fig.~\ref{fssh3d} is from $0.12$ to $1.92$ or so, more than sixteen times. The big value sits on the boundary of the FSS regime, as pointed out in previous sections, while the small value can be further lowered, we believe, by using smaller $R$. However, exactly at $R=0$, the static case, the usual FSS ought to be recovered. Accordingly, the crossover seems discontinuously. How the two behaviors connect exactly with each other is yet to be studied. In addition, as pointed out in the theory presented in Sec.~\ref{theory}, the same $\sigma$ also collapses well the $\langle m \rangle$ curves because it is transformed into $-\chi$ through Eq.~(\ref{msigma}). We note that it is remarkable that both $\chi$ and $\langle m \rangle$ can be collapsed by a single new exponent.

As pointed out in Ref.~\cite{Yuan} and listed in Table~\ref{tab1}, in two dimensions, we found $\sigma=2.75\pm0.15$, which could be $(2\gamma-6\beta)/\nu$ or $(\gamma+8\beta)/\nu$ and even others using Eq.~(\ref{law}) from the exact 2D critical exponents. However, these expressions yield very different values using the 3D critical exponents listed in Sec.~\ref{model}. Therefore, $\sigma$ is likely a new exponent different from any extant ones but produces correctly both the 2D and the 3D values.

\begin{figure}
  \centerline{\epsfig{file=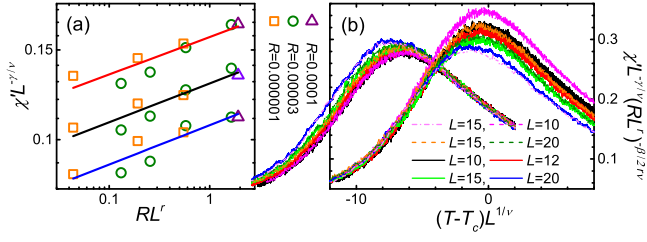,width=1\columnwidth}}
  \caption{\label{fssc3d} (Color online) (a) $\chi' L^{-\gamma/\nu}$ at fixed values of $(T-T_c)L^{1/\nu}=-7$, $-8$, and $-9$ (from up to down) versus $RL^r$ for the three $R$ values listed vertically in the middle in double logarithmic scales and (b) $\chi' L^{-\gamma/\nu}(RL^r)^{-\sigma/r}$ versus $(T-T_c)L^{1/\nu}$ with $\sigma=\beta/2\nu$ for $R=0.000~1$ (chain line), $0.000~03$ (solid lines), and $0.000~01$ (dashed lines) on various lattice sizes given in the legend for FSS in heating. The curves on the left are $\chi' L^{-\gamma/\nu}$ versus $(T-T_c)L^{1/\nu}$ in identical axes and scales with the curves within (b) but shifted by $(T-T_c)L^{1/\nu}=-6$ for clarity. Solid lines in (a) are linear fits to the data points, whose error bars are estimated to be about $0.01$, about twice the symbol sizes. The slopes of the three fitted lines in (a) are $0.074(32)$, $0.076(28)$, and $0.077(33)$ from up to down.}
\end{figure}
For FSS in cooling, the Bressy exponent $\sigma$ and its resultant collapse have been given in Fig.~3(b) of Ref.~\cite{Yuan}. Here, we add data of two more rates, $R=0.000~1$ and $0.000~01$, raising the range of $RL^r$ from $0.13$--$1.64$ up to $0.044$--$1.92$, more than three and a half times bigger. In Fig.~\ref{fssc3d}(a), the power-law dependence of $\chi' L^{-\gamma/\nu}$ at several fixed values of $(T-T_c)L^{1/\nu}$ on $RL^r$ is seen within the error bars, though the errors of the data are a bit large because of the small rates. This shows that $\chi'$ is the primary observable. Indeed, as exhibited in Fig.~\ref{fssc3d}(b), $\sigma=\beta/2\nu$ collapses all the curves in the ordered phase rather well, as compared to the original scaling collapse (the left curves), which resembles the curves in Figs.~\ref{fssc}(f) and~\ref{fssc403}(b) with less rates. Moreover, $\sigma/r\approx0.0711$ agrees with the slopes in Fig.~\ref{fssc3d}(a), though the latter have relatively large errors due to the small rates. Since this exponent is rather small, it can be expected that the secondary observable, $\langle |m| \rangle$ exhibits a good scaling collapse in Figs.~\ref{fssc}(b) and~\ref{fssc403}(a) and hence has appeared with parentheses in Table~\ref{tab1}. Conversely, this implies that $\langle |m| \rangle$ can only be secondary.

\begin{figure}
  \centerline{\epsfig{file=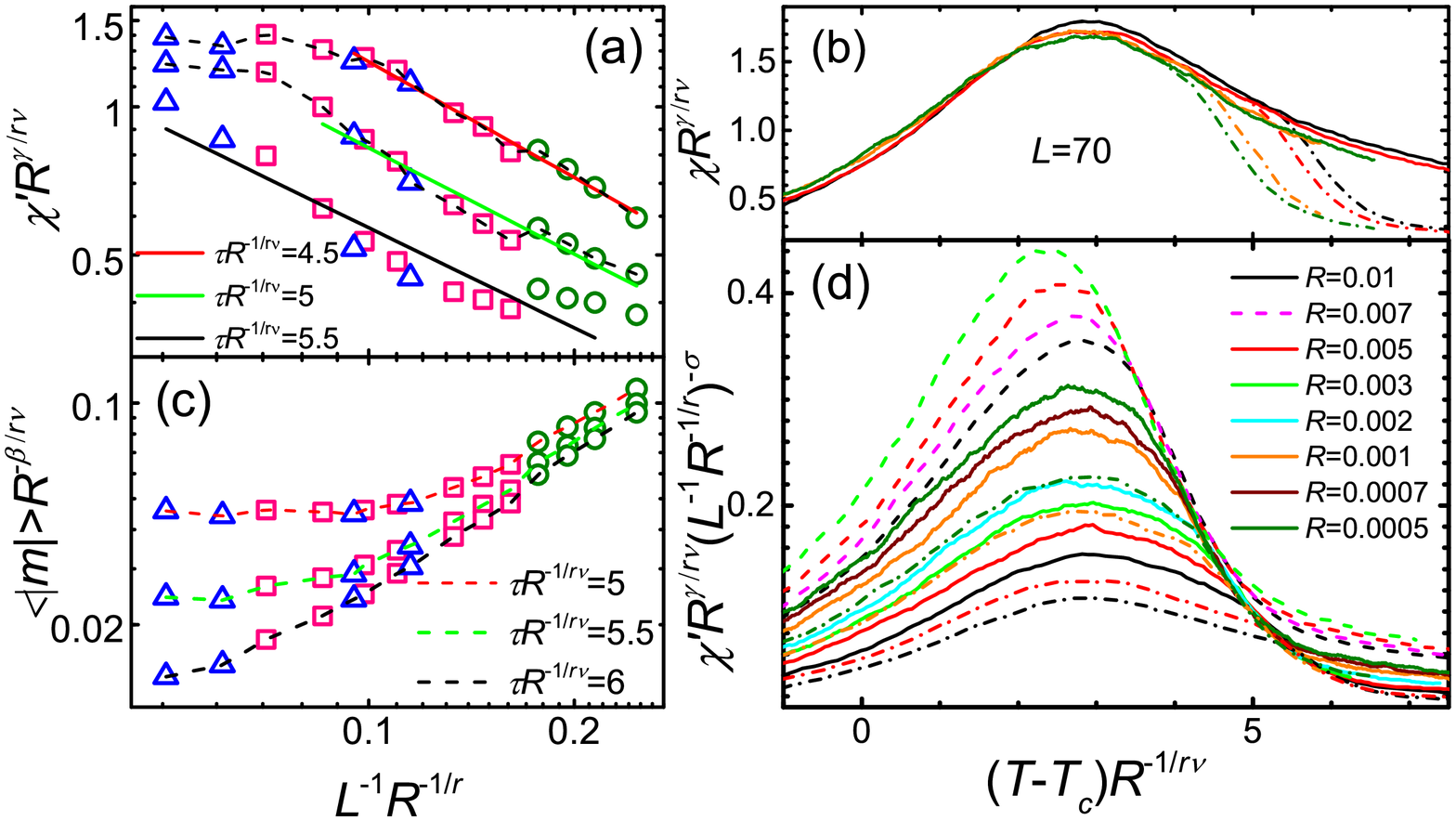,width=1\columnwidth}}
  \caption{\label{ftsh3d} (Color online) (a) $\chi' R^{\gamma/r\nu}$ and (c) $\langle |m| \rangle R^{-\beta/r\nu} $ on $L=20$ (circles), $L=50$ (squares), and $L=70$ (triangles) lattices at fixed values of $\tau R^{-1/r\nu}=(T-T_c)R^{-1/r\nu}$ given in their respective legends versus $L^{-1}R^{-1/r}$ in double logarithmic scales for FTS in heating. (b) $\chi R^{\gamma/r\nu}$ versus $(T-T_c)R^{-1/r\nu}$ in heating on $L=70$ simple cubic lattice. (d) $\chi' R^{\gamma/r\nu}(L^{-1}R^{-1/r})^{-\sigma}$ versus $(T-T_c)R^{-1/r\nu}$ with $\sigma=-(\gamma-2\beta)/\nu$ for the rates $R$ listed on $L=20$ (dashed lines), $L=50$ (solid lines), and $L=70$ (chain lines) lattices. In (a) and (c), dashed lines connecting symbols are only a guide to the eyes. Solid lines in (a) are linear fits to the data points within the ranges covered by the lines. In (b), chain lines denoting $\chi'$ are those displayed in (d) while solid lines stand for $\chi$. In (b) and (d), curves of identical colors have identical rates. The error bars are estimated to be about the sizes of the symbols. The slopes of the three fitted lines in (a) are $-0.79(7)$, $-0.72(12)$, and $-0.68(16)$ from up to down.}
\end{figure}
Next, we study FTS in heating. In Fig.~\ref{ftsh3d}(a), we show the dependence on $L^{-1}R^{-1/r}$ of $\chi' R^{\gamma/r\nu}$ at fixed values of $\tau R^{-1/r\nu}$. Different from Figs.~\ref{fssh3d}(a) and~\ref{fssc3d}(a), the data appeared not to fit a simple power law. However, as can be seen from Fig.~\ref{ftsh3d}(b) for $L=70$ and Fig.~\ref{ftsh}(f) for $L=50$, the $\chi$ and $\chi'$ curves do not yet separated from each other for the two small fixed values of $\tau R^{-1/r\nu}$ at large rates, corresponding to the data on the left. Accordingly, these data must follow the usual FTS and hence hardly depend on $L^{-1}R^{-1/r}$. Indeed, the left four data of $\tau R^{-1/r\nu}=4.5$ and three data of $\tau R^{-1/r\nu}=5$ are almost horizontal. Upon excluding these data, linearity emerges, though the errors in the slopes are relatively large. One reason comes from other sub-leading and/or corrections terms. One sees from Fig.~\ref{ftsh3d}(b), together with Figs.~\ref{ftsh}(d) and~\ref{ftsh}(f), that the large rate curves somehow deviate off the rescaled curves. For the same reason, we omit the rightmost data of $\tau R^{-1/r\nu}=5.5$ in the fit. On the other hand, from the inset of Fig.~\ref{ftsh}(e), at least for $R=0.001$ downwards on the $L=50$ lattice, $\langle |m| \rangle$ and $\langle m \rangle$ have already separated. Consequently, the behavior on the left in Fig.~\ref{ftsh3d}(c) arises from Eq.~(\ref{msigmat}) rather than the independence on $L^{-1}R^{-1/r}$ of the usual FTS as the case of $\chi'$ just noted. Therefore, $\langle |m| \rangle$ is the secondary observable and $\chi'$ is the primary one. Indeed, Fig.~\ref{ftsh3d}(d) shows that for $\sigma=-(\gamma-2\beta)/\nu\pm0.05$ the curves collapse quite well on the high-temperature slope of the peaks, which we have roughly termed as the disordered phase. This $\sigma\approx0.93$, albeit a little bigger, is compatible with the slopes within the errors in Fig.~\ref{ftsh3d}(a), noting the large errors in the slope as mentioned. Comparing with other Bressy exponents listed in Table~\ref{tab1}, we find the range of $\sigma$ that produces an acceptable collapse is somehow large. $(\gamma-2\beta)/\nu$ turns the leading exponent of $\chi'$ in the disordered phase to $2\beta/r\nu$, the exponent of $\langle m^2 \rangle$. However, in the 2D model, we have found a different expression that invalidates this conclusion~\cite{Yuan}. Therefore, like the case of FSS in cooling, $\sigma$ here is again likely a new exponent.

In the present case, the crossover from the usual FTS regime to the Bressy regime in which the extrinsic self-similarity is important occurs in finite values of $L^{-1}R^{-1/r}$ and $\tau R^{-1/r\nu}$. As just seen in Fig.~\ref{ftsh3d}(a), for small enough values of $L^{-1}R^{-1/r}$ and $\tau R^{-1/r\nu}$, the two sets of curves do not separate and hence the phases fluctuations are absent. As a result, the usual FTS works well and the Bressy exponent is not needed.

We emphasize again that it is the breaking of the self-similarity of the phases fluctuations that leads to the violation of the scaling, rather than the separation of the two sets of curves, though we have invoked the latter to exclude some data above. The separation of the two sets of curves is a transparent exhibition of the phases fluctuations, which is only a necessary condition, because both sets of curves exhibit good scaling collapses as seen in Fig.~\ref{ftshlr} once $L^{-1}R^{-1/r}$ is fixed.

Finally, we turn to FTS in cooling. As seen in Fig.~\ref{ftsclr}, once $L^{-1}R^{-1/r}$ is fixed, all FTS of all the four observables considered are good provided that corrections to scaling can be ignored. Even the completely bad scaling of Eq.~(\ref{FTSM}) shown in Figs.~\ref{FTS_30}(e) and~\ref{FTS_50}(a) recovers. As pointed out in Sec.~\ref{theory}, this indicates that the bad scaling originates from $L^{-1}R^{-1/r}$, which is to ensure the vanishing fluctuations in the thermodynamic limit, or the central limit theorem. Indeed, the revised scaling form, Eq.~(\ref{rescaling_huang}), describes well the behavior, as demonstrated in Figs.~\ref{FTS_30}(g) and~\ref{FTS_50}(c). Since keeping self-similarity of the phases fluctuations rectifies the scaling, we regard the exponent $d/2$ in Eq.~(\ref{ftfts}) as the Bressy exponent $\sigma$ in the case of FTS in cooling, as Eq.~(\ref{ftfts}) is similar to Eq.~(\ref{bressyexp}) for $F_T$. The meaning of this similarity is twofold. One is the appearance of the two relations and the other is that the curves collapse well with such extra scalings. However, we note one difference. The extra scaling for the revised scaling~(\ref{rescaling_huang}) makes the scaling collapse well in both phases except for yet another exponent to be considered shortly, whereas that for Eq.~(\ref{bressyexp}) works only for one phase. In other words, the former does not change the leading behaviors of the two phases whereas the latter does.

\begin{figure}
  \centerline{\epsfig{file=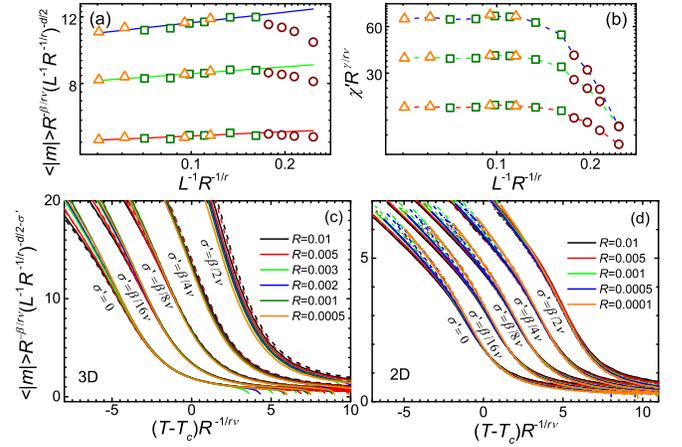,width=1\columnwidth}}
  \caption{\label{ftsc3d} (Color online) (a) $\langle |m| \rangle R^{-\beta/r\nu}(L^{-1}R^{-1/r})^{-d/2} $ and (b) $\chi' R^{\gamma/r\nu}$ on $L=20$ (circles), $L=50$ (squares), and $L=70$ (triangles) simple cubic lattices at fixed values of $(T-T_c)R^{-1/r\nu}=-3$, $-4$, and $-5$ (from down to up) versus $L^{-1}R^{-1/r}$ in double logarithmic scales for FTS in cooling. $\langle |m| \rangle R^{-\beta/r\nu}(L^{-1}R^{-1/r})^{-d/2-\sigma'} $ versus $(T-T_c)R^{-1/r\nu}$ with various $\sigma'=\sigma-d/2$ values marked near the curves for the rates $R$ listed on (c) 3D $L=50$ (solid lines) and $L=70$ (dashed lines) lattices and (d) 2D $L=50$ (dashed double dotted lines), $L=70$ (dashed dotted lines), $L=100$ (dashed lines), and $L=150$ (solid lines). Solid lines in (a) are linear fits to the left nine data points. Their slopes are $0.037(16)$, $0.061(23)$, and $0.092(31)$ from down to up. Dashed lines in (b) connecting symbols are only a guide to the eyes. In (a) and (b), the error bars are estimated to be about the sizes of the symbols. In (c) and (d), each curve with a finite $\sigma'$ is displaced by $1.5$ to the right with respect to its preceding one.}
\end{figure}
Apart from the $d/2$, one sees from Figs.~\ref{FTS_30}(c),~\ref{FTS_30}(g), and~\ref{FTS_50}(b)--\ref{FTS_50}(d) that the rescaled curves generally do not collapse as well as those in Figs.~\ref{ftsclr}(g)--\ref{ftsclr}(j) in the ordered state. As displayed in Fig.~\ref{ftsc3d}(a), $\langle |m| \rangle R^{-\beta/r\nu}(L^{-1}R^{-1/r})^{-d/2} $ at fixed values of $(T-T_c)R^{-1/r\nu}$ increases first with $L^{-1}R^{-1/r}$ and then decreases for large values of $L^{-1}R^{-1/r}$. After excluding the five largest values of $L^{-1}R^{-1/r}$ on the right, the linearity is quite good, though the slope increases with $|T-T_c|R^{-1/r\nu}$. This indicates that here $\langle |m| \rangle R^{-\beta/r\nu}(L^{-1}R^{-1/r})^{-d/2} $ is the primary observable. Consequently, $\chi' R^{\gamma/r\nu}$ at fixed values of $(T-T_c)R^{-1/r\nu}$ ought to behave according to Eq.~(\ref{xsigmat}). This is consistent with the behavior shown in Fig.~\ref{ftsc3d}(b): For small values of $|T-T_c|R^{-1/r\nu}$, the slope $\sigma'=\sigma-d/2$ is small and $\chi' R^{\gamma/r\nu}$ is almost a constant for small values of $L^{-1}R^{-1/r}$. As $|T-T_c|R^{-1/r\nu}$ increases, $\sigma'$ becomes bigger and $\chi' R^{\gamma/r\nu}$ decreases at relatively larger $L^{-1}R^{-1/r}$, excluding the five largest data.

A unique feature different from all the previous three cases is that here the slope in Fig.~\ref{ftsc3d}(a) increases steadily from $0.037(16)$ at $(T-T_c)R^{-1/r\nu}=-3$ to $0.19(7)$ at $-10$, with smaller errors above $(T-T_c)R^{-1/r\nu}=-6$ and only a milder increase around $-10$. Since the slope is not fixed and $(T-T_c)R^{-1/r\nu}=-10$ appears too far away from $T_c$, we depict in Fig.~\ref{ftsc3d}(c) the scaling collapses of $\langle |m| \rangle R^{-\beta/r\nu}(L^{-1}R^{-1/r})^{-d/2-\sigma'} $ for a series of $\sigma'$ up to $\beta/2\nu$, including the one with $\sigma'=0$ for comparison. The curves of the five largest data in Figs.~\ref{ftsc3d}(a) and~\ref{ftsc3d}(b) are absent since their $L^{-1}R^{-1/r}$ is large and does not collapse well at large negative $(T-T_c)R^{-1/r\nu}$. In accordance with the slopes found in Fig.~\ref{ftsc3d}(a), the collapses become better at larger values of $|T-T_c|R^{-1/r\nu}$, noticing that each curve has been displaced to right by $1.5$ with respect to its preceding one. In fact, $\beta/8\nu\approx0.0648$, close to the slope at $(T-T_c)R^{-1/r\nu}=-4$ in Fig.~\ref{ftsc3d}(a). However, the best value for a globally acceptable collapse within the range of ordinate shown seems $\beta/4\nu$, though the collapse at smaller $|T-T_c|R^{-1/r\nu}$ is not as good as those of the smaller $\sigma'$ ones. This is the difficulty of the present case. In all previous cases, it is $\chi$ that is the primary observable. It is a peak that enables us to collapse the curves onto its slopes, as Figs.~\ref{fssh3d}(c),~\ref{fssc3d}(b), and~\ref{ftsh3d}(d) demonstrate. Here, we lack such a reference and fall into difficulty in selecting a proper $\sigma$. The secondary observable $\chi'$ with a peak can only be collapsed through $\langle |m| \rangle$ by Eq.~(\ref{xsigmat}). For comparison, we depict the results of the 2D Ising model in Fig.~\ref{ftsc3d}(d) in the same way. Its cuts behave similarly to the 3D case displayed in Figs.~\ref{ftsc3d}(a) and~\ref{ftsc3d}(b), including the increasing slopes of Fig.~\ref{ftsc3d}(a). For the 2D model, the slopes are $0.038(15)$, $0.051(19)$, $0.064(21)$, and $0.072(22)$ at $(T-T_c)R^{-1/r\nu}=-2$, $-3$, $-4$, and $-5$, respectively. The best value appears at $\sigma'=\beta/2\nu$, which is $1/16=0.0625$ using the exact critical exponents $\beta=1/8$ and $\nu=1$ for the model and is close to the slope at $(T-T_c)R^{-1/r\nu}=-4$. However, the collapse appears again not as good as those of smaller $\sigma'$ values. Accordingly, we append large relative errors to the two estimated values given in Table~\ref{tab1}.

Therefore, the apparent values of $\sigma$ for the 2D and 3D Ising models appear to be different, though they are small with large errors and there exists no reference for a reliable estimate. In addition, these values appear to have no simple explanation similar to the heating cases and hence do not rule out the possibility of a new single exponent that yields the two values in the two spatial dimensions. Interestingly, the two numerical values may be both about $0.064$ within the estimated large errors, a situation which appears also in FTS and FSS, both in heating. One may then wonder whether the only left case of FSS in cooling should also have similar numerical values or not. However, the numerical value of the 3D $\beta/2\nu$ appears to have no peers due to the small 2D $\beta$ even if $\nu$ is separated similar to the case of FTS in heating.

We have seen that the violations of scaling during heating both for FSS and FTS are prominent and their Bressy exponents are quite big. Yet, that during cooling is less pronounced and the corresponding Bressy exponents are small if the revised scaling and its exponent of $d/2$ are excluded. Although the violation of the 3D FSS in cooling shown in Figs.~\ref{fssc}(f) and~\ref{fssc403}(b) are also evident and appear not to stem from corrections to scaling, its 2D counterpart appears good~\cite{Yuan}. This difference may arise from the small 2D value of $\beta/2\nu=1/16=0.062~5$ in comparison with the same 3D value of about $0.259$~\cite{Yuan}. However, one might suspect that larger lattice sizes might remove the violations. For the FTS in cooling, the reduced Bressy exponent $\sigma'=\sigma-d/2$ is even difficult to be determined, as exhibited in Fig.~\ref{ftsc3d}, because of no proper reference. Nevertheless, we emphasize that at least the heating results are definite.

\section{\label{FTSwh}FTS in the presence of external field in cooling}
In this section, we will apply an external field to study its effects on the phases fluctuations in FTS during cooling, since the behavior of FTS in cooling appears intriguing, while that in heating---which will be touched on briefly in Sec.~\ref{ftshl} for comparison---seems clear. It might be expected that an external field could break the symmetry of the two ordered dissymmetric phases and select a particular phase and thus suppress the phases fluctuations. This has indeed been found for a sufficiently large field in the scaling exactly at $T_c$~\cite{Huang}. However, for a small field, the mixed scaling regime persists between the two end regimes of FSS and FTS, though how the different behaviors cross over was not clear~\cite{Huang}. In this connection, whether or how such a small field affects the phases fluctuations are worth investigating. As the scaling of the whole driving process has exhibited far richer phenomena than that just at $T_c$, we study the whole cooling process in the presence of an external field. We note in passing that, in phase ordering, an external field accelerates exponentially the ordering and reduces the scaling regime~\cite{Bray+h}.

In order to clarify the effect of the external field, we apply the external field to the system in three different protocols:
\begin{itemize}
  \item Protocol A---the external field is switched on at $T_c$ till the end;
  \item Protocol B---the external field is switched on at the beginning of evolution and off at $T_c$;
  \item Protocol C---the external field is switched on at the beginning of evolution till the end.
\end{itemize}

We will consider two fixed values of $HR^{-\beta\delta/{r\nu}}$, from which $H$ is computed for a series of given $R$ from $0.0005$ to $0.01$. For comparison, we also consider a case in which $L^{-1}R^{-1/r}$ is further fixed to about $0.2687$, the value studied in Sec.~\ref{ftsofls}. In this case, $R$ has to be computed first from the given $L$. After investigating sequentially the scaling of the three protocols in Sec.~\ref{hatTca},~\ref{hatTcb}, and~\ref{hatTcc}, we consider the crossover and the violations of scaling and their rectifications in Sec.~\ref{cross}. Finally, we study the properties of the FTS in a field on large lattices in Sec.~\ref{ftshl}.

\subsection{\label{hatTca}Protocol A}
In this protocol, the external field is applying only below $T_c$. This may assume an effect similar to that of removing the unsaturated samples in Sec.~\ref{ftsofls} for a sufficiently large $H$, though the ordered regions of the former have a particular direction which is absent in the latter.

\begin{figure}
  \centerline{\epsfig{file=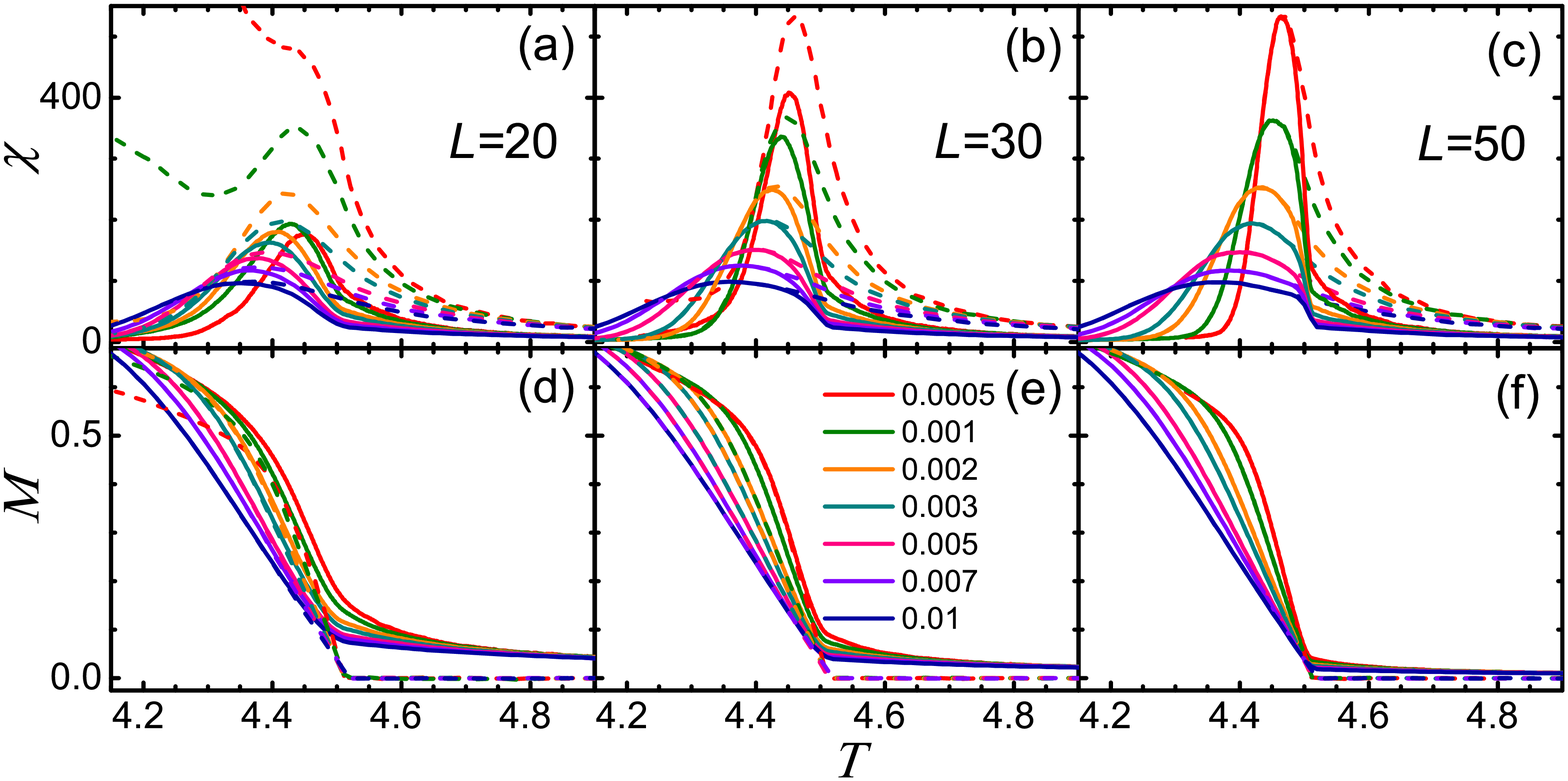,width=1.0\columnwidth}}
  \centerline{\epsfig{file=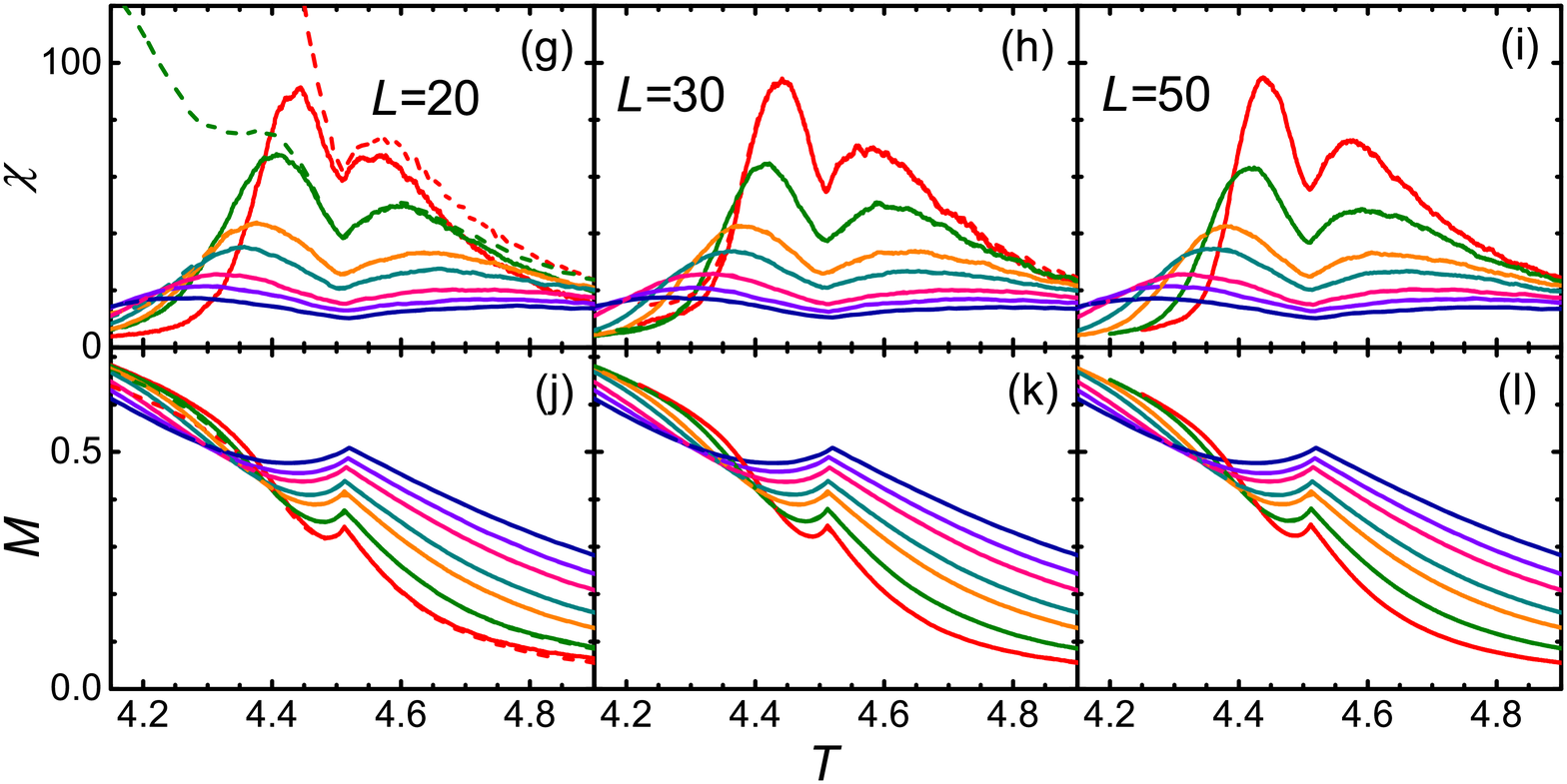,width=1.0\columnwidth}}
  \centerline{\epsfig{file=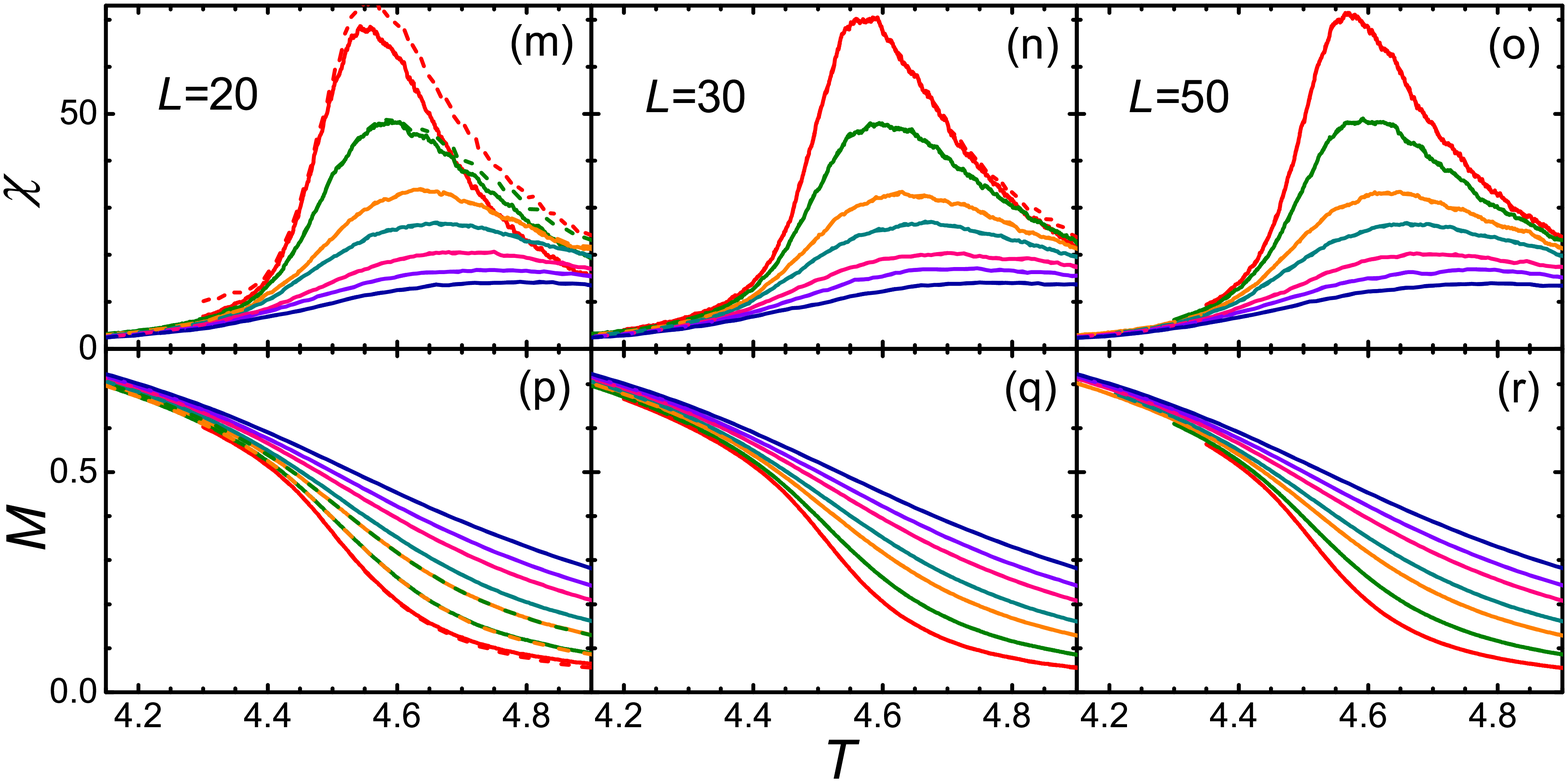,width=1.0\columnwidth}}
    \caption{\label{coola2} (Color online) $\chi$ and $M$ with fixed $HR^{-\beta\delta/{r\nu}}=2$ in (a)--(f) protocol A, (g)--(i) protocol B, and (m)--(r) protocol C on simple cubic lattices of size $L=20$ (left column), $L=30$ (middle column), and $L=50$ (right column). Solid lines denote results of $\chi'$ and $\langle |m|\rangle$ while dashed lines of $\chi$ and $\langle m\rangle$. All panels have the same abscissas and share the same legend listing the rates used.}
\end{figure}
Figures~\ref{coola2}(a)--\ref{coola2}(f) show the evolution of $\chi$ and $M$ with $T$ for fixed $HR^{-\beta\delta/{r\nu}}=2$ in protocol A. It can be seen that as $L$ and $R$ increase, the deviations of $\chi$ and $\langle m\rangle$ from their corresponding $\chi'$ and $\langle |m|\rangle$ are reduced, in consistence with the results in the absence of the external field in Sec.~\ref{ftsofls}. Owing to the applied field, now $\chi$ (dashed lines) exhibits a peak except for the two smallest values of $R$ on $L=20$ and its peak height behaves normally in that the peak increases as $R$ decreases. Similar behavior is seen for $\chi'$ (solid lines) except again for the two smallest values of $R$ on $L=20$, which show a reverse dependence similar to Fig.~\ref{hcxm}. Large deviations of $\chi$ and $\langle m\rangle$ from $\chi'$ and $\langle |m|\rangle$ and even the upturns in $\chi$ appear for $R<0.003$ on $L=20$, or $L^{-1}R^{-1/r}\approx 0.246$ and $R=0.0005$ on $L=30$, or $L^{-1}R^{-1/r}\approx 0.275$. In the absence of $H$, the system lies in the FSS regime, either shallow or deep, at these rates, according to Secs.~\ref{fssafr} and~\ref{ftsofls} on the relation between $\xi_R$ and $L$. These might result in the blowup of $\chi$ at low temperatures for the two smallest values of $R$ on $L=20$ that are relatively deep in the FSS regime, since phases fluctuations appear strong irrespective of $H$, as can be seen from the relatively large deviations of $\langle m\rangle$ from $\langle |m|\rangle$ in Fig.~\ref{coola2}(d). However, at the fixed $HR^{-\beta\delta/{r\nu}}=2$, $\xi_H$ varies from about $2.68$ to $6.10$ for the $R$ range used, which are invariably much shorter than $L$ and thus the system is, in fact, controlled by the field instead of $L$. In the disordered phase all spins fluctuate freely and equally along the two directions and $\langle m\rangle=0$ for $T>T_c$ as no external field is applied, whereas $\langle |m|\rangle$ is finite for finite $L$. As the external field is applied at and below $T_c$, regions of the phase opposite to the field transform to the other phase and hence $\langle m\rangle$ approaches $\langle |m|\rangle$ while the reduced fluctuation $\chi'$ tends to the original $\chi$ when the transformation is completed. The upturns of the two smallest rates in Fig.~\ref{coola2}(a) indicate there still exist samples whose $m$ is opposite to the field for the small $L$, similar to the case in the absence of the external field, albeit less prominent. By contrast, for large $L$ or $R$, the number of such samples is small since the number of regions of sizes $\xi_H$ is large and thus the opposite phase is likely to be averaged out.

\begin{figure}
  \centerline{\epsfig{file=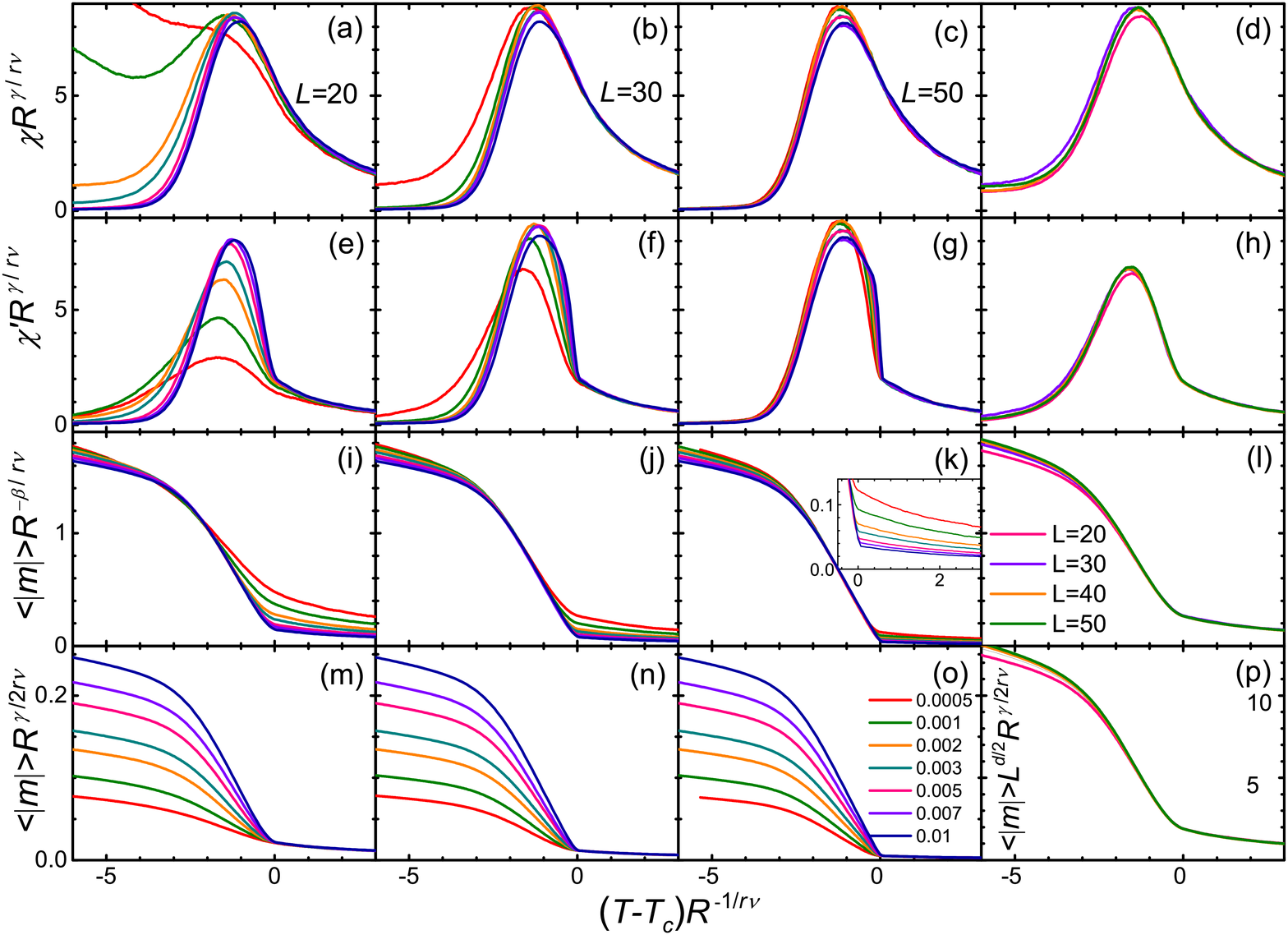,width=1\columnwidth}}
  \caption{\label{coola2f} (Color online) FTS of (a)--(d) $\chi$ (first row), (e)--(h) $\chi'$ (second row), and (i)--(l) $\langle |m|\rangle$ (third row), and (m)--(p) revised FTS of $\langle |m|\rangle$ (fourth row) with fixed $HR^{-\beta\delta/{r\nu}}=2$ on simple cubic lattices of sizes $L=20$ (first column), $L=30$ (second column), $L=50$ (third column), and four different values of $L$ that fix $L^{-1}R^{-1/r}\approx0.2687$ (fourth column) in protocol A. The inset in (k) magnifies the curves in the disordered phase. All panels except the last column share the same legend in (o) listing the rates used, while the last column shares the same legend in (l) listing the lattice sizes. Note that we have included the $L^{d/2}$ factor in the vertical axis according to Eq.~(\ref{rescaling_huang}) in (p) and the corresponding scale is different from those in the last row.}
\end{figure}
In Fig.~\ref{coola2f}, we show the FTS of the results in Figs.~\ref{coola2}(a)--\ref{coola2}(f). The results of $\langle m\rangle$ are not shown because they are similar to those of $\langle |m|\rangle$, as can be anticipated from Figs.~\ref{coola2}(d)--\ref{coola2}(f). It can be seen from Fig.~\ref{coola2f} that the FTS of $\chi$ for the rates in the FTS regime get better as $L$ increases, similar to the case in the absence of the field. While the scalings of $\chi'$ are poor on the high-temperature side of the peaks below $T_c$ for all the three $L$ values shown, they resemble those of $\chi$ on the low-temperature side of the peaks for the $R$ in the FTS regime, in accordance with results displayed in Figs.~\ref{coola2}(a)--\ref{coola2}(c). However, the scaling of $\langle |m|\rangle$ depends on whether the field is applied or not. On the one hand, when the field is present below $T_c$, the usual FTS according to Eq.~(\ref{FTSM}) becomes better as $L$ increases, though the revised FTS, Eq.~(\ref{rescaling_huang}), is simply invalid. On the other hand, when the field is absent, the usual scaling is poor while the revised one is good. At first sight, the usual FTS above $T_c$ might seem better as $L$ increases. However, the inset in Fig.~\ref{coola2f}(k) manifest itself that this is only apparent. In the absence of the external field, only the revised scaling is good as expected. Nevertheless, all scalings are good when $L^{-1}R^{-1/{r\nu}}$ is further fixed. This indicates that the poor scaling near $T_c$ for $\chi'$ results from this term rather than the initial condition of applying a field at $T_c$. The latter should have followed FTS because the initial length scale $\xi_R$ arisen from the cooling above $T_c$ is longer than $\xi_H$ for $HR^{-\beta\delta/{r\nu}}=2$ or $\xi_R\approx1.32\xi_H$ and thus it is this existing long length scale rather than the short scale that controls the dynamics scaling~\cite{Feng}. We note in passing that a scaling plot using $H$ as the scaling factor cannot remove the poor scaling though $\xi_H$ is the shortest long length scale. This is because $HR^{-\beta\delta/{r\nu}}$ is fixed and thus such a scaling plot is equivalent to the FTS, similar to the relation of FSS and FTS shown in Sec.~\ref{fsf}.

\begin{figure}
  \centerline{\epsfig{file=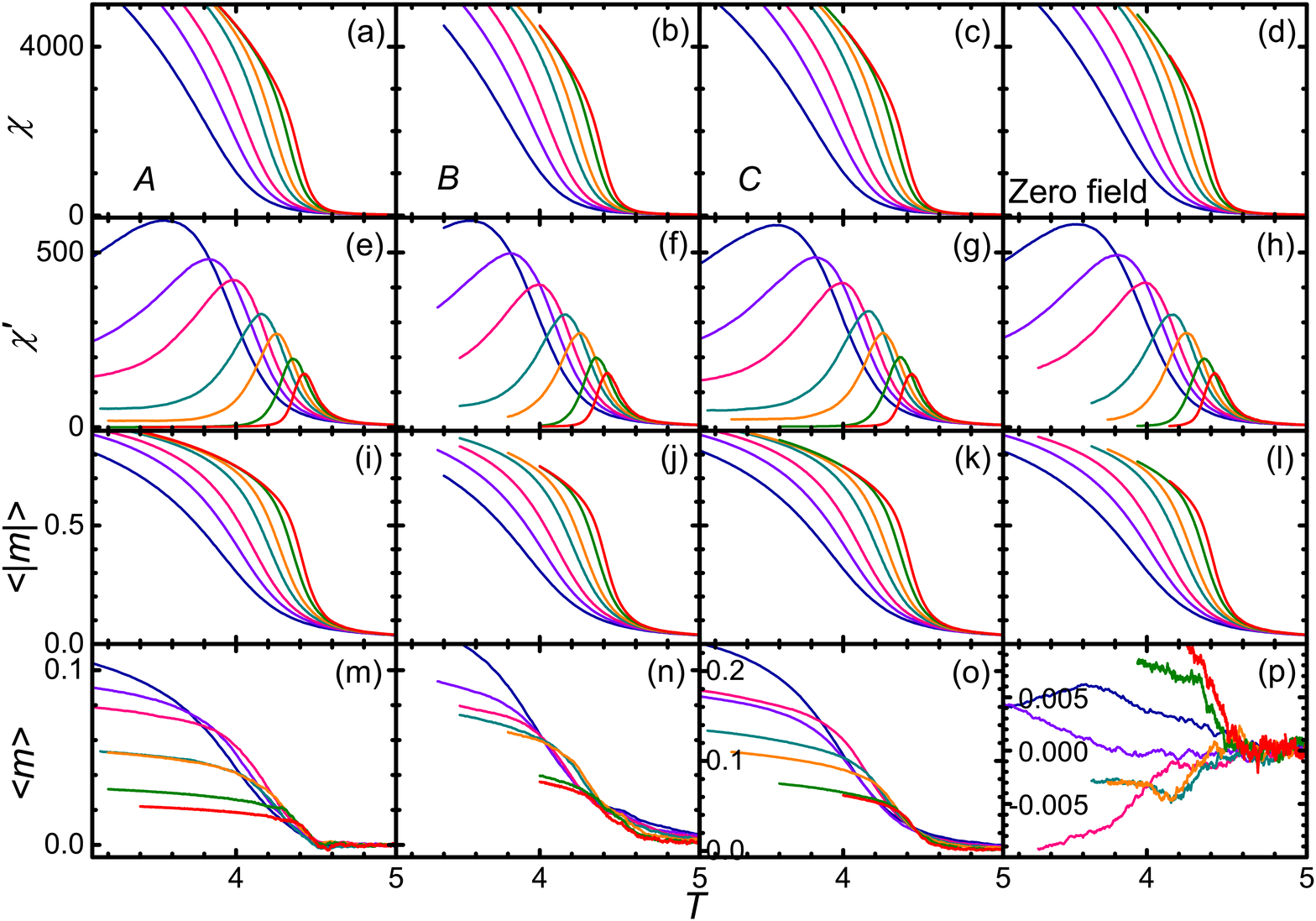,width=1\columnwidth}}
  \centerline{\epsfig{file=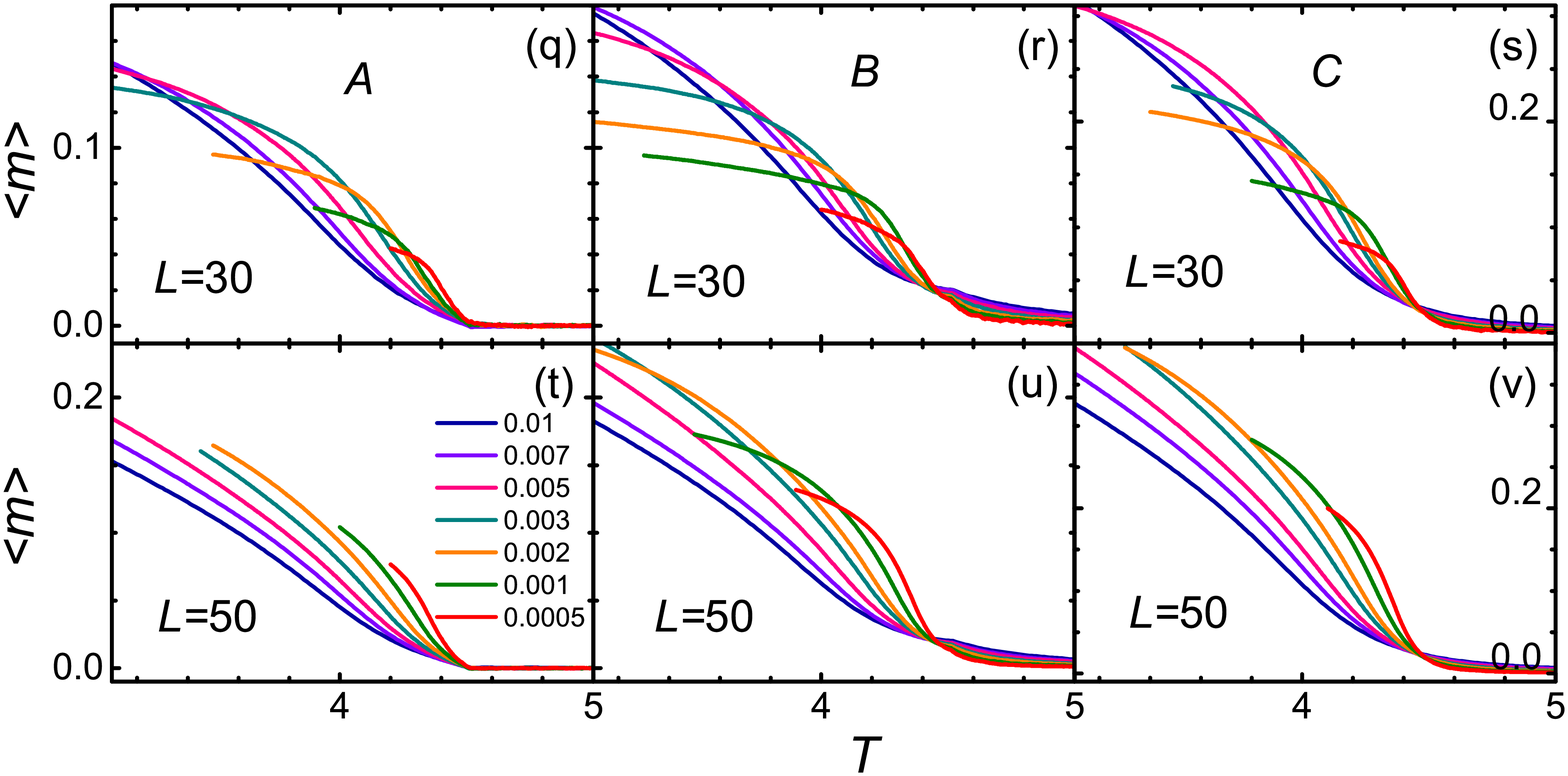,width=1.0\columnwidth}}
    \caption{\label{coolabc} (Color online) (a)--(d) $\chi$ (first row), (e)--(h) $\chi'$ (second row), (i)--(l) $\langle |m|\rangle$ (third row), and (m)--(p) $\langle m\rangle$ (fourth row) with fixed $HR^{-\beta\delta/{r\nu}}=0.05$ for an applied field in protocol A (first column), $B$ (second column), $C$ (third column), and zero field (fourth column) on $20\times20\times20$ simple cubic lattices. The last two rows, (q)--(v), depict $\langle m\rangle$ for the same $HR^{-\beta\delta/{r\nu}}$ in protocol A (first column), $B$ (second column), and $C$ (last column) on simple cubic lattices of size $L=30$ and $L=50$ marked. (o), (p), (s), and (v) possess ordinates different from their respective rows. All panels share the same legend listing the rates used.}
\end{figure}
In the first (leftmost) column of the upper part in Fig.~\ref{coolabc}, we show the evolution of $\chi$ and $M$ with $T$ for another fixed $HR^{-\beta\delta/{r\nu}}=0.05$ on $L=20$ lattices in protocol A. This choice yields $\xi_H\approx3.34\xi_R$ and thus the applied field is largely irrelevant. This is clearly seen by comparing the results in the first column with the corresponding ones in the absence of the field in the last column in Fig.~\ref{coolabc}. The only difference is $\langle m\rangle$ in Fig.~\ref{coolabc}(m) as compared with Fig.~\ref{coolabc}(p). $\langle m\rangle$ is now finite albeit small, indicating that the phase along the field is now invariably dominant. It increases with the lattice size $L$ as is seen in Figs.~\ref{coolabc}(q) and~\ref{coolabc}(t), since there are more regions of size $\xi_R$ similar to the case of $HR^{-\beta\delta/{r\nu}}=2$ above. However, the other three observables display similar behaviors to $L=20$ except for stronger fluctuations and hence later transitions---the peaks move to lower temperatures. Thus, we do not show them here, though we shall display their scaling collapses in the following.

\begin{figure}
  \centerline{\epsfig{file=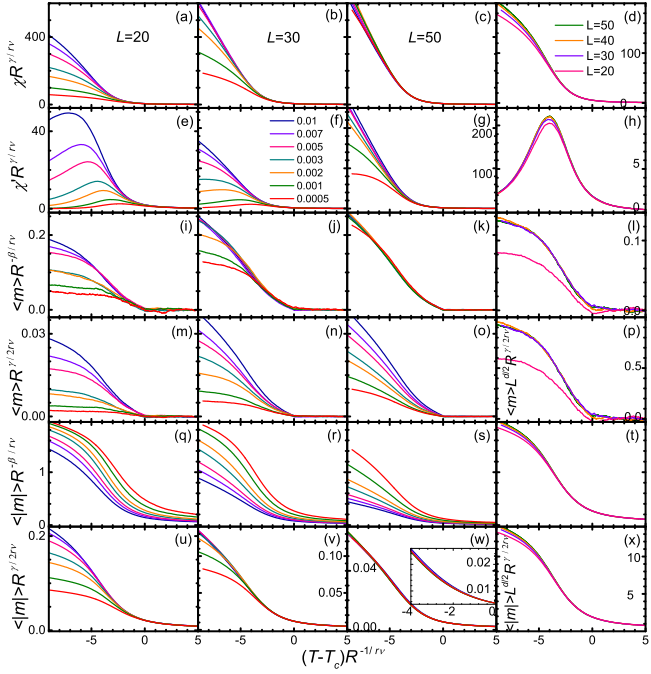,width=1\columnwidth}}
  \caption{\label{coola05f} (Color online) FTS of (a)--(d) $\chi$ (first row), (e)--(h) $\chi'$ (second row), (i)--(l) $\langle m\rangle$ (third row), and (q)--(t) $\langle |m|\rangle$ (fifth row), and revised FTS of (m)--(p) $\langle m\rangle$ (fourth row) and (u)--(x) $\langle |m|\rangle$ (sixth row) with fixed $HR^{-\beta\delta/{r\nu}}=0.05$ on simple cubic lattices of sizes $L=20$ (first column), $L=30$ (second column), $L=50$ (third column), and four different values of $L$ that fix $L^{-1}R^{-1/r}\approx0.2687$ (fourth column) in protocol A. The inset in (w) magnifies the curves below $T_c$. All panels except the last column share the same legend in (f) listing the rates used, while the last column shares the same legend in (d) listing the lattice sizes. Note that, in (p) and (x), we have included the $L^{d/2}$ factor in the vertical axis according to Eq.~(\ref{rescaling_huang}) and the corresponding scales are different from those in their respective rows. Also, the ordinates in (d), (h), (l), (v), and (w) are also different from those in their respective rows. In addition, (f) and (g) share identical ordinates labeled in (g).}
\end{figure}
In Fig.~\ref{coola05f}, we show the FTS for $HR^{-\beta\delta/{r\nu}}=0.05$. Stronger fluctuations and later transitions with larger lattice sizes as mentioned are manifest from Figs.~\ref{coola05f}(a),~\ref{coola05f}(b),~\ref{coola05f}(e), and~\ref{coola05f}(f). It can be seen that the scalings of $\chi$, $\chi'$, and $\langle |m|\rangle$ according to Eq.~(\ref{rescaling_huang}), the revised FTS, are similar to those shown in Figs.~\ref{FTS_30} and~\ref{FTS_50}: They get better as $L$ increases, even though there exists now an applied external field. Also, when $L^{-1}R^{-1/r}$ is fixed, even $R\lesssim0.0022$ on $L=20$ displays impressive scaling, as seen in the last column in Fig.~\ref{coola05f}. Moreover, the poor scalings near $T_c$ in Figs.~\ref{coola2f}(e)--\ref{coola2f}(g) are absent. The revised FTS of $\langle |m|\rangle$ below $T_c$ in Fig.~\ref{coola05f}(w) appears rather acceptable in sharp contrast with that in Fig.~\ref{coola2f}(o) for $HR^{-\beta\delta/{r\nu}}=2$. Meanwhile, the usual FTS of $\langle |m|\rangle R^{-\beta/r\nu}$ in Fig.~\ref{coola05f}(s) is manifestly poor, again in stark contrast with that in Fig.~\ref{coola2f}(k). However, the inset in Fig.~\ref{coola05f}(w) shows some slight dispersion as compared with, e.g., Fig.~\ref{coola05f}(x). We will return to this later on in protocol C. Nevertheless, the results confirm the validity of Eqs.~(\ref{FTSX}) and~(\ref{FTSM}) in general and Eq.~(\ref{rescaling_huang}) in particular in the presence of a sufficiently weak field at least for some lattice sizes.

What is remarkable from Fig.~\ref{coola05f} is the scaling behavior of $\langle m\rangle$ in comparison with that of $\langle |m|\rangle$. They are completely opposite. Whereas the usual scaling with $\beta$ according to Eq.~(\ref{FTSM}) is invalid and the revised scaling with $\gamma$ according to Eq.~(\ref{rescaling_huang}) gets apparently better with increasing $L$ for the latter, the former shows quite good scaling with $\beta$ for the large $L$ values and simply cannot be described by Eq.~(\ref{rescaling_huang}). This indicates that Eqs.~(\ref{FTSM}) and~(\ref{rescaling_huang}) work for $\langle m\rangle$ and $\langle |m|\rangle$, respectively, in cooling under weak external fields, since, in heating and in cooling with strong external fields, Eq.~(\ref{FTSM}) holds well for $\langle |m|\rangle$ as seen in Fig.~\ref{ftsh} and discussed about Fig.~\ref{coola2f}.

Figures~\ref{coola05f}(l) and~\ref{coola05f}(p) would appear to indicate that the FTS of $\langle m\rangle$, Eq.~(\ref{FTSM}), did not hold for $L=20$ even $L^{-1}R^{-1/r}$ is fixed, though the other observables work rather well. However, this is again attributed to the strong phases fluctuations for small $R$ and $L$. The small applied field is too weak for small $L$ to order sufficiently for the sample size used. We have checked that $\langle m\rangle$ varies with sample sizes considerably, though again the other observables show little appreciable effects since $\langle m\rangle$ is small compared with $\langle |m|\rangle$ and $\langle m^2\rangle$. This fluctuation of $\langle m\rangle$ for small $R$ on $L=20$ can also be seen from Figs.~\ref{coolabc}(m) and~\ref{coola05f}(i), where the curves of $R=0.002$ and $R=0.003$ almost overlap.

\subsection{\label{hatTcb}Protocol B}
Contrary to protocol A, in this part, we apply the external field at the beginning of evolution and switch it off at $T_c$.

The evolution of $\chi$ and $M$ with $T$ for fixed $HR^{-\beta\delta/{r\nu}}=2$ in protocol B is shown in Figs.~\ref{coola2}(g)--\ref{coola2}(l). Because of the external field applying from the beginning, now $\langle m\rangle$ is finite and increases with the strength of the field, which increases with $R$. Only for the smallest rates on the smallest $L$ can $\langle m\rangle$ differentiate from $\langle |m|\rangle$ and thus $\chi$ from $\chi'$. For these rates, the magnetization can reverse after the field is turned off and leads to the upturn in $\chi$. Again due to the field, the fluctuations $\chi$ are substantially reduced as seen from the peak heights; and their first (right) peaks characterizing the transition temperature shift to high temperatures, which are even higher than the original $T_c$ for the chosen field that assists ordering. After the field is switched off, the fluctuations are enhanced, giving rise to the second peaks. The positions of the two peaks exhibit opposite $R$ dependences, though they both move closer to $T_c$ as $R$ is lowered, because $\xi_R$ or $\xi_H$ both increase with decreasing $R$ or $H$. The $M$ curves also get closer to $T_c$ for smaller $R$ in the presence to the field, similar to the behavior in Fig.~\ref{ftsclr}(a).

\begin{figure}
   \centerline{\epsfig{file=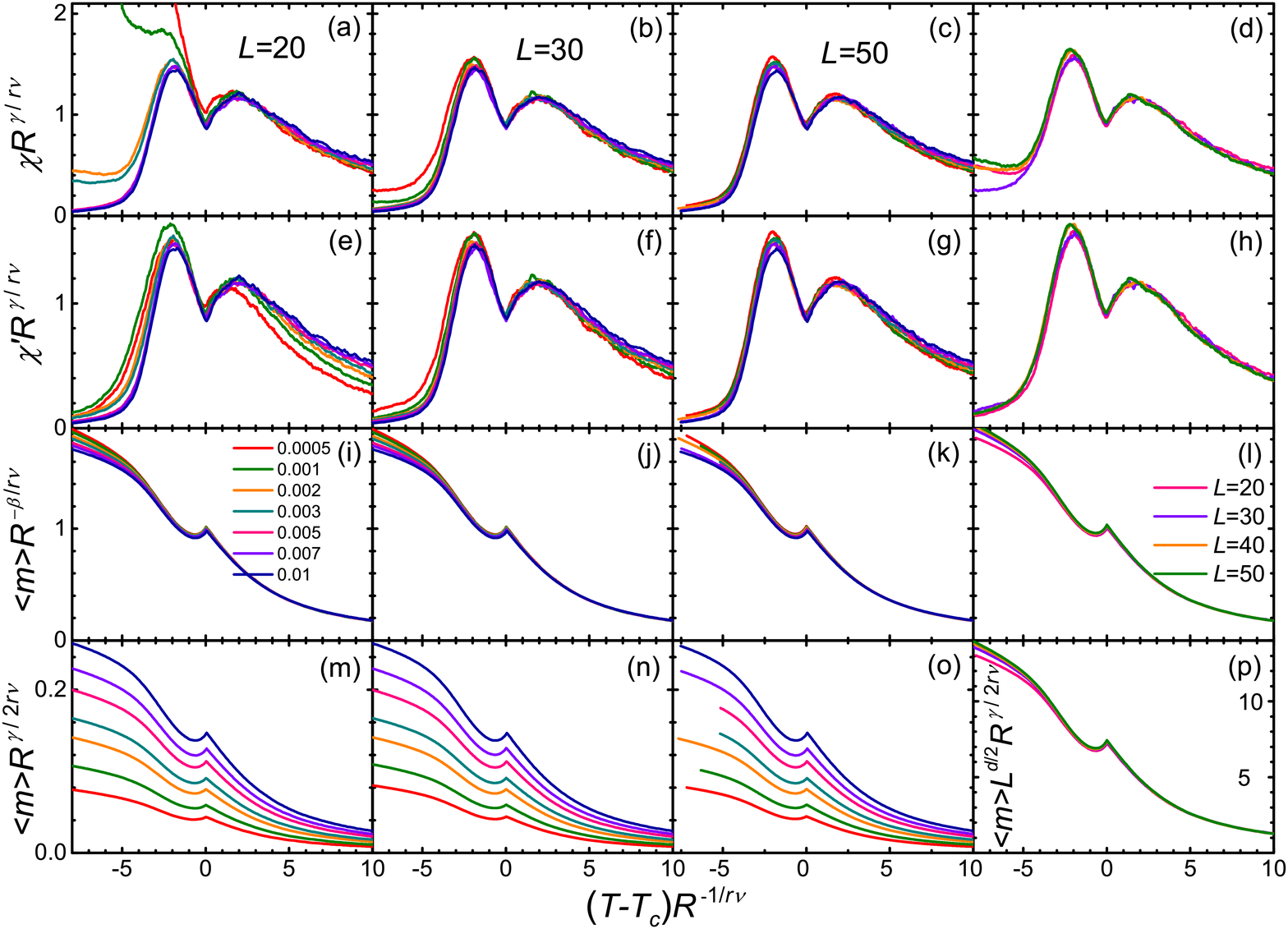,width=1\columnwidth}}
  \caption{\label{coolb2f} (Color online) FTS of (a)--(d) $\chi$ (first row), (e)--(h) $\chi'$ (second row), and (i)--(l) $\langle |m|\rangle$ (third row), and (m)--(p) revised FTS of $\langle |m|\rangle$ (fourth row) with fixed $HR^{-\beta\delta/{r\nu}}=2$ on simple cubic lattices of sizes $L=20$ (first column), $L=30$ (second column), $L=50$ (third column), and four different values of $L$ that fix $L^{-1}R^{-1/r}\approx0.2687$ (fourth column) in protocol B. All panels except the last column share the same legend in (i) listing the rates used, while the last column shares the same legend in (l) listing the lattice sizes. Note that we have included the $L^{d/2}$ factor in the vertical axis according to Eq.~(\ref{rescaling_huang}) in (p) and the corresponding scale is different from those in the last row.}
\end{figure}
The FTS for $HR^{-\beta\delta/{r\nu}}=2$ in protocol B is shown in Fig.~\ref{coolb2f}. Different from the protocol A, here, both $\chi$ and $\chi'$ as well as $M$ show good scaling once $L$ is large enough. In addition, $M$ behaves according to Eq.~(\ref{FTSM}) rather than Eq.~(\ref{rescaling_huang}). Except for the smallest $R$ and $L$ values, the ordering induced by the field above $T_c$ makes the transition similar to heating. This indicates that, for such a strong field that the transition occurs before $T_c$ is reached, the phases fluctuations are so substantially suppressed by the field in  protocol B that the revised FTS is not needed even though no field is applying below $T_c$. However, we will see shortly that a weak field has no such an effect.

When $HR^{-\beta\delta/{r\nu}}=0.05$, the evolution of $\chi$ and $M$ on $L=20$ lattices are depicted in the second column of the upper part in Fig.~\ref{coolabc}. Except for $\langle m\rangle$, the other observables are almost identical with their corresponding ones in protocol A. The applied field renders $\langle m\rangle$ finite above $T_c$ and larger than its corresponding one in protocol A. Similar to protocol A, larger lattice sizes have larger $\langle m\rangle$, as seen in Fig.~\ref{coolabc}(r) and~\ref{coolabc}(u).

\begin{figure}
  \centerline{\epsfig{file=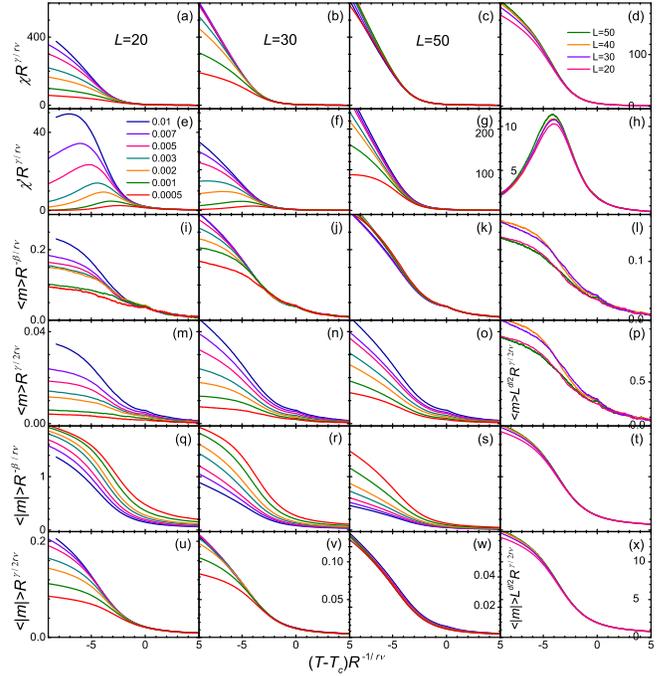,width=1\columnwidth}}
  \caption{\label{coolb05f} (Color online) FTS of (a)--(d) $\chi$ (first row), (e)--(h) $\chi'$ (second row), (i)--(l) $\langle m\rangle$ (third row), and (q)--(t) $\langle |m|\rangle$ (fifth row), and revised FTS of (m)--(p) $\langle m\rangle$ (fourth row) and (u)--(x) $\langle |m|\rangle$ (sixth row) with fixed $HR^{-\beta\delta/{r\nu}}=0.05$ on simple cubic lattices of sizes $L=20$ (first column), $L=30$ (second column), $L=50$ (third column), and four different values of $L$ that fix $L^{-1}R^{-1/r}\approx0.2687$ (fourth column) in protocol B. All panels except the last column share the same legend in (e) listing the rates used, while the last column shares the same legend in (d) listing the lattice sizes. Note that, in (p) and (x), we have included the $L^{d/2}$ factor in the vertical axes according to Eq.~(\ref{rescaling_huang}) and the corresponding scales are different from those in their respective rows. Also, the ordinates in (d), (h), (l), (v), and (w) are also different from those in their respective rows. In addition, (f) and (g) share identical scales marked in (g).}
\end{figure}
The rescaled curves are shown in Fig.~\ref{coolb05f}. It can be seen that the scalings of $\chi$ and $\chi'$ are almost identical with their corresponding ones in Fig.~\ref{coola05f} for protocol A, even though the field is absent below $T_c$ in protocol B. $\langle m\rangle$ and $\langle |m|\rangle$ also appear to behave similarly to their corresponding ones in protocol A, including their opposite scaling results. However,in contrast with protocol A, their scalings on $L=50$ lattices are not so good and the scaling of $\langle m\rangle$ for fixed $L^{-1}R^{-1/r}$ is even quite bad, including the disordered phase at $T>T_c$ where the external field is applied. We attribute the bad scaling of $\langle m\rangle$ of the three differences to the fact that the $\langle m\rangle$ curves vary appreciably with the samples of size $30~000$. Large lattices allow for the magnetization to flip in the absence of the external field. Accordingly, small lattices can still show appreciable scaling collapses but large lattices are likely to reflect the random fluctuations of the order parameter in the absence of the applied field. To the contrast, in protocol A, the scaling of $\langle m\rangle$ are believed to be obeyed for sufficiently large sample sizes because of the presence of the external field below $T_c$, although the order parameter curves depend also on the sample size. The other two differences will be discussed in protocol C and solved in Sec.~\ref{cross} below.

\subsection{\label{hatTcc}Protocol C}
When the external field is applied during the whole evolution in protocol C, the evolution of $\chi$ and $M$ with $T$ for fixed $HR^{-\beta\delta/{r\nu}}=2$ are displayed in Figs.~\ref{coola2}(m) to~\ref{coola2}(r). It is seen that now $\chi$ is quite similar to that in heating in Fig.~\ref{hcxm} in that fluctuations increase with lowering $R$. Also, the order parameter curves above $T>T_c$ resembles those in protocol B correctly. Moreover, the peak heights and positions of $\chi$ trace the right peaks in protocol B. The differences between $\chi$ and $\chi'$ and $\langle m\rangle$ and $\langle |m|\rangle$ are only appreciable for the small $R$ and $L$. The peculiar behavior of $\chi$ for small $R$ and $L$ in protocols A and B disappears and thus the strong phases fluctuations are substantially suppressed.

\begin{figure}
  \centerline{\epsfig{file=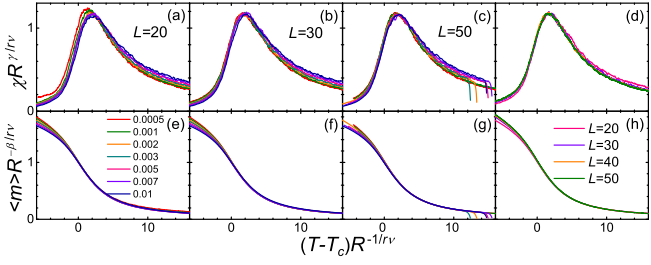,width=1\columnwidth}}
  \caption{\label{coolc2} (Color online)  FTS of (a)--(d) $\chi$ (first row), and (e)--(h) $\langle m\rangle$ (second row) with fixed $HR^{-\beta\delta/{r\nu}}=2$ on simple cubic lattices of sizes $L=20$ (first column), $L=30$ (second column), $L=50$ (third column), and four different values of $L$ that fix $L^{-1}R^{-1/r}\approx0.2687$ (fourth column) in protocol C. All panels except the last column share the same legend in (e) listing the rates used, while the last column shares the same legend in (d) listing the lattice sizes.}
\end{figure}
The scalings of $\chi$ and $\langle m \rangle$ are depicted in Fig.~\ref{coolc2}. Those of $\chi'$ and $\langle |m| \rangle$ are similar and absent. We do not show the scaling of $\langle |m| \rangle R^{\gamma/2r\nu}$ because it is no doubt bad. Except for the smallest rates on the $L=20$ lattice which lie outside the FTS regime, as pointed out in Sec.~\ref{ftsofls}, the scaling near $T_c$ is quite good for $\chi$ and extends even to quite high temperatures for $\langle m \rangle$. Once $L^{-1}R^{-1/r}$ is also fixed, both $\chi$ and $\langle m \rangle$ exhibit better scaling to high temperatures, as seen in the last column of Fig.~\ref{coolc2}.

For $HR^{-\beta\delta/{r\nu}}=0.05$, the evolution of $\chi$ and $M$ on $L=20$ lattices are shown in the third column of the upper part in Fig.~\ref{coolabc}. Again, except for $\langle m\rangle$, the other observables are almost identical with their corresponding ones in protocols A and B. As the field is applied all the way during the evolution, $\langle m\rangle$ is substantially larger than its corresponding ones in protocols A and B. Again, similar to protocols A and B, larger lattice sizes have larger $\langle m\rangle$, as seen in Fig.~\ref{coolabc}(s) and~\ref{coolabc}(v).

\begin{figure}
  \centerline{\epsfig{file=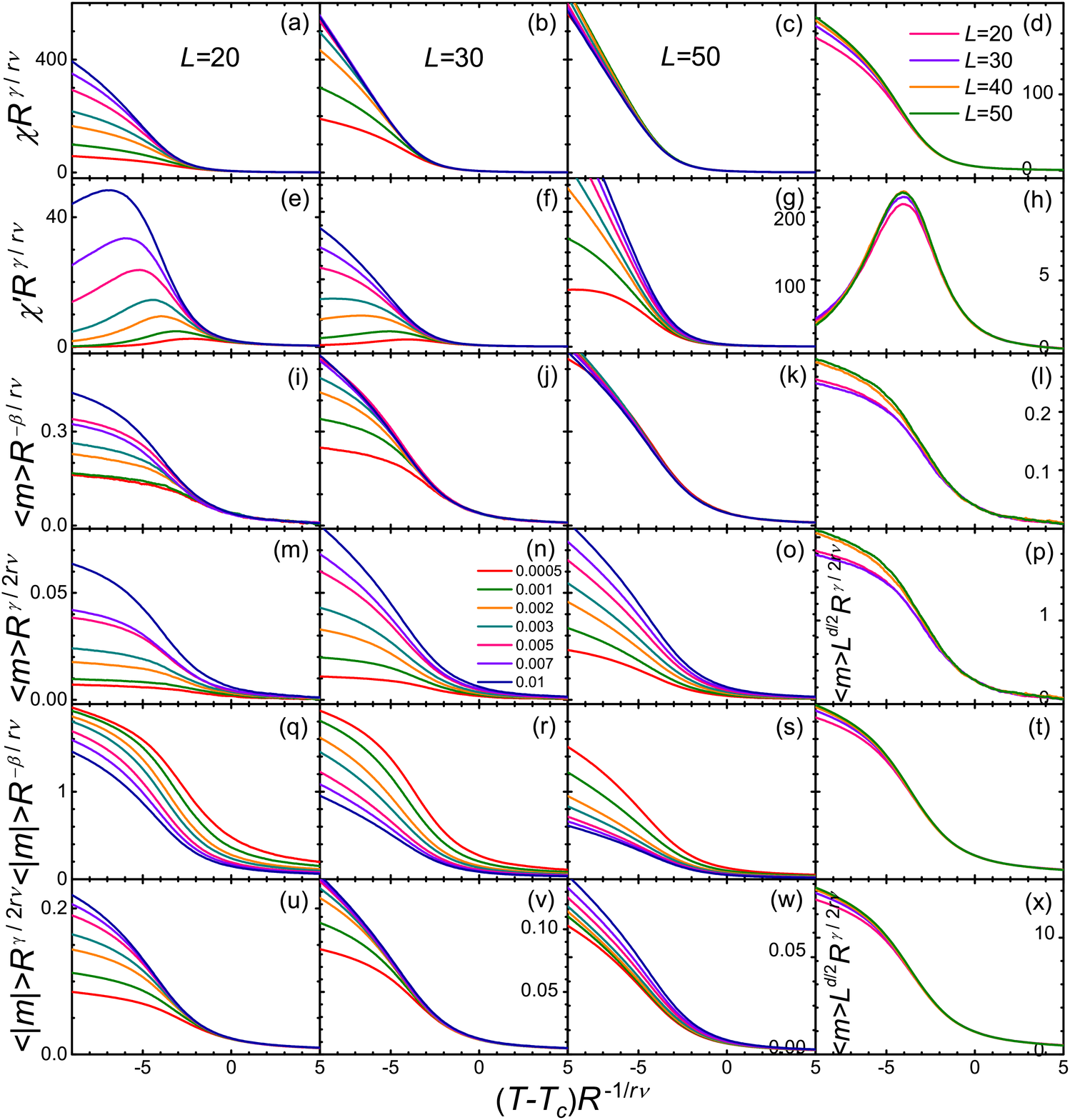,width=1\columnwidth}}
  \caption{\label{coolc05f} (Color online) FTS of (a)--(d) $\chi$ (first row), (e)--(h) $\chi'$ (second row), (i)--(l) $\langle m\rangle$ (third row), and (q)--(t) $\langle |m|\rangle$ (fifth row), and revised FTS of (m)--(p) $\langle m\rangle$ (fourth row) and (u)--(x) $\langle |m|\rangle$ (sixth row) with fixed $HR^{-\beta\delta/{r\nu}}=0.05$ on simple cubic lattices of sizes $L=20$ (first column), $L=30$ (second column), $L=50$ (third column), and four different values of $L$ that fix $L^{-1}R^{-1/r}\approx0.2687$ (fourth column) in protocol C. All panels except the last column share the same legend in (n) listing the rates used, while the last column shares the same legend in (d) listing the lattice sizes. Note that, in (p) and (x), we have included the $L^{d/2}$ factor in the vertical axes according to Eq.~(\ref{rescaling_huang}) and the corresponding scales are different from those in their respective rows. Also, the ordinates in (d), (h), (l), (v), and (w) are also different from those in their respective rows. In addition, (f) and (g) share identical scales marked in (g).}
\end{figure}
The rescaled curves are displayed in Fig.~\ref{coolc05f}. The scalings of $\chi$ and $\chi'$ are again almost identical with their corresponding ones in Figs.~\ref{coola05f} and~\ref{coolb05f} in protocols A and B, respectively. Note that the rescaled $\chi$ and $\chi'$ in the three figures, Figs.~\ref{coola05f},~\ref{coolb05f} and~\ref{coolc05f}, have identical abscissas and ordinates. The only appreciable difference is $\chi'$ of the large $R$ values on the $L=50$ lattices. Those curves move to higher temperatures as compared with their counterparts in protocols A and B. We will see in Sec.~\ref{cross} below that this reflects the peculiar scaling on this lattice size. $\langle m\rangle$ and $\langle |m|\rangle$ again appear to behave similarly to their corresponding ones in protocols A and B, including their opposite scaling results. However, on the one hand, different from protocol B, the scalings of $\langle m\rangle$ on $L=50$ lattices and for fixed $L^{-1}R^{-1/r}$ are good similar to protocol A, since both the latter are subjected to external fields. On the other hand, in contrast with protocol A and even protocol B, the scaling of $\langle |m|\rangle$ on $L=50$ lattices is simply bad, including the disordered phase in $T>T_c$, which is of course identical with protocol B, though those for smaller lattices and for fixed $L^{-1}R^{-1/r}$ are reasonable similar to protocols A and B. In addition, compared with Fig.~\ref{coolc2}, the scalings get better as the strength of the external field increases. Of course, if the field is so strong that $\xi_H$ is too short, the scaling is expected to deteriorate.

The bad scaling according to Eq.~(\ref{rescaling_huang}) of $\langle |m|\rangle$ on $L=50$ lattices in Fig.~\ref{coolc05f}(w) is rather surprising, noticing that the same scaling in protocol A as shown in Fig.~\ref{coola05f}(w) is good and the scalings on $L=20$ and $L=30$ are reasonable. Note also that $\langle |m|\rangle$ has almost no sample size dependence different from $\langle m\rangle$. However, as pointed out above, the same scaling in protocol B in Fig.~\ref{coolb05f}(w) is also not good, especially in the disordered phase in which the external field in applied. Therefore, whether the scaling of $\langle |m|\rangle$ on $L=50$ lattices is good or not depends on the disordered phase. On the one hand, when the scaling in the disordered phase above $T_c$ is good, as in protocol A in which no external field is applied, the scaling below $T_c$ is also good. On the other hand, when the scaling in the disordered phase above $T_c$ is poor, the scaling below is also poor, as is the cases in protocols B and C.
The problem is now why the scaling of $\langle |m|\rangle$ on $L=50$ lattices in the disordered phase above $T_c$ is not good in the presence of a small external field. Note that the same scaling on $L=30$ and $L=20$ lattices is rather good. This rules out the possibility that the special scaling form, Eq.~(\ref{rescaling_huang}), is only valid in the absence of the external field and invalid in the weak field.

\subsection{\label{cross}Crossover, Bressy exponents and beyond}
In this subsection, we will deal with the cases in the above three subsections that violate the scaling. We will find that crossover from the revised FTS regime to an FTS regime in a field is responsible for the bad scaling of $\langle|m|\rangle$ in Fig.~\ref{coolc05f}(w) and also Fig.~\ref{coolb05f}(w). Most unexpectedly, we will find that the Bressy exponent responsible for the FTS in \emph{heating} applies also to the self-similarity breaking in \emph{field cooling}. This indicates that the real difference between heating and cooling is the symmetry of the system states.

\begin{figure}
  \centerline{\epsfig{file=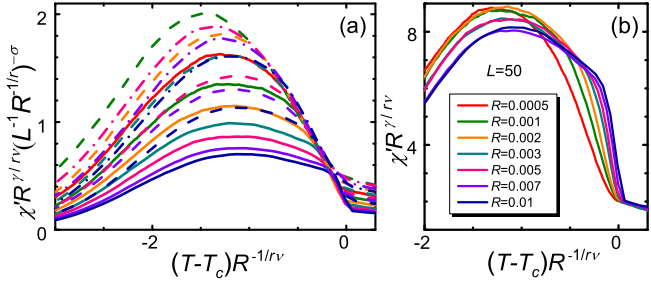,width=1\columnwidth}}
  \caption{\label{coola2ftsh} (Color online) (a) $\chi' R^{\gamma/r\nu}(L^{-1}R^{-1/r})^{-\sigma}$ versus $(T-T_c)R^{-1/r\nu}$ with $\sigma=(\gamma-2\beta)/\nu=0.5842/\nu$ for the rates $R$ listed in (b) on $L=20$ (dashed dotted lines), $L=30$ (dashed lines), and $L=50$ (solid lines) simple cubic lattices for $X=2$ in protocol A. (b) enlarged view of Fig.~\ref{coola2f}(g) for comparison.}
\end{figure}
Before tackling the apparent bad scaling of $\langle|m|\rangle$, we first consider the explicit breaking of scaling for $\chi'$ in Figs.~\ref{coola2f}(e)--\ref{coola2f}(g).
That fixing $L^{-1}R^{-1/{r\nu}}$ remedies the poor scaling of $\chi'$ near $T_c$ in Fig.~\ref{coola2f}(h) reminds us of the self-similarity of the phases fluctuations. This indicates that the poor scaling may arise from breaking of the self-similarity. From the first two rows of Fig.~\ref{coola2}, we see that the separation of the two sets of the observables appears even after the field is applied. The field breaks the symmetry of the up and down phases. Only $\chi'$ appears to violate scaling. These indicate that the situation here is similar to the FTS in heating. We show in Fig.~\ref{coola2ftsh}(a) the collapse using the Bressy exponent of FTS in heating. For comparison, we magnify in Fig.~\ref{coola2ftsh}(b) the original FTS collapse of the curves on the $L=50$ lattice in Fig.~\ref{coola2f}(g) only. It is the best one compared with the other two smaller lattice sizes, as clearly seen from Figs.~\ref{coola2f}(e)--\ref{coola2f}(g). Upon comparing Fig.~\ref{coola2ftsh}(a) with Fig.~\ref{coola2ftsh}(b), the good collapses of all the three lattice sizes on the high-temperature slope of the $\chi'$ peak for the two apparently distinct cases are marvellous! The violation of FTS upon cooling in a field that operates at and below $T_c$ is unexpectedly described by the same Bressy exponent as that of heating in the absence of field noting that cooling in the absence of the field is another matter! This indicates that what matters in the difference between heating and cooling is in fact whether the symmetry is broken or not and hence whether the system is ordered or disordered. Conversely, it also provides a distinct source to the Bressy exponents and thus corroborates their validity.

We note that, for $X=0.05$ in protocol A, the field is too small to induce a sufficiently large $\langle|m|\rangle$ and suppress the phase fluctuations. Accordingly, the kink at $T_c$ that is vulnerable to the self-similarity breaking is hardly visible, as seen is Fig.~\ref{coola05f}(g). Therefore, no anomalous scaling appears and hence no Bressy exponent is needed here, even though we will see in Figs.~\ref{coolc05ftsh} and~\ref{coolb05ftshc} below for $X=0.05$ in protocols C and B, respectively, that the same weak field being applied from the disordered phase produces appreciable effects.

\begin{figure}
  \centerline{\epsfig{file=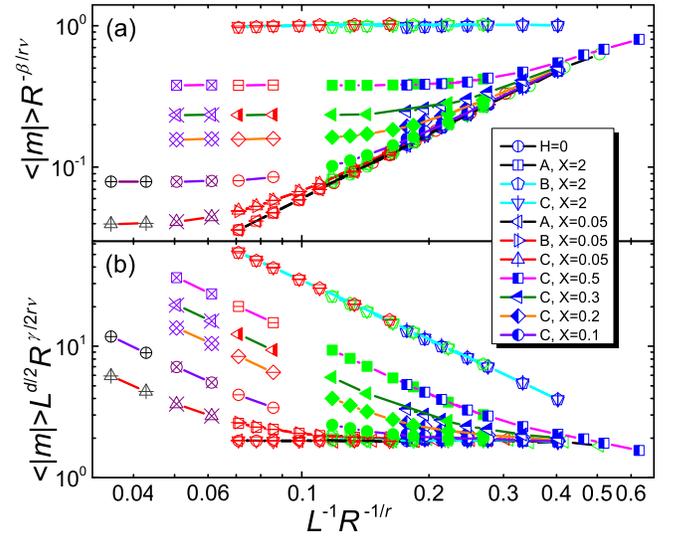,width=1.0\columnwidth}}
  \caption{\label{lram} (Color online) (a) $\langle|m|\rangle R^{-\beta/r\nu}$ and (b) $\langle|m|\rangle L^{d/2}R^{\gamma/2r\nu}$ at $T_c$ versus $L^{-1}R^{-1/r}$ for several fixed values of $X\equiv HR^{-\beta\delta/{r\nu}}$ and the three protocols including $H=0$ on 3D simple cubic lattices. The legend lists the symbols for $L=20$, which are blue symbols with vertical lines or divisions. The other lattice sizes have the same symbol types, only their details differ: Green open or filled symbols denote $L=30$, red symbols with minuses represent $L=50$, violet symbols with crosses stand for $L=70$, and black symbols with pluses indicate $L=100$. The lowest and highest oblique lines in (a) and (b) have slopes approximately $d/2$ and $-d/2$, respectively. Lines connecting symbols are only a guide to the eyes. Identical $X$ value shares identical line color.}
\end{figure}
Next we turn to the scaling of $\langle|m|\rangle$. To find out the reasons for the bad scaling in Fig.~\ref{coolc05f}(w), we draw in Fig.~\ref{lram} the dependence on $L^{-1}R^{-1/r}$ of $\langle|m|\rangle$ exactly at $T_c$ rescaled separately by $R^{-\beta/r\nu}$ and $L^{d/2}R^{\gamma/2r\nu}$. Results for larger lattice sizes and other $X$ are also presented. This is the kind of plots studied in detail in Ref.~\cite{Huang}. First, note the lowest (black) lines (with circles) with $H=0$, the lines which are most visible from the left of the two panels and are the straightest in both panels. At least for $L^{-1}R^{-1/r}<0.3$ in the FTS regime, $\langle|m|\rangle R^{-\beta/r\nu}$ would be a horizontal line according to the standard FTS, Eq.~(\ref{FTSM}). Yet, it is oblique with a slope of about $d/2$, as Fig.~\ref{lram}(a) shown. It is this reason that leads to the introduction of Eq.~(\ref{rescaling_huang}). As a consequence, it becomes indeed a horizontal line in Fig.~\ref{lram}(b), signalling the leading behavior is $L^{-d/2}R^{-\gamma/2r\nu}$ instead of $R^{\beta/r\nu}$. For large $L^{-1}R^{-1/r}\gg0.3$, the two lines will have slopes of $\beta/\nu\approx0.5182$ and $-\gamma/2\nu\approx-0.9817$, respectively, in the FSS regime~\cite{Huang} and thus they will slightly bend downwards. When a strong enough external field is applied, $\langle|m|\rangle R^{-\beta/r\nu}$ becomes a usual horizontal line and $\langle|m|\rangle L^{d/2}R^{\gamma/2r\nu}$ changes correctly to an oblique line with a slope of $-d/2$ accordingly. This is the case of the (cyan) lines (with inverted triangles) with $X=2$ in protocol C, though the lines of different $L$ values do not collapse exactly onto each other, possibly because of corrections to scaling that possibly become important for such a large $X=2$ but have not been taken into account. The only other case that has lines (with pentagons) which overlap these lines is again the case with $X=2$ but in protocol B. This must be true because the field is just switched off at $T_c$ in protocol B and hence this protocol must be identical with protocol C by $T_c$, as pointed out above. For the other case with $X=2$ in protocol A, the lines (with squares) must coincide with the case with no applied field as seen in Fig.~\ref{lram} because the field is applied at and below $T_c$.

We now turn to the cases with $X=0.05$ and will see that the crossover from the special revised FTS regime governed by Eq.~(\ref{rescaling_huang}) to the usual FTS regime following Eq.~(\ref{FTSM}) is responsible for the bad scaling. Similar to those with $X=2$, for protocol A, the results exactly at $T_c$ are identical with those in the absence of the field, while those in protocols B and C coincide, as seen from the lines with left triangles for protocol A and right and up triangles for protocols B and C in Fig.~\ref{lram}. However, it is clearly seen from Fig.~\ref{lram} that the (red) lines in protocols B and C deviate from those in the absence of the field and bend upwards for small $L^{-1}R^{-1/r}$. With the four up triangles for $L=100$ and $L=70$ and $R=0.01$ and $R=0.005$ in protocol C, one sees clearly that the curves in protocols B and C turn horizontal in Fig.~\ref{lram}(a) and oblique with a slope of $-d/2$ in Fig.~\ref{lram}(b), viz., enter the FTS regime for smaller values of $L^{-1}R^{-1/r}$. Therefore, the bends lie just in the crossover of the two regimes and exhibit bad scaling.

From Fig.~\ref{lram}, the deviation away from the revised FTS regime to the FTS regime in a field depends on $X=HR^{-\beta\delta/r\nu}$. The smaller the $X$ is, the smaller the crossover value of $L^{-1}R^{-1/r}$. This implies that $L$ is a fixed times of the field length $\xi_H\sim H^{-\nu/\beta\delta}$ when the crossover begins, since different $R$ ought to cross over at the same $L^{-1}R^{-1/r}$ for a fixed $X$. In fact, this in turn reflects the competition of $L$- and $\xi_H$-dominant regime determined by $LH^{\nu/\beta\delta}$. Accordingly, the crossover occurs where $LR^{1/r}\propto (HR^{-\beta\delta/r\nu})^{-\nu/\beta\delta}$ satisfying $L^{-1}R^{-1/r}\rightarrow0$ for $X=0$. For $X=0.05$, the crossover starts at $L^{-1}R^{-1/r}\approx0.16$. In other words, the crossover begins at $L=0.05^{\nu/\beta\delta}/0.16\approx1.87\xi_H$. Indeed, this value gives back the crossover beginning at about $0.21$, $0.28$, $0.33$, and $0.40$ for $X=0.1$, $0.2$, $0.3$, and $0.5$, respectively, approximately consistent with the real values of crossover in Fig.~\ref{lram}. The same relation ought to hold for the end of the crossover. For $X=0.05$, Fig.~\ref{lram} gives the FTS regime in a field begins at about $L^{-1}R^{-1/r}\approx 0.04$, which corresponds to $L\approx7.5\xi_H$ roughly. This yields the crossover ends at $L^{-1}R^{-1/r}\approx 0.04$, $0.05$, $0.07$, $0.08$, and $0.1$ for $X=0.05$, $0.1$, $0.2$, $0.3$, and $0.5$, respectively, appearing to agree roughly with the expected values in the figure too.

Therefore, the crossover from the revised FTS regime to the FTS regime in a field falls within a range of roughly $1.87\xi_H\lesssim L\lesssim7.5\xi_H$ for a fixed $X$. For a larger $L$, the usual FTS appears (but see Sec.~\ref{ftshl} below), whereas for a smaller $L$ (but still larger than $\xi_R$), the revised FTS still persists even though is applying an external field which breaks the up-down symmetry. Conversely, given a sufficiently large fixed $L$, for sufficiently large $R$ satisfying $\xi_R\ll L$ but not too large so that corrections to scaling can be neglected, the system exhibits the revised FTS behavior for an applied external field smaller than $4.72L^{-\beta\delta/\nu}$ or so, or its corresponding $\xi_H$ larger than about $L/1.87$. This includes, of course, the case of $H=0$. On the other hand, when the field is so strong that its corresponding $\xi_H$ is smaller than $L/7.5$ or so, the usual FTS behavior comes out. This is true even when $\xi_H$ is smaller than $\xi_R$ and hence the controlling factor is $H$ instead of $R$, provided that $HR^{-\beta\delta/r\nu}$ is fixed. This is the case of $X=2$.

\begin{table*}
\caption{\label{tab} Summary of the behavior of $\langle|m|\rangle$ in various protocols in cooling.}
\begin{ruledtabular}
\begin{tabular}{llccccccccccc}
&&$H=0$ &\multicolumn{4}{c}{Protocol A}& \multicolumn{3}{c}{Protocol B} & \multicolumn{3}{c}{Protocol C}\\\cline{4-7}\cline{8-10}\cline{11-13}
$X\equiv HR^{-\frac{\beta\delta}{r\nu}}$&&&\multicolumn{2}{c}{$X=2$} & \multicolumn{2}{c}{$X=0.05$}&$X=2$ & \multicolumn{2}{c}{$X=0.05$} &$X=2$ & \multicolumn{2}{c}{$X=0.05$}\\\cline{4-5}\cline{6-7}\cline{9-10}\cline{12-13}
\rule{0pt}{10pt}&&  & $T<T_c$&$T>T_c$ &$T<T_c$&$T>T_c$ &  & $L=50$  & $L=30$ &  & $L=50$  & $L=30$\\
\hline
$\langle |m|\rangle R^{-\frac{\beta}{r\nu}}$ vs $L^{-1}R^{-\frac{1}{r}}$ & slope &$d/2$ & \multicolumn{2}{c}{$d/2$} &\multicolumn{2}{c}{$d/2$} & $0$  & curved & $d/2$ & $0$ & curved & $d/2$\\
$\langle |m|\rangle R^{-\frac{\beta}{r\nu}}$ vs $\tau R^{-\frac{1}{r\nu}}$& overlap & $\times$& $\checkmark$&$\times$ &$\times$ &$\times$& $\checkmark$&$\times$&$\times$& $\checkmark$ &$\times$& $\times$\\
$\langle |m|\rangle L^{\frac{d}{2}}R^{\frac{\gamma}{2r\nu}}$ vs $L^{-1}R^{-\frac{1}{r}}$ & slope&$0$ & \multicolumn{2}{c}{$0$} &\multicolumn{2}{c}{$0$} & $-d/2$ & curved & $0$ & $-d/2$ & curved & $0$ \\
$\langle |m|\rangle L^{\frac{d}{2}}R^{\frac{\gamma}{2r\nu}}$ vs $\tau R^{-\frac{1}{r\nu}}$&overlap&$\checkmark$&$\times$&$\checkmark$ &$\sim\checkmark$ &$\checkmark$& $\times$&$\times$&$\checkmark$& $\times$ &$\times$& $\checkmark$\\
\end{tabular}
\end{ruledtabular}
\end{table*}
We note that Fig.~\ref{lram} appears to show good scaling in the sense that data of different lattices for a fixed $X$ collapse well even for the bends in the crossover, which enters the FTS regime for smaller values of $L^{-1}R^{-1/r}$~\cite{Huang}. This is true because the results obey both Eqs.~(\ref{FTSM}) and~(\ref{rescaling_huang}) at $T_c$ and thus $\langle|m|\rangle R^{-\beta/r\nu}$ and  $\langle|m|\rangle L^{d/2}R^{\gamma/2r\nu}$ are functions of $L^{-1}R^{-1/r}$, provided that other corrections to scaling can be ignored. However, this good scaling does not ensure good scaling collapses when the entire curves rather than the values exactly at $T_c$ are considered. Indeed, in the special scaling regime in which $L^{-1}R^{-1/r}$ plays an important role and thus cannot be neglected as a small perturbation, the scaling collapse of $\langle|m|\rangle R^{-\beta/r\nu}$ versus $\tau R^{-1/r\nu}$ is destined to be bad. Only when it is negligible can the scaling collapses well. This is the case when the leading behavior is caught and thus a horizontal line appears. Thus, protocol C with $X=2$ and protocol B with $X=2$ and above $T_c$ have good collapses in terms of $\langle|m|\rangle R^{-\beta/r\nu}$ versus $\tau R^{-1/r\nu}$. So do cooling in zero field, protocol A above $T_c$, and protocols B and C with $X=0.05$ and sufficiently large $L^{-1}R^{-1/r}$ in terms of $\langle|m|\rangle L^{d/2}R^{\gamma/2r\nu}$ versus $\tau R^{-1/r\nu}$. These are summarized in Table~\ref{tab}, in which the first column indicates the information of the row, the second further specifies either the results in the row are the slopes of the lines in Fig.~\ref{lram} or whether the scaling collapses in the row are good or bad, represented by $\checkmark$ or $\times$, respectively. Note that the values of the slope in Table~\ref{tab} are the theoretical ones generic in the regimes classified by the simple characteristics given in the columns. For example, in the two columns labelled by $L=30$, the system can enter the FSS regime with slopes different from the listed ones for rates far smaller than those we used. In addition, appropriate rates on $L=20$ that is not listed in Table~\ref{tab} can exhibit similar behavior to $L=30$, as can be seen from Figs.~\ref{coolb05f},~\ref{coolc05f}, and~\ref{lram}.

There are three entries in Table~\ref{tab} that run beyond the rules summarized above. All are related to $T<T_c$ in protocol A. This is because at $T_c$ the slopes in protocol A satisfy the rule of zero field as seen in Table~\ref{tab} but the field is applied below it. Accordingly, the scaling collapses should follow the behaviors in protocol C. This is true for the strong fields with $X=2$, see Table~\ref{tab}, and also the weak fields with $X=0.05$ on the two small lattices, as Figs.~\ref{coola05f}(u) and~\ref{coola05f}(v) display. However, for $L=50$, the overlap in Fig.~\ref{coola05f}(w) and its inset is not as poor as the corresponding one in protocol C, Fig.~\ref{coolc05f}(w), and even in protocol B, Fig.~\ref{coolb05f}(w). This results in the special entry meaning ``almost good'' in Table~\ref{tab}. The reason for the almost good collapse on $L=50$ lattices is the good collapse in $T>T_c$ in contrast to the poor ones in protocols B and C for $X=0.05$. This ``initial'' condition prevents the curves to separate appreciably below $T_c$.

\begin{figure}[b]
  \centerline{\epsfig{file=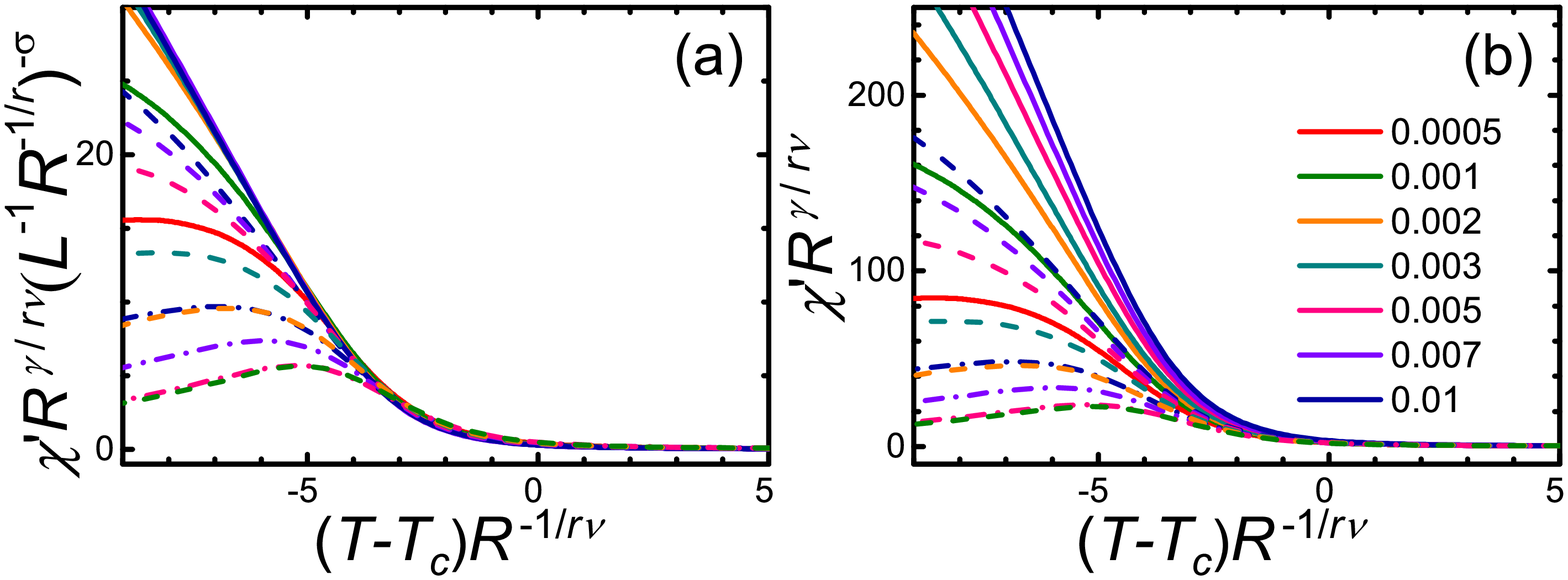,width=1\columnwidth}}
  \caption{\label{coolc05ftsh} (Color online) (a) $\chi' R^{\gamma/r\nu}(L^{-1}R^{-1/r})^{-\sigma}$ versus $(T-T_c)R^{-1/r\nu}$ with $\sigma=(\gamma-2\beta)/\nu=0.5842/\nu$ and (b) $\chi' R^{\gamma/r\nu}$ versus $(T-T_c)R^{-1/r\nu}$ for the rates $R$ listed in (b) on $L=20$ (dashed dotted lines), $L=30$ (dashed lines), and $L=50$ (solid lines) simple cubic lattices for $X=0.05$ in protocol C. (b) is a combination of the three curves in Fig.~\ref{coolc05f}(e), six curves in Fig.~\ref{coolc05f}(f), and all curves in Fig.~\ref{coolc05f}(g) for comparison.}
\end{figure}
We have clarified the origin of the two cases that do not exhibit good scaling in both the usual FTS and revised FTS forms listed in Table~\ref{tab}, viz., $L=50$ with $HR^{-\beta\delta/r\nu}=0.05$ in protocols B and C. However, as shown in Figs.~\ref{coolb05f}(t),~\ref{coolb05f}(x),~\ref{coolc05f}(t), and~\ref{coolb05f}(x), once $L^{-1}R^{-1/r}$ is fixed, good scalings are regained. This again reminds us of the importance of the extrinsic self-similarity. Similar to Fig.~\ref{coola2ftsh}, in Fig.~\ref{coolc05ftsh}(a), we use again the heating FTS Bressy exponent $\sigma=(\gamma-2\beta)/\nu$ to collapse the $\chi'$ curves in Figs.~\ref{coolc05f}(e)--(g) in protocol C. The result is again wonderful comparing with Fig.~\ref{coolc05ftsh}(b), even though the field here is far smaller than that in Fig.~\ref{coola2ftsh}.

\begin{figure}[b]
  \centerline{\epsfig{file=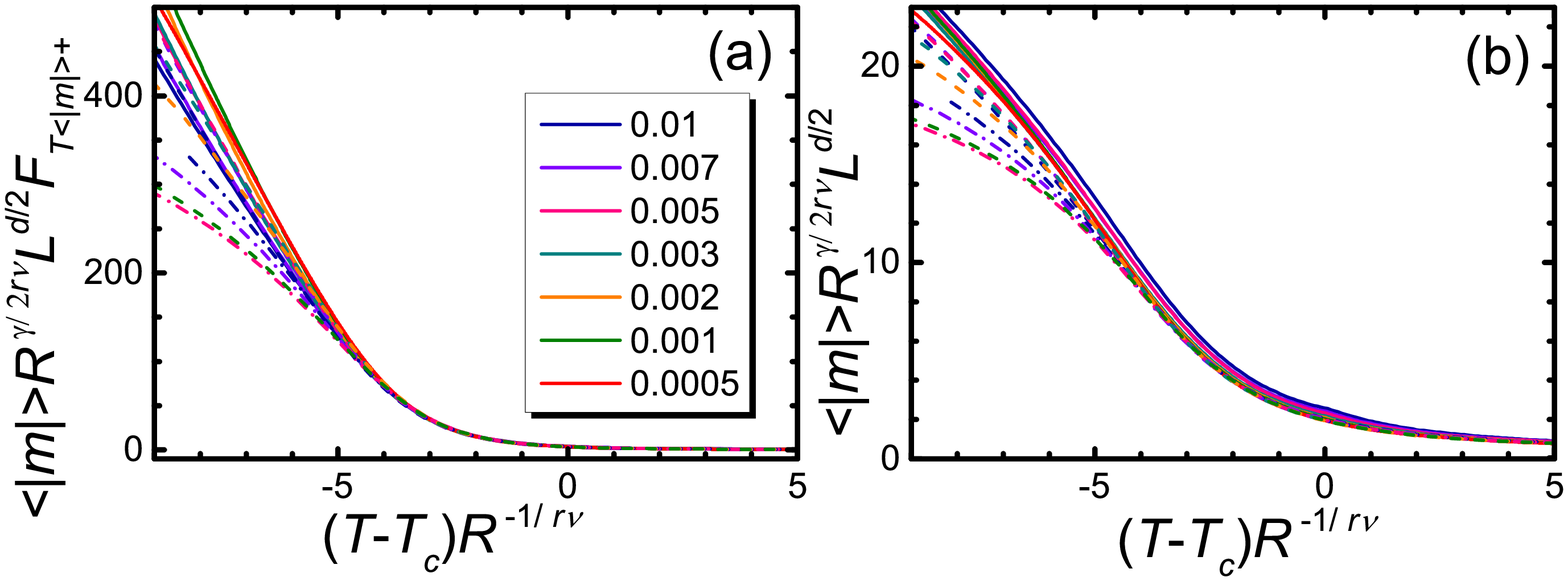,width=1\columnwidth}}
  \caption{\label{coolb05ftshc} (Color online) (a) $\langle|m|\rangle R^{\gamma/2r\nu}L^{d/2}F_{T\langle|m|\rangle+}$ versus $(T-T_c)R^{-1/r\nu}$ with $F_{T\langle|m|\rangle+}$ given by Eq.~(\ref{msigmat}) and (b) $\langle|m|\rangle R^{\gamma/2r\nu}L^{d/2}$ versus $(T-T_c)R^{-1/r\nu}$ for the rates $R$ listed in (a) on $L=20$ (dashed dotted lines), $L=30$ (dashed lines), and $L=50$ (solid lines) simple cubic lattices for $X=0.05$ in protocol B. (b) is a combination of the three curves in Fig.~\ref{coolb05f}(u), six curves in Fig.~\ref{coolb05f}(v), and all curves in Fig.~\ref{coolb05f}(w) multiplied by their respective $L^{d/2}$ factor for comparison.}
\end{figure}
The case of $L=50$ with $HR^{-\beta\delta/r\nu}=0.05$ in protocols $B$ is subtle. For $T>T_c$ on the one hand, the field is being applied and the situation is similar to the case of protocol C. The primary observable is $\chi'$ and $\langle|m|\rangle$ is the secondary observable obeying Eq.~(\ref{msigmat}). However, a plot similar to Fig.~\ref{coolc05ftsh} is not good. This is because for $T\leq T_c$ on the other hand, the field is absent and $\langle|m|\rangle$ is the primary observable satisfying $\langle|m|\rangle R^{-\beta/r\nu}\sim (L^{-1}R^{-1/r})^{d/2+\beta/4\nu}$. In Fig.~\ref{coolb05ftshc}(a), we plot $\langle|m|\rangle R^{\gamma/2r\nu}F_{T\langle|m|\rangle+}$ versus $(T-T_c)R^{-1/r\nu}$ with $F_{T\langle|m|\rangle+}$ given by Eq.~(\ref{msigmat}). Comparing with the original rescaled curves in Fig.~\ref{coolb05ftshc}(b), one sees that the collapse is substantially improved for $(T-T_c)R^{-1/r\nu}>-5$. Noting that $F_{T\langle|m|\rangle+}$ has already contained the Bressy exponent $\sigma$ for FTS in heating, we have, however, omitted the reduced exponent $\sigma'=\beta/4\nu$ for FTS in cooling. Including this exponent indeed puts the curves at large $-(T-T_c)R^{-1/r\nu}$ closer to each other. Yet, the good collapse at the remaining part deteriorates. Both are not as good as the case shown in Fig.~\ref{ftsc3d}. Accordingly, the Bressy exponent for FTS in cooling appears not needed. This is understandable since the applied field up to $T_c$ must affect the evolution below it, which is apparent by comparing Fig.~\ref{coolb05f}(w) in an applied field with Fig.~\ref{FTS_50}(c) in the absence of the field and also the almost good collapse in Fig.~\ref{coola05f}(w) of protocol A, even though the field is small and the $\langle|m|\rangle$ curves themselves appear similar.

We have seen that the quality of the scaling above $T_c$ determines that below $T_c$ in the three protocols for the small $X$ on $L=50$ lattices. In protocol A on the one hand, the applied field below $T_c$ does not invalidate the somehow good revised scaling even though it should, according to Fig.~\ref{lram} and the results of protocol C. In protocol B on the other hand, the absence of the applied field below $T_c$ does not guarantee the revised scaling owing to the applied field above $T_c$ that invalidates both the FTS in a field and the revised FTS there. These indicate that the weak field below $T_c$ in protocol A hardly break the revised scaling while that above $T_c$ in protocol B deteriorates it below $T_c$. Moreover, for the large $X$, the presence of the strong field above $T_c$ simply eradicates the revised FTS in the whole process in protocol B given that the field raises the practical transition temperature above $T_c$, even though in protocol A the revised FTS below $T_c$ is also eliminated due to the rather strong field in contrast to the case of the weak field. Therefore, the phases fluctuations at and just above $T_c$ are more important than those below it, even though in the small $X$ case the practical transition temperature characterized by the $\chi$ peak lies below $T_c$. Moreover, an order at $T_c$ is important to suppress the phases fluctuations in cooling and acts like the heating with an ordered state. In addition, these results demonstrate the advantage of applying an external field in different protocols.

\subsection{\label{ftshl}FTS regime in a field on large lattices}
In the previous section, we have referred to the regime on the left beyond the end of the revised FTS regime in Fig.~\ref{lram} as ``the FTS regime in a field". In this section, we will show that this cooling FTS regime in a field is different from the heating one: Whereas the heating one is the usual FTS regime, the cooling FTS regime in a field is somewhat special in the sense that the scaling function of the order parameter $M$ is odd in $X=HR^{-\beta\delta/r\nu}$, the scaled variable proportional to the field. Accordingly, for fixed $X$, $M$ is asymptotically proportional to $R^{-\beta/r\nu}$, the usual one. This is the characteristic exhibited by the horizontal lines on the left of Fig.~\ref{lram}(a) and listed in Table~\ref{tab}. However, for different small $X$, $M\sim R^{-\beta/r\nu}X^{-1}$.

In the thermodynamic limit $L=\infty$, the behavior of a driven system subject to a constant external field is theoretically simple. According to Eq.~(\ref{FTSM}), $M\sim R^{\beta/r\nu}$ in the FTS regime for $HR^{-\beta\delta/r\nu}\ll1$ and $M\sim H^{1/\delta}$ in the field-governed regime for $HR^{-\beta\delta/r\nu}\gg1$ with a crossover between the two regimes. Therefore, the FTS regime in the graph of $MR^{-\beta/r\nu}$ at $T_c$ versus $HR^{-\beta\delta/r\nu}$ is a horizontal line independent on $HR^{-\beta\delta/r\nu}$ and changes to an inclined line with a slope of $1/\delta$ in the field-governed regime.

The above picture contradicts manifestly the feature of the FTS regime on large finite lattices shown in Fig.~\ref{lram}, where different $HR^{-\beta\delta/r\nu}$ values have different values of $MR^{-\beta/r\nu}$ even for rather small $HR^{-\beta\delta/r\nu}$ and $L^{-1}R^{-1/r}$. In fact, the present picture exhibited in Fig.~\ref{lram} has appeared when several confinements
and hence several length scales interplay~\cite{Cao}. Figure~\ref{lram}(a) is the projections of the intersections of the planes of constant $HR^{-\beta\delta/r\nu}$ and the surface of $F_T(0, L^{-1}R^{-1/r}, HR^{-\beta\delta/r\nu})$, the scaling function of $M$ at $T_c$, onto the plane $MR^{-\beta/r\nu}$ versus $L^{-1}R^{-1/r}$. On the basis of it, one can thus reconstruct the surface, on which the bottom of the revised FTS bevel demarcated by the surface $L^{-1}R^{-1/r}\approx1/1.87 HR^{-\beta\delta/r\nu}$ rises. However, this rise for small $HR^{-\beta\delta/r\nu}$ does not yet reach the crossover to the field-dominated regime in the thermodynamic limit and its slope of about $1$ is far bigger than $1/\delta\approx0.21$. What happens?

\begin{figure}
\centering
\centerline{\epsfig{file=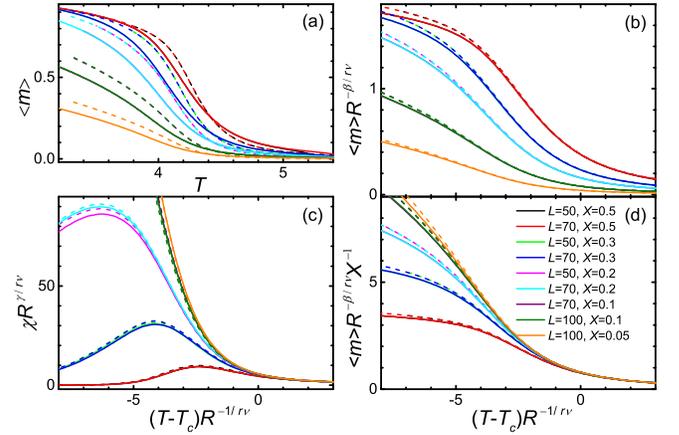,width=1.0\columnwidth}}
\caption{\label{ftsch}(Color online) (a) $\langle m\rangle$ and FTS of (b) $\langle m\rangle$ and (c) $\chi$ for cooling at $R=0.01$ (solid lines) and $R=0.005$ (dashed lines) and various $X=HR^{-\beta\delta/r\nu}$ on 3D simple cubic lattice of sizes $L$ given in the legend in (d), which is a revised form of (b).}
\end{figure}
To uncover the secret, we plot in Fig.~\ref{ftsch}(a) $\langle m\rangle$ versus $T$ for $L\geq50$. The dependence of $\langle m\rangle$ on $L$ and $X$ is similar to previous results. The size independence of $\langle m\rangle$ on $L$ for a fixed $X$ is clearly seen from the coincidence of the curves. We do not show the results for the other set of the observables because they exhibit similar behavior. This does not mean that $\langle m\rangle$ and $\langle |m|\rangle$ do not differ. In fact, they start to differ appreciably for $X=0.05$ and $L=70$ and then $X=1$ and $L=50$, both at $R=0.005$. Note that the difference between the two sets of the observables increases for smaller lattices and smaller rates in conformity with the results in Sec.~\ref{ftsofls}, besides for smaller $X$. In Fig.~\ref{ftsch}(b) and~\ref{ftsch}(c), we show the FTS of $\langle m\rangle$ and $\chi$, respectively. One sees from Fig.~\ref{ftsch}(b) that the FTS of $\langle m\rangle$ for different $X$ values is totally bad, although the FTS of $\chi$ is good for $T>T_c$. Note, however, that for a given $X$, the FTS of $\langle m\rangle$ for different $R$ and $L$ values is quite good. This indicates that the theory of FTS as is embodied in the scaling function $F_T$ describes the scaling well. The bad scaling only means that $X$ in $F_T$ cannot be ignored. This may be because either $X$ is not small enough or it cannot be ignored at all, similar to the role of $L^{-1}R^{-1/r}$ in cooling in the absence of $H$.
In Fig.~\ref{ftsch}(d), one sees that a further factor $X^{-1}$ collapses the curves well. This means that $F_T$ does not depend on $X$ singularly. What special is only that the first term in the expansion of $F_T$ with respect to $X$ is not a constant but rather a linear term. This in turn is reasonable since the field select the preferred direction and $\langle m\rangle$ must be proportional to $H$. From the collapses in Figs.~\ref{ftsch}(c) and~\ref{ftsch}(d), it is clear that the smaller the $X$ values are, the better the collapses. Therefore, $\chi$ is standard; only $M$ that is proportional to $H$ is special. For $X=0$, $\langle m\rangle=0$ too.

The above exponent of $-1$ provides an example in support of our conclusion that the Bressy exponents found in Sec.~\ref{bressy} ought to be new. A combination of the 3D Ising model exponents that is close to $1$ is $\nu/2\beta$, which is about $0.97$ and can collapse the $\langle m\rangle$ curves quite well. For the 2D Ising model, this combination yields $4$. However, there exist several other combinations that give rise to $1$ given the exactly known 2D critical exponents. Accordingly, we see that we could have different expressions of the exponent for the 2D and the 3D models. However, the real exponent is in fact a new one, which is just that of the linear term in the expansion!

\begin{figure}
\centering
\centerline{\epsfig{file=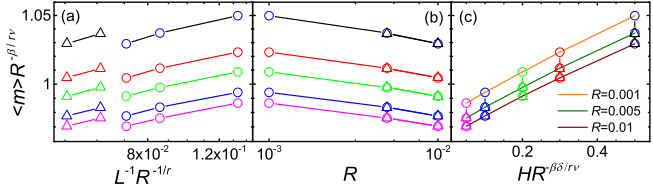,width=1.0\columnwidth}}
\caption{\label{hmlrh} (Color online) Dependence of $\langle m\rangle R^{-\beta/r\nu}$ at $T_c$ on (a) $L^{-1}R^{-1/r}$, (b) $R$, and (c) $HR^{-\beta\delta/{r\nu}}$ for the Ising model subject to heating at $R=0.001$, $R=0.005$, and $R=0.01$ [from up to down in (c)] on 3D simple cubic lattice of sizes $L=50$ (circles) and $L=70$ (triangles) with fixed $HR^{-\beta\delta/{r\nu}}=0.5$ (black), $0.3$ (red), $0.2$ (green), $0.1$ (blue), and $0.05$ (magenta) [from up to down in (a) and (b)]. Double logarithmic scales are used in (a) and (b) while linear scale is used in the abscissa of (c). Lines connecting symbols are only a guide to the eyes.}
\end{figure}
\begin{figure}
\centering
\centerline{\epsfig{file=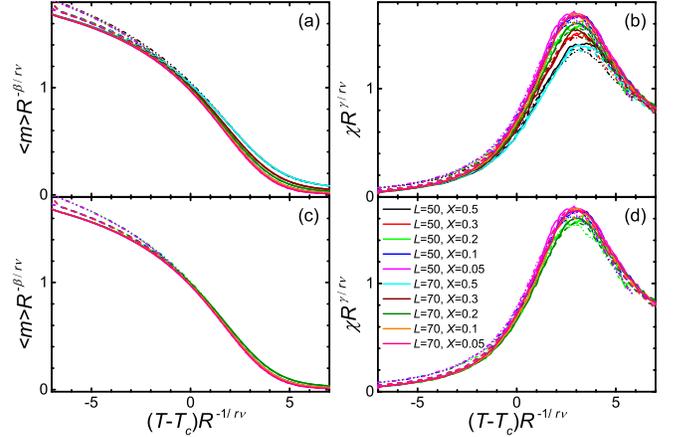,width=1.0\columnwidth}}
\caption{\label{ftshh}(Color online) FTS of (a) $\langle m\rangle$ and (b) $\chi$ for heating at $R=0.01$ (solid lines), $R=0.005$ (dashed lines), and $R=0.001$ (chain lines) and various $X=HR^{-\beta\delta/r\nu}$ on 3D simple cubic lattice of sizes $L$ given in the legend in (d). (c) and (d) are (a) and (b), respectively, in the absence of the curves with $X=0.5$ and $X=0.3$.}
\end{figure}

For comparison, we draw Fig.~\ref{hmlrh} for heating, a graph which is similar to Fig.~\ref{lram}(a). We employ $\langle m\rangle$ instead of $\langle |m|\rangle$ since they differ only at high temperatures for the chosen parameters. Similar feature appears in heating. Both the apparent independence and the apparent dependence of $MR^{-\beta/r\nu}$ at $T_c$ on $HR^{-\beta\delta/r\nu}$ are displayed in Fig.~\ref{hmlrh}(b) and~\ref{hmlrh}(c), respectively. Only the variation with $L^{-1}R^{-1/r}$ shows difference even for $L=70$, which may be still not large enough. For heating, the dependence on $HR^{-\beta\delta/{r\nu}}$ appears even exponential as seen in Fig.~\ref{hmlrh}(c), though the three lines there for the three rates are in fact not that straight. In Fig.~\ref{ftshh}(a) and~\ref{ftshh}(b), we display again the FTS of $\langle m\rangle$ and $\chi$, respectively, for heating. Here, the FTS for $\langle m\rangle$ is not as bad as that in cooling, Fig.~\ref{ftsch}(a). Interestingly, one observes three curves at low temperatures while four curves at high temperatures. They correspond to the three rates at low temperatures and the five $X$ at high temperatures with the two curves for the smallest $X$ values being almost overlapped. A similar trend can also be perceived for $\chi$ in Fig.~\ref{ftshh}(b). Accordingly, it seems impossible to find an extra factor similar to that for cooling to produce a better collapse. In fact, when the curves for the two largest $X$ values are omitted, the scaling collapses become better as illustrated in Figs.~\ref{ftshh}(c) and~\ref{ftshh}(d). This is again reasonable since the initial condition of an ordered state has selected a preferred direction. As a consequence, here the constant term in the expansion survives in contrast to cooling. This is another qualitatively different behavior in heating and cooling in the presence of an external field.

\section{\label{nr of ordering}FTS in heating with different dynamic critical exponents}
In this section, we study the issue of different dynamic critical exponent $z$ estimated in heating and cooling in the absence of an external field. Only heating is considered since the estimated cooling $z$ agree well with other estimations. We employ the usual definitions of $M$ and $\chi$ without using absolute values, viz., the first equations in Eqs.~(\ref{M}) and~(\ref{modelII}), in this section.

\begin{figure}
\centerline{\epsfig{file=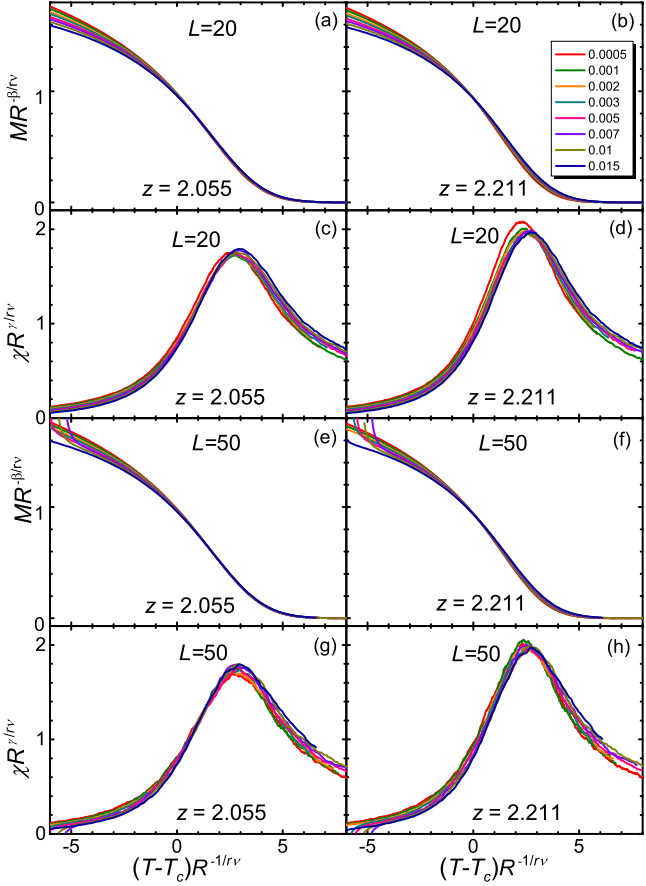,width=1.0\columnwidth}}
\caption{(Color online) FTS of (a), (b), (e), and (f), $M$ and (c), (d), (g), and (h), $\chi$ using $z=2.055$ (left column) and $z=2.211$ (right column) for the 3D Ising model on $20\times20\times20$, (a)--(d), and $50\times50\times50$, (e)--(f), lattices for a series of rate $R$ given in the legend, which is shared by all the panels.}
\label{heatL}\end{figure}
In Ref.~\cite{Huang}, the dynamical critical exponent $z$ was estimated through fitting the expansions of the scaling functions in different regimes right at $T_c$. The best fitted $z$ was found to be $2.211(2)$ and $2.055(5)$ in heating and cooling, respectively, for the 3D Ising model using $\langle |m|\rangle$ and $\langle m^2\rangle$. However, the scaling collapse of $\chi'$ at $T_c$ versus $L^{-1}R^{-1/r}$ in heating for the 3D model requires a $z$ which is, to the contrast, smaller than that in cooling, a situation which is different from the 2D model (see below). Corrections to scaling and the fact that $T_c$ lies in the ordered phase in heating whereas it sits in the disordered phase in cooling as seen from Figs.~\ref{ftshlr} and~\ref{ftsclr}, as well as a small sample size of no more than 10 thousands, were suggested as possible reasons for the different $z$ in heating and cooling. Indeed, Figs.~\ref{ftshlr} and~\ref{ftsclr} show that the scaling exactly at $T_c$ in heating is poor than that in cooling. This may give rise to the difference values of the estimated $z$. Here, we compare the scaling collapses of the two exponents in the whole critical regime rather than just at $T_c$ itself.

\begin{figure}
\centerline{\epsfig{file=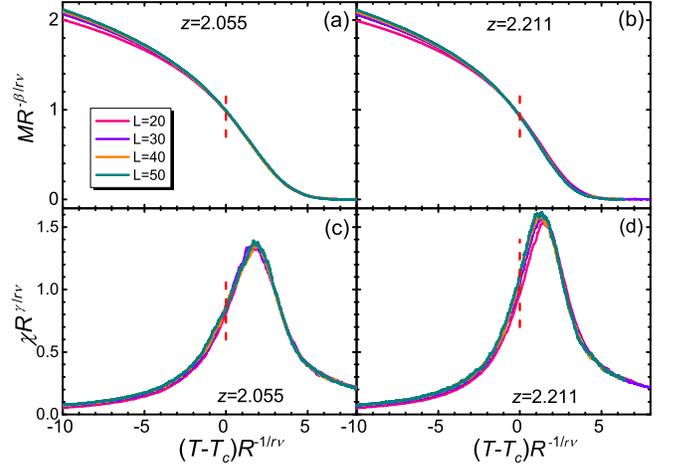,width=1.0\columnwidth}}
\caption{(Color online) FTS of (a) and (b) $M$ and (c) and (d) $\chi$ with fixed $L^{-1}R^{-1/r}\approx0.3541$ for $z=2.055$, (a) and (c), and $z=2.211$, (b) and (d), for the 3D Ising model. Panels in both rows and columns share the same scales and labels. The vertical dashed lines mark $T=T_c$. The legend specifies the lattice sizes used for all the panels.}
\label{heatz}
\end{figure}

We show the scaling collapses of $M$ and $\chi$ on fixed lattice sizes in Fig.~\ref{heatL} and those for fixed $L^{-1}R^{-1/r}$ in Fig.~\ref{heatz}. Note that the rescaled $\chi$ peaks rise with $z$ because of the bigger $r$ from Eq.~(\ref{rzv}). One sees that the collapses for both values of $z$ become better as $L$ increases similar to those in Fig.~\ref{ftsh}, although they are both worse than those for fixed $L^{-1}R^{-1/r}$ in Fig.~\ref{heatz}. Since the curves from the small lattice sizes are absent, the scaling collapses in Figs.~\ref{heatz}(a) and~\ref{heatz}(c) appear far better than Figs.~\ref{ftshlr}(e) and~\ref{ftshlr}(f), all have identical fixed $L^{-1}R^{-1/r}$. It is clear from Figs.~\ref{heatL} and~\ref{heatz} that the collapses of both $M$ and $\chi$ with $z=2.055$ are better than those with $z=2.211$. Therefore, $z$ appears to be around $2.055$ for both heating and cooling from the scaling collapses in the 3D Ising model.

\begin{figure}
\centerline{\epsfig{file=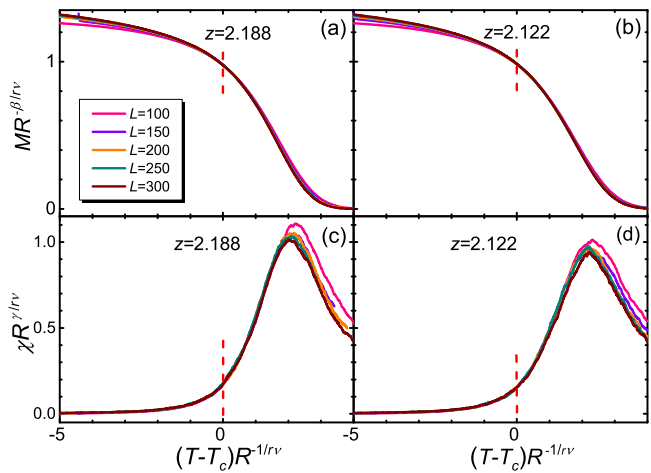,width=1.0\columnwidth}}
\caption{(Color online) FTS of (a) and (b) $M$ and (c) and (d) $\chi$ with fixed $L^{-1}R^{-1/r}\approx0.0758$ for $z=2.188$, (a) and (c), and $z=2.122$, (b) and (d), for the 2D Ising model. Panels in both rows and columns share the same scales and labels. The vertical dashed lines mark $T=T_c$. The legend specifies the lattice sizes used for all the panels.}
\label{heatz2d}
\end{figure}
For comparison, we show in Fig.~\ref{heatz2d} the same collapses for the 2D Ising model. For the 2D model, the $z$ values estimated from the collapses of $\chi'$ at $T_c$ agree with results obtained from $\langle |m|\rangle$ and $\langle m^2\rangle$, which give rise to $z=2.188(2)$ for cooling and $z=2.1223(9)$ for heating. Note the different trends in heating and cooling for the 2D and 3D models. One sees from Fig.~\ref{heatz2d} that the difference between the two $z$ values is only slight, possibly because the two $z$ values differ less than $0.1$. Therefore, from the curve collapses, the 3D $z$ values in heating and cooling appearq
 identical but the 2D ones different.

\section{\label{sum}Summary}
We have studied in detail the dynamic scaling of the Ising model driven by heating or cooling through its critical point on finite-size lattices. Two sets of observable quantities, $\langle m\rangle$ and $\chi$ versus $\langle|m|\rangle$ and $\chi'$, are investigated. We have studied FTS on fixed lattice sizes and FSS at fixed driving rates as well as their full scaling forms. For the 3D Ising model, the FTS is found to be good for $L^{-1}R^{-1/r}<0.3$ or so while the FSS behaves well for about $RL^r<1.6$. Fluctuations in heating and cooling in the FTS regime exhibit qualitatively different behavior. More importantly, we have shown that the two sets of the observables exhibit distinctive scaling behavior. Only one set of the observables displays good FTS on fixed lattice sizes or FSS at fixed driving rates in their respective scaling regimes according to the standard theory, provided that other sub-leading terms and corrections to scaling are ignored. The other set violates FTS or FSS even in their respective scaling regimes. However, when the scaled variables $L^{-1}R^{-1/r}$ or $RL^r$ is fixed, the violated scalings are completely restored.

In order to explain these observations, we need to endow large clusters with physics and separate critical fluctuations into magnitude fluctuations that associated with the forming of the large clusters of the spin-up and the spin-down phases constituting the ordered phase and the so-called phases fluctuations which are the flipping of the large clusters. In equilibrium and in the thermodynamics limit, the large clusters are of the size of the correlation length on average and have roughly the equilibrium magnetization. Thus, at $T\geq T_c$, their magnetization is roughly zero and the picture is the usual picture of fluctuations on all scales as depicted by Kogut~\cite{Kogut}. However, for $T<T_c$ but near $T_c$, they have predominantly up or down spins and hence a finite magnetization, thus acting as the up or down phases. Yet, they are not frozen in but are fluctuating unless $T$ is low. Under driving or on finite-sized lattices, the size of the phases clusters may be controlled by these additional length scales and hysteresis or advance occur. A most direct evidence of the phases fluctuations comes from the FSS regime in which the whole finite system is just one large cluster on average because the correlation length is longer than the system size at least deep inside the regime. In this case the system is no doubt in one dissymmetric phase and changes to the other with time back and forth, which is a transparent picture of the phases fluctuations. Accordingly, $\langle m\rangle$ can be vanishingly small and $\chi$ be large because of the phases fluctuations, whereas $\langle|m|\rangle$ can be finite and $\chi'$ be relatively small because the flipping of the clusters are removed and thus both probe the magnitude fluctuations, although the phases fluctuations also contribute to the magnitude fluctuation. Also, $\langle m\rangle$ may be finite at some range of temperatures depending on initial conditions. Therefore, phases fluctuations lead to the difference between the two sets of the observables. It is the origin of the Kibble-Zurek mechanism since the boundary between the clusters of different phases are just the topological defects.

The scaled variables $L^{-1}R^{-1/r}$ and $RL^r$ must be the only origin of the violated scalings, because once they are fixed, the violations are completely restored, provided that other sub-leading terms and corrections to scaling are negligible. On the one hand, fixing $L^{-1}R^{-1/r}$ in FTS means that lattices of different sizes are driven by different rates $R$ so that every lattice contains the same number of phases clusters of the driven length $R^{-1/r}$ on average. This is the spatial self-similarity of the phases fluctuations. On the other hand, fixing $RL^r$ in FSS indicates that the survival time $L^z$ of the fluctuating phases on every lattice is a fixed fraction of the driven time $R^{-z/r}$ on average. This is the temporal self-similarity of the phases fluctuations. Therefore, fixing the scaled variables is just to maintain the spatial or temporal self-similarity of the phases fluctuations. This is an extrinsic self-similarity that is guaranteed by external conditions such as the system size and the driving rates in contrast to the intrinsic self-similarity controlled by the critical point. Accordingly, it is the breaking of the extrinsic self-similarity of the phases fluctuations that results in the violations of FTS and FSS. In addition, it also gives rise to the qualitatively different behavior in heating and cooling shown in Fig.~\ref{hcxm} because they behave similarly if $L^{-1}R^{-1/r}$ is fixed.

Our numerical results have revealed that the violated scalings can also be rectified even the extrinsic self-similarity of the phases fluctuations is broken. This is achieved by introducing breaking-of-(extrinsic-)self-similarity, abbreviated as Bressy, exponents to collapse the separated rescaled curves. In this regard, the two observables that violate the scaling exhibit different behaviors. One is a primary observable whose scaling function is a power-law of the scaled variable $L^{-1}R^{-1/r}$ or $RL^r$. The other is referred to as the secondary observable whose scaling function possesses a regular term and may thus appear not to violate the scaling. This is why two secondary observables appear parenthesized in Table~\ref{tab1}. Because the scalings are violated only in either the ordered phase or the disordered phase, the Bressy exponents then lead to different leading behavior in the two phases, in stark contrast with the equilibrium critical phenomena in which the leading behaviors of the two phases are identical, only their amplitudes differ, though with universal amplitude ratios. The Bressy exponents, whose values are summarized in Table~\ref{tab1} along with the phases and the observables that violate the scaling, are found to be different for heating and cooling and in FTS and FSS, except possibly for FSS in cooling. They have also different expressions in the 2D and the 3D Ising models, except again for FSS in cooling. This implies that they are most likely new exponents that produce the 2D and 3D values found here. At present, the Bressy exponents of heating for both FSS and FTS are manifest and their absolute values are relatively large, while those of cooling are not so definite and their values are relatively small.

The regime in which Bressy exponents are needed is limited and there exists crossover from this regime to the usual scaling regime. The collapses in finding the Bressy exponents are limited to a certain range of the scaled variable $L^{-1}R^{-1/r}$ or $RL^r$. First, if $RL^r$ ($L^{-1}R^{-1/r}$) is too large, crossover from the FSS (FTS) regime to the FTS (FSS) regime occurs and the FSS (FTS) fails. The collapses cannot therefore be good. Then, if $R=0$ or $L=\infty$, the usual FSS or FTS, respectively, must recover. Except these special points, the Bressy-exponent dominated regime is believed to be valid for arbitrarily small $RL^r$ ($L^{-1}R^{-1/r}$) for the FSS in both heating and cooling (the FTS in cooling). Accordingly, the crossovers from this regime to the usual FSS and FTS are believed to occur at $RL^r=0$ and $L^{-1}R^{-1/r}=0$ abruptly. For FTS in heating, the crossover appears to happen abruptly at a finite $L^{-1}R^{-1/r}$ as seen in Fig.~\ref{ftsh3d}(a).

We have also studied cooling in an external field whose magnitude fixes $X=HR^{-\beta\delta/r\nu}$ in order to study its effect on the phases fluctuations. The field is applied in three protocols, viz., only below $T_c$ (A), only above $T_c$ (B), and during the whole process (C). The most unexpected result discovered is that the Bressy exponent of FTS in heating can describe the three cases in which scaling is poor, $X=2$ in protocol A and $X=0.05$ in both protocol B and protocol C. The reason for these surprising heating-exponent-describing-cooling-processes are that the field breaks the up-down symmetry of the two low-temperature phases just as the initial condition does in heating. This shows that the essence of the difference between heating and cooling is the symmetry of the system states, namely whether they are ordered or disordered. It also indicates that the Bressy exponents are not just applicable to the four cases of FTS and FSS in heating and cooling but can have a wider application. Conversely, this result provides another distinctive source to the Bressy exponents and thus further confirms their validity.

The scalings of the three different protocols show that the phases fluctuations are more important at and just above $T_c$ than those below it. For the large $X$ on the one hand, the revised FTS is not needed in the whole process if the field is applied above $T_c$ in protocol B, while it is only needed above $T_c$ if the field is applied below $T_c$. For the small $X$ on the other hand, the small applied field below $T_c$ in protocol A hardly alter the scaling there while that above $T_c$ in protocol B changes the behavior below it. This, together with the fact that the zero-field heating Bressy exponent describes well the field-cooling processes, shows that an ordered state can suppress the phases fluctuations in cooling and acts like the heating with an ordered state, which leads to the qualitative difference between heating and cooling.

We have also verified that there exists a revised FTS regime described by Eq.~(\ref{rescaling_huang}) in between the two end regimes of FTS and FSS even in the presence of an external field. It is the crossover that is responsible from the poor scalings of the large lattice size in protocols B and C. The range of the crossover from this regime to the FTS regime in a field is estimated to be roughly $1.87\xi_H\lesssim L\lesssim7.5\xi_H$ for a fixed $X$. For larger values of $L$, the FTS in a field shows, while for smaller $L$ but not yet reaching the FSS regime, the revised FTS regime comes out. In other words, the revised FTS regime appears when $\xi_H$ is about the size of $L$, both being larger than $\xi_R$. In addition, the end FTS regime in a field on large lattices in cooling has also been shown to exhibit a special feature different from that in heating. In cooling, the scaling function of the order parameter must be proportional to $X$, the scaled variable of the field, whereas it can be a constant when the field is vanishingly small in heating. Yet another result is that the revised FTS is never needed for $\langle m\rangle$ in the presence of a small external field in the revised FTS regime.

Besides, in order to investigate whether heating and cooling have a different dynamic critical exponent $z$, we have shown FTS collapses in heating in the absence of an externally applied field using different values of $z$. The quality of curve collapses shows that the 3D $z$ values in heating and cooling appear identical but the 2D ones different, possibly because of the difference in 2D is too small to be resolved by the method.

Half a century have passed since the renormalization-group theory in which critical phenomena culminate was set up. Still, we have seen that, through a simple driving, a series of driven nonequilibrium critical phenomena centring on the violations of the usual FSS and FTS in one phase only and hence different leading behavior of the two phases emerge with novel ideas of the phases fluctuations and self-similarity breaking together with its associated critical exponents. Yet, these are just the tip of an iceberg and a lot of problems are yet to be resolved such as how the self-similarity breaking results in the breaking of self-similarity (Bressy) exponents, what are the Bressy exponents, why in some cases it is one set while in other cases it is the other set of the observables that violate the scaling and the similar questions about the primary and secondary observables and the phase where scaling is violated, how exactly the crossover between the usual scaling and the self-similarity-breaking controlled regimes occurs and so on.

In retrospect, that FTS in general and the KZ scaling in particular and even the revised FTS involve only the equilibrium critical exponents is specific and rather fortunate, given that critical phenomena are generically nonlinear owing to the generically non-integer critical exponents and that nonequilibrium occurs under driving~\cite{Zhong06}. We have demonstrated that when the lattice size and an external field are considered, new exponents are generally required for the scaling in the whole driving process. Although these exponents and many other problems are yet to be understood, our findings open a door in critical phenomena and suggest that much is yet to be explored in driven nonequilibrium critical phenomena.

\begin{acknowledgments}
This work was supported by the National Natural Science Foundation of China (Grant No. 11575297).
\end{acknowledgments}




\begin{thebibliography}{99}
\bibitem{Greiner}M. Greiner, O. Mandel, T. Esslinger, T. W. H\"{a}nsch, and I. Bloch, Quantum phase transition from a superfluid to a Mott insulator in a gas of ultracold atoms, Nature {\bf 415}, 39 (2002).
\bibitem{Kinoshita}T. Kinoshita, T. Wenger, and D. S. Weiss. A quantum Newton's cradle, Nature {\bf 440}, 900 (2006).
\bibitem{Hofferberth}S. Hofferberth, I. Lesanovsky, B. Fischer, T. Schumm, and J. Schmiedmayer, Non-equilibrium coherence dynamics in one-dimensional Bose gases, Nature {\bf 449}, 324 (2007).
\bibitem{Zhang}X. Zhang, C-L. Hung, S-K. Tung, and C. Chin, Observation of quantum criticality with ultracold atoms in optical lattices, Science {\bf 335}, 1070 (2012).
\bibitem{Feng} B. Feng, S. Yin, and F. Zhong, Theory of driven nonequilibrium critical phenomena, Phys. Rev. B {\bf 94}, 144103 (2016).
\bibitem{KZ1} T. W. B Kible, Topology of cosmic domains and strings, J. Phys. A {\bf 9}, 1387 (1976).
\bibitem{Kibble2} T. Kibble, Phase-transition dynamics in the lab and the universe, Phys. Today {\bf 60} (9), 47 (2007).
\bibitem{KZ2} W. H. Zurek, Cosmological experiments in superfluid helium? Nature (London) {\bf 317}, 505 (1985).
\bibitem{KZ3} W. H. Zurek, Cosmological experiments in condensed matter systems, Phys. Rep. {\bf 276}, 177 (1996).
\bibitem{KZp1} P. Laguna and W. H. Zurek, Density of kinks after a quench: When symmetry breaks, how big are the pieces? Phys. Rev. Lett. {\bf 78}, 2519 (1997).
\bibitem{KZp2} A. Yates and W. H. Zurek, Vortex formation in two dimensions: When symmetry breaks, how big are the pieces? Phys. Rev. Lett. {\bf 80}, 5477 (1998).
\bibitem{KZp3} N. D. Antunes, L. M. A. Bettencourt, and W. H. Zurek, Is domain formation decided before or after the transition? Phys. Rev. Lett. {\bf 82}, 2824 (1999).
\bibitem{KZp4} G. J. Stephens, E. A. Calzetta, B. L. Hu, and S. A. Ramsey, Defect formation and critical dynamics in the early Universe, Phys. Rev. D {\bf 59}, 045009 (1999).
\bibitem{KZp5}  S. Suzuki, Cooling dynamics of pure and random Ising chains, J. Stat. Mech. (2009) P03032.
\bibitem{delcamp} A. del Campo, Universal statistics of topological defects formed in a quantum phase transition, Phys. Rev. Lett. {\bf 121}, 200601 (2018).
\bibitem{Gomez}F. J. G\'{o}mez-Ruiz, J. J. Mayo, and A. del Campo, Full counting statistics of topological defects after crossing a phase transition, arXiv:1912.04679 (2019).

\bibitem{inexper4} A. del Campo and W. H. Zurek, Universality of phase transition dynamics: Topological defects from symmetry breaking, Int. J. Mod. Phys. A {\bf 29}, 1430018 (2014).
\bibitem{Bray} A. J. Bray, Theory of phase-ordering kinetics, Adv. Phys. {\bf 43}, 357 (1994).
\bibitem{Biroli}G. Biroli, L. F. Cugliandolo, and A. Sicilia, Kibble-Zurek mechanism and infinitely slow annealing through critical points, Phys. Rev. E. {\bf 81}, 050101(R) (2010).
\bibitem{fss}M. E. Fisher and M. N. Barber, Scaling theory for finite-size effects in the critical region, \prl{\bf 28,} 1516 (1972).
\bibitem{Barber}M. N. Barber, Finite-size scaling, in {\it Phase Transitions and Critical Phenomena}, edited by C. Domb and J. Lebowitz (Academic, New
York, 1983), Vol. 8.
\bibitem{Cardy}J. Cardy, ed. {\it Finite Size Scaling} (North-Holland, Amsterdam, 1988).
\bibitem{Privman}V. Privman, ed. {\it Finite Size Scaling and Numerical Simulations of Statistical Systems} (World Scientific, Singapore, 1990).
\bibitem{Zhong1} S. Gong, F. Zhong, X. Huang, and S. Fan, Finite-time scaling via linear driving, New J. Phys. {\bf 12}, 043036 (2010).
\bibitem{Zhong2} F. Zhong, Finite-time Scaling and its Applications to Continuous Phase Transitions, in \textit{Applications of Monte Carlo Method in Science and Engineering}, edited by S. Mordechai (Intech, Rijeka, Croatia, 2011), p. 469. Available at
    http://www.dwz.cn/B9Pe2
\bibitem{Zhong06}F. Zhong, Probing criticality with linearly varying external fields: Renormalization group theory of nonequilibrium critical dynamics under driving, Phys. Rev. E {\bf 73}, 047102 (2006).
\bibitem{Yin} S. Yin, X. Qin, C. Lee, and F. Zhong, Finite-time scaling of dynamic quantum criticality, arXiv: 1207.1602 (2012).
\bibitem{Yin3} S. Yin, P. Mai, and F. Zhong, Nonequilibrium quantum criticality in open systems: The dissipation rate as an additional indispensable scaling variable, Phys. Rev. B {\bf 89}, 094108 (2014).

\bibitem{Huang} Y. Huang, S. Yin, B. Feng, and F. Zhong, Kibble-Zurek mechanism and finite-time scaling, Phys. Rev. B {\bf 90}, 134108 (2014).
\bibitem{Liu} C.-W. Liu, A. Polkovnikov, and A. W. Sandvik, Dynamic scaling at classical phase transitions approached through nonequilibrium quenching, Phys. Rev. B {\bf 89}, 054307 (2014).


\bibitem{Liupre}C. W. Liu, A. Polkovnikov, A. W. Sandvik, and A. P. Young, Universal dynamic scaling in three-dimensional Ising spin glasses, Phys. Rev. E {\bf 92}, 022128 (2015).
\bibitem{Liuprl}C. W. Liu, A. Polkovnikov, and A. W. Sandvik, Quantum versus classical annealing: Insights from scaling theory and results for spin glasses on 3-regular graphs, Phys. Rev. Lett. {\bf 114}, 147203 (2015).

\bibitem{Pelissetto}A. Pelissetto and E. Vicari, Off-equilibrium scaling behaviors driven by time-dependent external fields in three-dimensional $\mathrm{O}(N)$ vector models, Phys. Rev. E {\bf 93}, 032141 (2016).
\bibitem{Xu}N. Xu, C. Castelnovo, R. G. Melko, C. Chamon, and A. W. Sandvik, Dynamic scaling of topological ordering in classical systems, Phys. Rev. B {\bf 97}, 024432 (2018).
\bibitem{Xue}M. Xue, S. Yin, and L. You, Universal driven critical dynamics across a quantum phase transition in ferromagnetic spinor atomic Bose-Einstein condensates, Phys. Rev. A {\bf 98}, 013619 (2018).
\bibitem{Cao}X. Cao, Q. Hu, and F. Zhong, Scaling theory of entanglement entropy in confinements near quantum critical points, Phys. Rev. B {\bf 98}, 245124 (2018).
\bibitem{Gerster}M. Gerster, B. Haggenmiller, F. Tschirsich, P. Silvi, and S. Montangero, Dynamical Ginzburg criterion for the quantum-classical crossover of the Kibble-Zurek mechanism, Phys. Rev. B {\bf 100}, 024311 (2019).
\bibitem{Li}Y. Li, Z. Zeng, and F. Zhong, Driving driven lattice gases to identify their universality classes, Phys. Rev. E {\bf 100}, 020105(R) (2019).
\bibitem{Mathey}S. Mathey and S. Diehl, Activating critical exponent spectra with a slow drive, Phys. Rev. Res. {\bf 2}, 013150 (2020).
\bibitem{Clark}L. W. Clark, L. Feng, and C. Chin, Universal space-time scaling symmetry in the dynamics of bosons across a quantum phase transition, Science {\bf 354}, 606 (2016).
\bibitem{Keesling}A. Keesling, A. Omran, H. Levine, H. Bernien, H. Pichler, S. Choi, R. Samajdar, S. Schwartz, P. Silvi, S. Sachdev, P. Zoller, M. Endres, M. Greiner, V. Vuleti\'{c}, and M. D. Lukin, Quantum Kibble–Zurek mechanism and critical dynamics on a programmable Rydberg simulator, Nature (London), {\bf 568}, 207 (2019).
\bibitem{Huang2} Y. Huang, S. Yin, Q. Hu, and F. Zhong, Kibble-Zurek mechanism beyond adiabaticity: Finite-time scaling with critical initial slip, Phys. Rev. B {\bf 93}, 024103 (2016).

\bibitem{Yuan} W. Yuan, S. Yin, and F. Zhong, Self-similarity breaking: Anomalous nonequilibrium finite-size scaling and finite-time scaling, Chin. Phys. Lett. {\bf 38}, 026401 (2021) [arXiv:2004.08041 (2020)].
\bibitem{Mandelbrot}B. B. Mandelbrot, {\it The Fractal Geometry of Nature} (Freeman, New York, 1983).
\bibitem{Meakin}P. Meakin, {\it Fractal, Scaling and Growth Far From Equilibrium} (Cambridge, Cambridge, 1998).

\bibitem{FSS1} E. Br\'ezin, An investigation of finite size scaling, J. de Phys. {\bf 43}, 15 (1982).
\bibitem{FSS2} E. Br\'ezin and J. Zinn-Justin, Finite size effects in phase transitions, Nucl. Phys. B {\bf 257}, 867 (1985).
\bibitem{Zhong} F. Zhong and Q. Z. Chen, Theory of the dynamics of first-order phase transitions: Unstable fixed points, exponents, and dynamical scaling, Phys. Rev. Lett. {\bf 95}, 175701 (2005).
\bibitem{Hohenberg} P. C. Hohenberg and B. I. Halperin, Theory of dynamic critical phenomena, Rev. Mod. Phys. {\bf 49}, 435 (1977).

\bibitem{Mask}S. K. Ma, {\it Modern Theory of Critical Phenomena} (W. A. Benjamin, Inc., Canada, 1976).

\bibitem{Fisher} M. E. Fisher, {\it Scaling, Universality and Renormalization Group Theory}, Lecture notes presented at the ``Advanced Course on Critical Phenomena" (The Merensky Institute of Physics, University of Stellenbosch, South Africa, 1982).
\bibitem{Cardyb}J. Cardy, {\it Scaling and Renormalization in Statistical Physics} (Cambridge University Press, Cambridge, 1996).
\bibitem{Wegner} F. W. Wegner, Corrections to scaling laws, Phys. Rev. B {\bf 5,} 4529 (1972).
\bibitem{Kogut} J. B. Kogut, An introduction to lattice gauge theory and spin systems, Rev. Mod. Phys. {\bf 51}, 659 (1979).
\bibitem{Suzuki} M. Suzuki, Phase transition and fractals, Prog. Theor. Phys. {\bf 69}, 65 (1983).

\bibitem{exponent2}  A. M. Ferrenberg and D. P. Landau, Critical behavior of the three-dimensional Ising model: A high-resolution Monte Carlo study, Phys. Rev. B {\bf 44}, 5081 (1991).
\bibitem{Tc}  A. Pelissetto and E. Vicari, Critical phenomena and renormalization-group theory, Phys. Rep. {\bf 368}, 549 (2002).
\bibitem{exponent1} H. Kleinert, Critical exponents from seven-loop strong-coupling ${\ensuremath{\varphi}}^{4}$ theory in three dimensions, Phys. Rev. D {\bf 60}, 085001 (1999).

\bibitem{z3d1} P. Grassberger, Damage spreading and critical exponents for model A Ising dynamics, Physica A. {\bf 214}, 547 (1995).
\bibitem{z3d2} M. Kikuchi and N. Ito, Statistical dependence time and its application to dynamical critical exponent, J. Phys. Soc. Japan {\bf 62}, 3052 (1993).
\bibitem{MC}N. Metropolis, A. W. Rosenbluth, M.N. Rosenbluth, A. M. Teller, and E. Teller, Equation of state calculations by fast computing machines, J. Chem. Phys. {\bf 21}, 1087 (1953).
\bibitem{Glauber} R. J. Glauber, Time-dependent statistics of the Ising Model, J. Math. Phys. {\bf 4}, 294 (1963).
\bibitem{landaubinder}D. P. Landau and K. Binder, {\it A Guide to Monte Carlo Simulations in Statistical
Physics} 2nd edn (Cambridge University Press, Cambridge, 2005).
\bibitem{Landau76} D. P. Landau, Finite-size behavior of the Ising square lattice, Phys. Rev. B {\bf 7}, 2997 (1976).
\bibitem{Bray+h} J. A. N. Filipe,  A. J. Bray, and  S. Puri, Phase-ordering kinetics with external fields and biased initial conditions, Phys. Rev. E {\bf 52}, 6082 (1995).

\end{thebibliography}
\end{document}